\documentclass[12pt,a4paper]{article}

\pdfoutput=1
\usepackage{cite}
\usepackage{amsmath,amssymb,amsfonts}
\usepackage{calc}
\usepackage{hyperref}
\usepackage[pdftex]{color,graphicx}
\usepackage{array}

\usepackage{bookmark}

\newcommand{\cR}{\mathcal{R}}

\def\no{\nonumber}

\numberwithin{equation}{section}
\newlength{\dinwidth}
\newlength{\dinmargin}
\newcommand{\bel}[1]{\be\label{#1}}

\newcommand{\be}{\begin{equation}}
\newcommand{\ee}{\end{equation}}

\def\beqn{\begin{eqnarray}}
\def\ba{\begin{array}{c}}
\def\bat{\begin{array}{cc}}
\def\bat{\begin{array}{cc}}
\def\ea{\end{array}}
\def\bat{\begin{array}{cc}}
\def\batt{\begin{array}{ccc}}
\def\eeqn{\end{eqnarray}}

\definecolor{nicered}{rgb}{1.0,0.0,0.2}

\definecolor{color1}{rgb}{0.9,.4,.2}

\definecolor{color2}{rgb}{0.3,.6,.7}

\definecolor{color3}{rgb}{0.7,.2,.7}

\usepackage{color}

\usepackage{multirow}

\usepackage{xcolor,colortbl}

\definecolor{Gray}{gray}{0.95}
\definecolor{RGray}{gray}{0.85}
\definecolor{CGray}{gray}{0.92}

\newcolumntype{a}{>{\columncolor{Gray}}c}
\newcolumntype{b}{>{\columncolor{white}}c}

\begin{document}

\title{
\vspace*{-1.5cm}
\begin{flushright}\vbox{\normalsize LMU-ASC 11/15 \\[-3pt] FTUV/15$-$0322 \\[-3pt] IFIC/15$-$12}
\end{flushright}\vskip 30pt
{\LARGE \bf Flavour-changing top decays\\[-2pt] in the aligned two-Higgs-doublet model}}
\bigskip

\author{Gauhar Abbas$^{1}$, Alejandro Celis$^{2}$, Xin-Qiang Li$^{3,4}$, Jie Lu$^{5}$ and Antonio Pich$^{1}$\\[15pt]
{$^1$\small IFIC, Universitat de Val\`encia -- CSIC, Apt. Correus 22085, E-46071 Val\`encia, Spain}\\[0.2cm]
 $^2$\small Ludwig-Maximilians-Universit\"at M\"unchen, Fakult\"at f\"ur Physik, \\
     \small Arnold Sommerfeld Center for Theoretical Physics, 80333 M\"unchen, Germany\\[0.2cm]
 $^3$\small Institute of Particle Physics and Key Laboratory of Quark and Lepton Physics~(MOE), \\
     \small Central China Normal University, Wuhan, Hubei 430079, China\\[0.2cm]
 $^4$\small State Key Laboratory of Theoretical Physics, Institute of Theoretical Physics, \\
     \small Chinese Academy of Sciences, Beijing 100190, China\\[0.2cm]
{$^5$\small Department of Physics, Shanghai University, Shanghai 200444, China}}

\date{}
\maketitle
\vspace{1.5cm}

\centerline{\large\bf Abstract}
\begin{quote}
We perform a complete one-loop computation of the two-body flavour-changing top decays $t \rightarrow c h$ and $t \rightarrow c V$ ($V= \gamma, Z$), within the aligned two-Higgs-doublet model. We evaluate the impact of the model parameters on the associated branching ratios, taking into account constraints from flavour data and measurements of the Higgs properties. Assuming that the $125$~GeV Higgs corresponds to the lightest CP-even scalar of the CP-conserving aligned two-Higgs-doublet model, we find that the rates for such flavour-changing top decays lie below the expected sensitivity of the future high-luminosity phase of the LHC. Measurements of the Higgs signal strength in the di-photon channel are found to play an important role in limiting the size of the $t \rightarrow c h$ decay rate when the charged scalar of the model is light.
\end{quote}

\renewcommand{\thefootnote}{\arabic{footnote}}\setcounter{footnote}{0}
\setcounter{page}{0}
\thispagestyle{empty}

\newpage


\section{Introduction }

The discovery of a new boson, with a mass close to $125$~GeV, by the ATLAS~\cite{Aad:2012tfa} and CMS~\cite{Chatrchyan:2012ufa} collaborations stands as a remarkable success of the Standard Model (SM) of electroweak interactions. The properties of this boson are so far in agreement with those of the SM Higgs, indicating that the new particle is indeed associated with the mechanism of electroweak symmetry breaking (EWSB). A characteristic feature of the SM is the absence of flavour-changing neutral-current (FCNC) interactions at tree-level. FCNCs are generated through quantum loop corrections in the SM, but they are strongly suppressed by the Glashow--Iliopoulos--Maiani (GIM) mechanism~\cite{Glashow:1970gm}.

A widely studied enlargement of the electroweak theory consists in adding a second scalar doublet to the SM field content. The so-called two-Higgs-doublet model (2HDM) represents a minimal extension of the SM scalar sector that easily accommodates electroweak precision data and leads to a very rich phenomenology~\cite{Branco:2011iw}. In the most general version of the 2HDM unwanted FCNCs appear at tree-level, which represents a major shortcoming of the model. The hypothesis of natural flavour conservation (NFC) is
the usual way out to this issue. By limiting the number of scalar doublets coupling to a given type of right-handed fermion to be at most one, the absence of dangerous FCNCs is guaranteed~\cite{Glashow:1976nt,Paschos:1976ay}. A more general solution is that of Yukawa alignment~\cite{Pich:2009sp}. The aligned two-Higgs-doublet model (A2HDM) assumes that the two Yukawa matrices coupled to the same type of right-handed fermion are aligned in flavour space, so that no FCNCs appear at tree level. Explicit models where a Yukawa aligned structure arises due to an underlying symmetry have been discussed in refs.~\cite{Serodio:2011hg,Cree:2011uy,Varzielas:2011jr,Celis:2014zaa,Botella:2015yfa}.   Interestingly, all different versions of the 2HDM with NFC are recovered as particular limits of the A2HDM. Constraints on the A2HDM from flavour and collider data have been analyzed in refs.~\cite{Jung:2010ik,Jung:2012vu,Celis:2012dk,Jung:2013hka,Li:2014fea,Dekens:2014jka} and~\cite{Altmannshofer:2012ar,Bai:2012ex,Celis:2013rcs,Barger:2013ofa,Lopez-Val:2013yba,Duarte:2013zfa,Celis:2013ixa,Wang:2013sha}, respectively, extracting relevant bounds on the model parameters.

In this work we study the flavour-changing top-quark decays $t\rightarrow ch$ and $t\rightarrow c V$ ($V=\gamma, Z$), within the framework of the A2HDM. They arise at the loop level and are strongly suppressed in the SM, due to the GIM mechanism. A significant enhancement can be achieved in alternative scenarios of EWSB, making these processes a suitable place to look for new physics beyond the SM. Early considerations of these effects were done in refs.~\cite{Eilam:1989zm,DiazCruz:1989ub,Eilam:1990zc,Hou:1991un,Atwood:1996vj,Mele:1998ag}.  A concise review of the flavour-changing top-decay phenomenology can be found in ref.~\cite{AguilarSaavedra:2004wm}. In the A2HDM these decays receive additional charged Higgs contributions at the one-loop level which could lift the decay rates.

Comprehensive analyses of flavour-changing top decays within 2HDMs with NFC have been done in refs.~\cite{Eilam:1990zc,Atwood:1996vj,Bejar:2000ub,Arhrib:2005nx}.   However, these rare processes have not been investigated yet within the more general setting of the A2HDM. Furthermore, since these studies were performed, considerable experimental progress in our understanding of the EWSB mechanism has been made, translating into tight constraints on possible extensions of the SM scalar sector.

Searches for flavour-changing top decays have been performed recently by the ATLAS~\cite{Aad:2012ij,Aad:2014dya} and CMS collaborations~\cite{Chatrchyan:2013nwa,CMS:2014qxa,CMS:2014hwa}, placing limits on the associated branching ratios. An overview of the current experimental status can be found in ref.~\cite{Goldouzian:2014xfa}. With the large amount of data that will be collected in the future LHC runs, it is expected that these bounds will be improved by at least one order of magnitude~\cite{AguilarSaavedra:2000aj,Agashe:2013hma}.

Our paper is organized as follows. In section~\ref{sec:frame} we introduce the A2HDM.  Flavour-changing top decays are discussed in section \ref{sec:topdecays}, which presents the results of our calculations. A phenomenological analysis of these processes is given in section~\ref{sec:pheno}, and our conclusions are finally summarized in section~\ref{concl}. Explicit analytical results for the relevant decay amplitudes are given in the appendices.

\section{Framework}
\label{sec:frame}

The 2HDM extends the SM scalar sector with an additional complex scalar doublet. In the Higgs basis, where only one doublet acquires vacuum expectation value, the scalar fields are parametrized by~\cite{Celis:2013rcs}
\begin{equation}  \label{Higgsbasisintro}
\Phi_1=\left[ \begin{array}{c} G^+ \\ \frac{1}{\sqrt{2}}\, (v+S_1+iG^0) \end{array} \right] \; ,
\qquad\qquad
\Phi_2 = \left[ \begin{array}{c} H^+ \\ \frac{1}{\sqrt{2}}\, (S_2+iS_3)   \end{array}\right] \; ,
\end{equation}
with $v = ( \sqrt{2} G_F )^{-1/2} \simeq 246$~GeV. Here $G^{0, \pm}$ correspond to the would-be Goldstone bosons, giving mass to the gauge vector bosons, while $H^{\pm}$ is a charged Higgs. The scalar spectrum also contains three neutral Higgs bosons $\varphi_j^0(x)=\{h(x),H(x),A(x)\}$, given by $\varphi_j^0 = \cR_{jk}  S_k $, where $\cR$ is an orthogonal matrix obtained after diagonalizing the mass terms in the scalar potential~\cite{Celis:2013rcs}. In general none of the neutral Higgs bosons are CP eigenstates.

\subsection{Scalar sector}

The most general scalar potential allowed by the electroweak gauge symmetry can be written as
\beqn \label{eq:potentialintro}
V & = &  \mu_1\; \Phi_1^\dagger\Phi_1\, +\, \mu_2\; \Phi_2^\dagger\Phi_2 \, +\, \left[\mu_3\; \Phi_1^\dagger\Phi_2 \, +\, \mu_3^*\; \Phi_2^\dagger\Phi_1\right]
\no\\[0.2cm] & + & \lambda_1\, \left(\Phi_1^\dagger\Phi_1\right)^2 \, +\, \lambda_2\, \left(\Phi_2^\dagger\Phi_2\right)^2 \, +\,
\lambda_3\, \left(\Phi_1^\dagger\Phi_1\right) \left(\Phi_2^\dagger\Phi_2\right) \, +\, \lambda_4\, \left(\Phi_1^\dagger\Phi_2\right) \left(\Phi_2^\dagger\Phi_1\right)
\no\\[0.2cm] & + & \biggl[  \left(\lambda_5\; \Phi_1^\dagger\Phi_2 \, +\,\lambda_6\; \Phi_1^\dagger\Phi_1 \, +\,\lambda_7\; \Phi_2^\dagger\Phi_2\right) \left(\Phi_1^\dagger\Phi_2\right)
\, +\, \mathrm{h.c.}\,\biggr]\,.
\eeqn
Due to the Hermiticity of the scalar potential, all parameters are real with the exception of $\mu_3, \lambda_5, \lambda_6$ and $\lambda_7$. The minimization conditions impose the relations
\bel{eq:minimumcond}
\mu_1\; =\; -\lambda_1\, v^2\, ,
\qquad\qquad\qquad
\mu_3\; =\; -\frac{1}{2}\,\lambda_6\, v^2\, .
\ee
The mass of the charged Higgs is given by
\bel{eq:MHp}
M_{H^\pm}^2\; =\; \mu_2 + \frac{1}{2}\,\lambda_3\, v^2 \,,
\ee
while those of the neutral scalars have been obtained in ref.~\cite{Celis:2013rcs} to first order in the CP-violating parameters. In the CP-conserving limit the neutral Higgs bosons are CP eigenstates. The CP-odd field $A$ corresponds to $S_3$ and the CP-even states are orthogonal combinations of $S_1$ and $S_2$:
\bel{eq:mixing}
\left(\ba h\\ H\ea\right)\; = \;
\left[\bat \cos{\tilde\alpha} & \sin{\tilde\alpha} \\ -\sin{\tilde\alpha} & \cos{\tilde\alpha}\ea\right]\;
\left(\ba S_1\\ S_2\ea\right) \, .
\ee
Here $M_h \leqslant M_H$ by convention and the mixing angle $\tilde \alpha$ is determined by
\beqn \label{eq:wclimit}
\sin 2 \tilde \alpha\; =\;   \frac{-2 \lambda_6 v^2  }{  M_H^2 - M_h^2   }    \,,
\qquad\qquad
\cos 2 \tilde \alpha\; =\;   \frac{ M_A^2 + 2 (\lambda_5 - \lambda_1) v^2   }{  M_H^2 - M_h^2   }  \,.
\eeqn
In the CP-conserving limit the masses of the neutral Higgs bosons are given by
\bel{eq:Higgs_masses}
M_h^2\; =\;\frac{1}{2}\,\left( \Sigma-\Delta\right) ,
\quad\quad
M_H^2\; =\;\frac{1}{2}\,\left( \Sigma+\Delta\right) ,
\quad\quad
M_A^2 \; =\; M_{H^\pm}^2\, +\, v^2\,\left(\frac{\lambda_4}{2} - \lambda_5\right) ,
\ee
where
\beqn\label{eq:Sig}
\Sigma & \;=\;&    M_{H^{\pm}}^{2}   +  \left( 2 \lambda_1 +  \frac{ \lambda_4 }{2}  + \lambda_5 \right)  v^2   \, ,
\\[0.2cm] \label{eq:Delta}
\Delta & \;=\;&\sqrt{\left[  M_A^2 + 2  (\lambda_5- \lambda_1) v^2    \right]^2 + 4 v^4 \lambda_6^2}\, .
\eeqn
By performing a phase redefinition of the CP-even fields one can restrict the mixing angle to the range $0 \leqslant \tilde \alpha < \pi$. Moreover, the Higgs basis of the CP-conserving 2HDM is defined up to a global rephasing of the second Higgs doublet $\Phi_2 \rightarrow - \Phi_2$~\cite{O'Neil:2009nr,Asner:2013psa}. Without loss of generality, one can then fix the sign of $\lambda_6$; by convention, we choose $\lambda_6 \leqslant 0$ so that $0 \leqslant   \tilde \alpha \leqslant \pi/2$.

\subsection{Yukawa sector}

In the A2HDM, the interactions of the physical scalar fields with fermions are described by~\cite{Pich:2009sp}
\beqn \label{lagrangianY}
 \mathcal L_Y & = &  - \frac{\sqrt{2}}{v}\; H^+ \Bigl\{ \bar{u} \left[ \varsigma_d\, V M_d \, P_R - \varsigma_u\, M_u^\dagger V  \, P_L \right]  d\, + \, \varsigma_l\, \bar{\nu} M_l \, P_R l \Bigr\}
\nonumber \\[0.2cm]
& & -\,\frac{1}{v}\; \sum_{\varphi^0_i, f}\, y^{\varphi^0_i}_f\, \varphi^0_i  \; \left[\bar{f}\,  M_f \, P_R  f\right]
\; + \;\mathrm{h.c.} \, ,
\eeqn
where $P_{L,R} = (1 \mp \gamma_5)/2$ are the chirality projectors, $M_{f=u,d,l}$ represent the diagonal fermion mass matrices, $V$ is the Cabibbo--Kobayashi--Maskawa (CKM)~\cite{Cabibbo:1963yz,Kobayashi:1973fv} matrix, and
\begin{equation}  \label{yukascal}
y_{d,l}^{\varphi^0_i} = \cR_{i1} + (\cR_{i2} + i\,\cR_{i3})\,\varsigma_{d,l}  \, ,
\qquad\qquad
y_u^{\varphi^0_i} = \cR_{i1} + (\cR_{i2} -i\,\cR_{i3}) \,\varsigma_{u}^* \, .
\end{equation}
The parameters $\varsigma_{f}$ ($f=u,d,l$) are family-universal complex quantities which introduce new sources of CP violation beyond the CKM matrix. For particular real values of these parameters, indicated in table~\ref{tab:NFC}, one recovers all different versions of the 2HDM with NFC.

\begin{table}[t]
\begin{center}
\caption{\it \small Two-Higgs-doublet models with natural flavour conservation.}
\vspace{0.2cm} \tabcolsep 0.15in
\begin{tabular}{|c|c|c|c|}
\hline \rowcolor{RGray}
Model & $\varsigma_d$ & $\varsigma_u$ & $\varsigma_l$  \\[0.1cm]
\hline
Type I  & $\cot{\beta}$ &$\cot{\beta}$ & $\cot{\beta}$ \\[0.1cm]   \rowcolor{Gray}
Type II & $-\tan{\beta}$ & $\cot{\beta}$ & $-\tan{\beta}$ \\[0.1cm]
Type X (lepton specific)  & $\cot{\beta}$ & $\cot{\beta}$ & $-\tan{\beta}$ \\[0.1cm] \rowcolor{Gray}
Type Y (flipped) & $-\tan{\beta}$ & $\cot{\beta}$ & $\cot{\beta}$ \\[0.1cm]
\hline
\end{tabular}
\label{tab:NFC}
\end{center}
\end{table}

The Yukawa alignment condition is not stable against quantum corrections~\cite{Pich:2009sp,Jung:2010ik,Ferreira:2010xe}. However, the flavour symmetries of the A2HDM constrain tightly the possible FCNC effects, keeping them well below present experimental bounds~\cite{Pich:2009sp,Braeuninger:2010td,Bijnens:2011gd,Cvetic:1997zd,Cvetic:1998uw}. The only FCNC local structures induced at one-loop take the form~\cite{Jung:2010ik}
\begin{align} \label{fcnc:ct}
\mathcal{L}_{\mbox{\scriptsize{FCNC}}} \;=\;  & \frac{\mathcal{C}}{  4 \pi^2 v^3 }\,   (1 + \varsigma_u^* \varsigma_d )     \sum_j  \, \varphi_j^0 \, \biggl\{    (  \mathcal{R}_{j2}  + i \mathcal{R}_{j3} ) (  \varsigma_d - \varsigma_u ) \Bigl[ \bar d_L \, V^{\dag}  M_u M_{u}^{\dag} V M_d \, d_R    \Bigr]   \nonumber \\
& - (\mathcal{R}_{j2} - i \mathcal{R}_{j3})  ( \varsigma_d^* - \varsigma_u^* ) \Bigl[  \bar u_L \, V M_d M_d^{\dag}   V^{\dag}  M_u u_R  \Bigr]  \biggr\} + \mathrm{h.c.} \,,
\end{align}
which vanishes exactly for the 2HDMs with NFC. In general, the size of the induced flavour-changing interactions is controlled by three powers of quark masses and the GIM mechanism.

The renormalization of the coupling constant $\mathcal{C}$ is determined, using dimensional regularization, to be~\cite{Li:2014fea}
\be
\mathcal{C} = \mathcal{C}_{R}(\mu)  + \frac{1}{2}  \left\{ \frac{  2 \mu^{D-4}}{D-4}   + \gamma_E - \ln(4\pi) \right\} \,,
\ee
where $\gamma_E\simeq 0.577$ is the Euler constant and $\mu$ is an arbitrary renormalization mass scale. The renormalized coupling satisfies
\be   \label{eq:reno}
 \mathcal{C}_R(\mu) = \mathcal{C}_R(\mu_0)  - \ln(\mu/\mu_0) \,.
\ee
Assuming Yukawa alignment to be exact at a given energy scale $\Lambda_A$, so that $\mathcal{C}_R(\Lambda_A)=0$, implies that $\mathcal{C}_R(\mu) = \ln(\Lambda_A/\mu)$.

\section{Flavour-changing top decays}
\label{sec:topdecays}

The flavour-changing top decays $t \rightarrow c h$ and $t \rightarrow c V$ ($V=\gamma, Z$) occur firstly at the one-loop level in the SM. The decay rate for these processes is not only suppressed by the loop factor, but receives in addition a strong CKM and GIM suppression~\cite{Haeri:1988jt,Eilam:1990zc,Mele:1998ag,AguilarSaavedra:2004wm}. Here we focus on final states with the charm quark because $\mathrm{Br}(t \rightarrow u X)/\mathrm{Br}(t\rightarrow c X) \simeq |V_{ub}/V_{cb}|^2 \sim 7 \times 10^{-3}$ in the SM, as well as in the A2HDM. Fixing the Higgs mass at $M_h \simeq 125$~GeV, one obtains the SM branching ratios: $\mathrm{Br}(t\rightarrow c h) \sim \mathcal{O}(10^{-15})$, $\mathrm{Br}(t\rightarrow c \gamma) \sim \mathcal{O}(10^{-14})$ and $\mathrm{Br}(t \rightarrow c Z) \sim  \mathcal{O}(10^{-14})$. Within the A2HDM these decay rates can be enhanced due to additional charged Higgs contributions at the loop level. For $t \rightarrow c \varphi_j^0$ decays, the counter-term piece in eq.~\eqref{fcnc:ct} would also contribute.

In the SM, the dominant decay mode of the top quark is the unsuppressed two-body decay $t \rightarrow W^+ b$, with $\Gamma(t \rightarrow W^+ b)/m_t \sim 1 \%$. To compute the relevant branching ratios we take
\be
\mathrm{Br}(t \rightarrow c X) \; =\; \dfrac{ \Gamma(t \rightarrow c X) }{\Gamma_{\mbox{\scriptsize{tot}}}(t) }  \,.
\ee
To a very good approximation, $\Gamma_{\mbox{\scriptsize{tot}}}(t) \simeq \Gamma(t \rightarrow W^+ b)$ holds also in the A2HDM, except when $M_{H^{\pm}}< m_t - m_b $; in this case, the additional decay mode $t \rightarrow H^+ b$ must be taken into account. The partial decay widths for $t \rightarrow W^+ b$ and $t \rightarrow H^+ b$ are calculated at leading order:
\begin{align}
\Gamma(t \rightarrow W^{+} b) \; =& \;  \frac{g^{2} \, |V_{tb}|^2}{64 \, \pi \, m_t^3}  \;\,   \lambda^{1/2}(m_t^2,m_b^2,M_W^2)\, \biggl[ m_{t}^{2} + m_{b}^2 + \frac{(m_t^2-m_b^2)^2}{M_W^2} - 2 M_W^2\biggr] \,, \label{eqWbm}
\end{align}
\begin{align}
\Gamma(t \rightarrow H^{+} b) \;  =& \;   \frac{|V_{tb}|^2}{16 \, \pi \, m_t^3 \, v^2}\; \,  \lambda^{1/2}(m_t^2, m_b^2, M_{H^{\pm}}^2 ) \, \biggl[ \left(  m_t^2 + m_b^2 - M_{H^{\pm}}^2  \right)\,\left( m_b^2 |\varsigma_d|^2 + m_t^2 |\varsigma_u|^2  \right)\quad   \nonumber \\
&\hskip 5.9cm\mbox{}
- 4 m_b^2 m_t^2 \,\mathrm{Re}\left(  \varsigma_d \varsigma_u^* \right) \biggr]\,.
\label{eqHbm}
\end{align}
Here $\lambda(x,y,z)\equiv x^2+y^2+z^2-2(xy+xz+yz)$, and $g$ is the $\mathrm{SU(2)}_L$ gauge coupling constant.

Due to the smallness of $m_{d,s}$ and the unitarity of the CKM matrix, the $t \rightarrow c \varphi_j^0$ and $t \rightarrow c V$ decay amplitudes turn out to be very sensitive to the bottom quark mass~\cite{AguilarSaavedra:2004wm}. The most adequate choice for the internal quark masses is the running $\overline{\rm MS}$ quark mass evaluated at a typical scale $\mathcal{O}(m_t)$~\cite{AguilarSaavedra:2004wm}; we follow this prescription. The external quark masses are taken as the on-shell pole masses.

In section~\ref{sec:t1} we describe the calculation of the $t \rightarrow c \varphi_j^0$ decay amplitudes. The decays $t \rightarrow c V$ ($V=\gamma,Z$) are discussed in section~\ref{sec:t2}. Explicit analytical results for these processes within the A2HDM are collected in the appendices.  All our results are presented in the Feynman gauge, corresponding to $\xi=1$ in the $R_{\xi}$ gauge. We have however checked the gauge independence of our results by additionally performing all calculations in the unitary gauge. We have also checked analytically that our results reproduce the corresponding SM predictions~\cite{Eilam:1989zm,Eilam:1990zc,Mele:1998ag}.

\subsection{\texorpdfstring{$\mathbf{\boldsymbol{t \rightarrow c\, \varphi_j^0}}$}{Lg} decays}
\label{sec:t1}

\begin{figure}[ht!]
\centering
\includegraphics[width=0.2\textwidth]{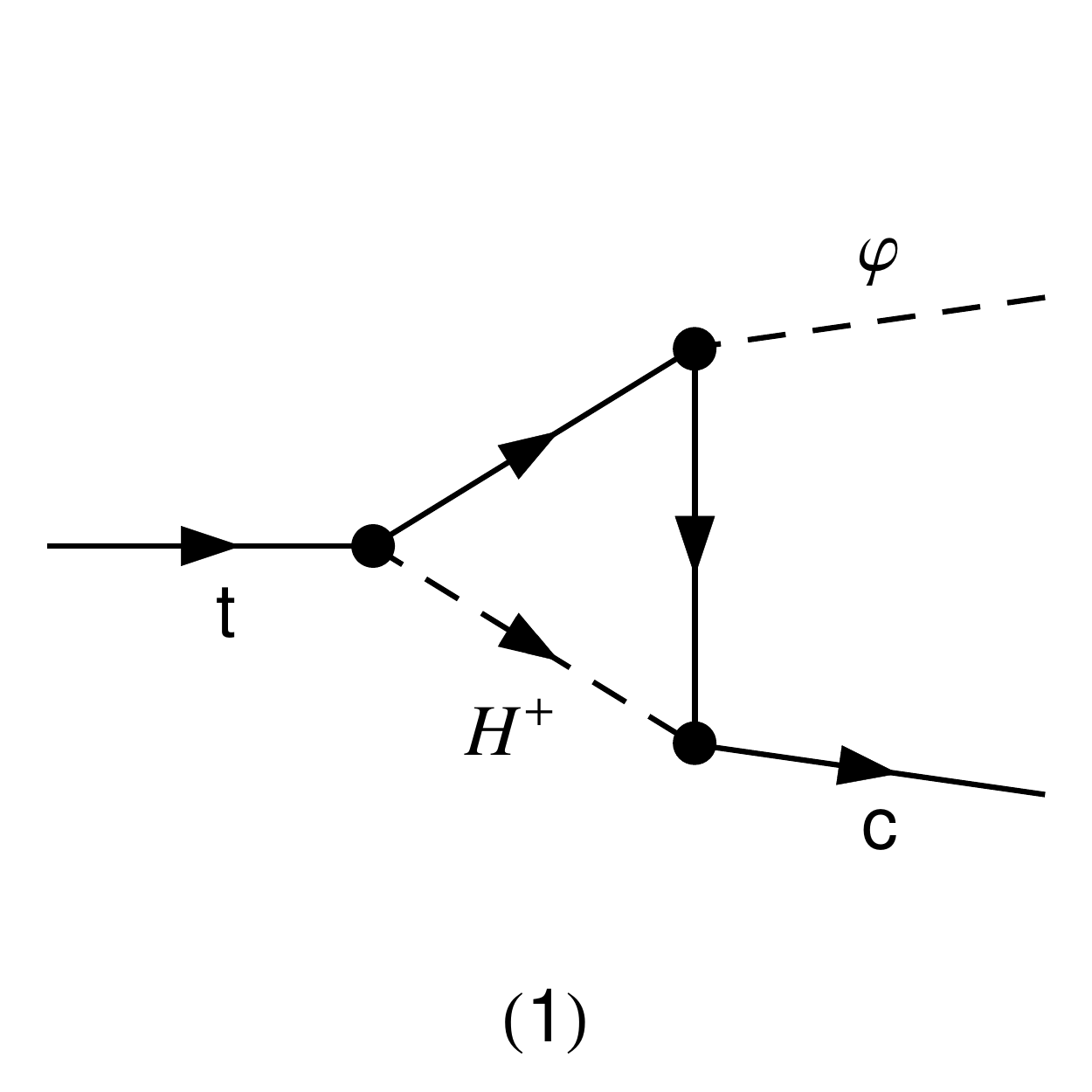}
~
\includegraphics[width=0.2\textwidth]{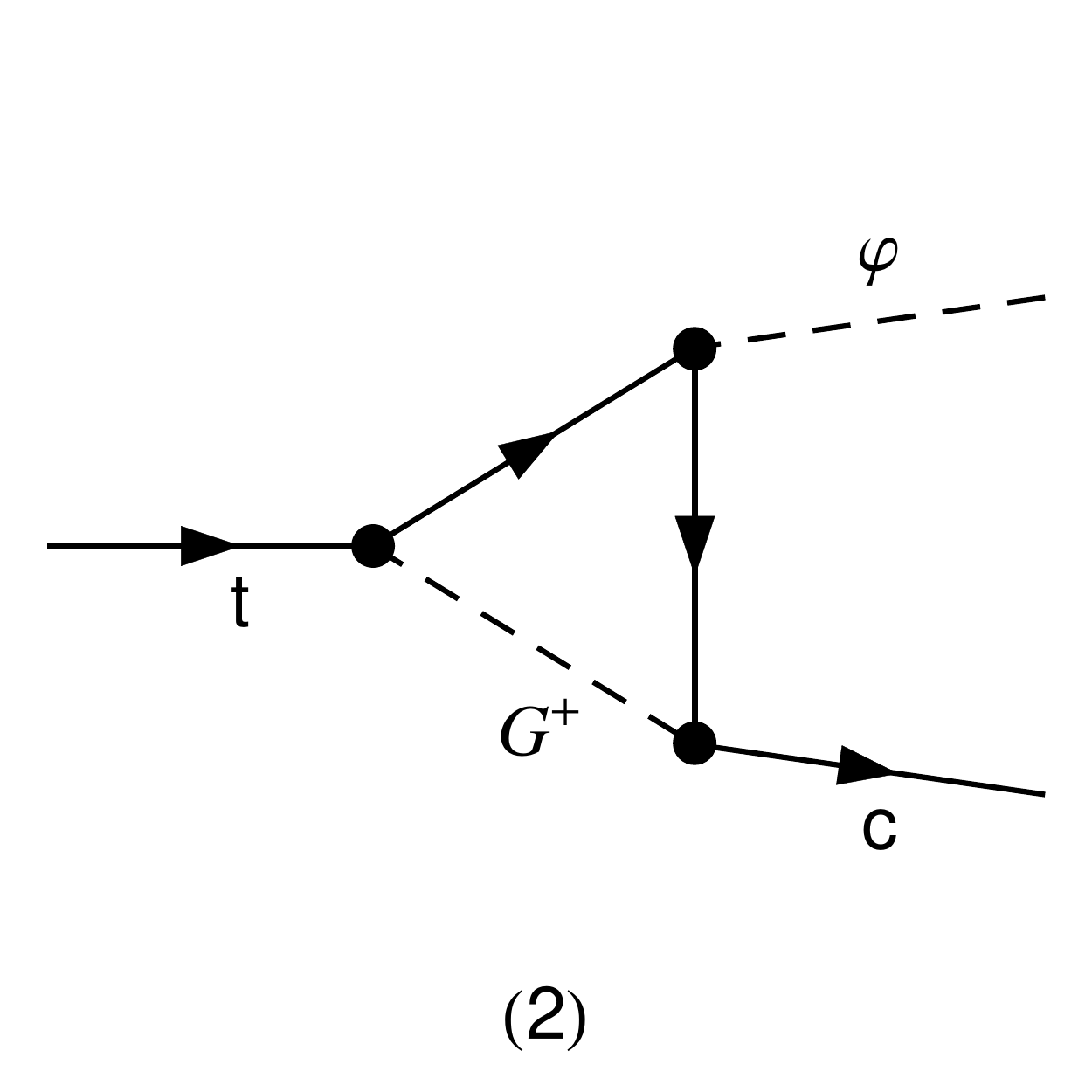}
~
\includegraphics[width=0.2\textwidth]{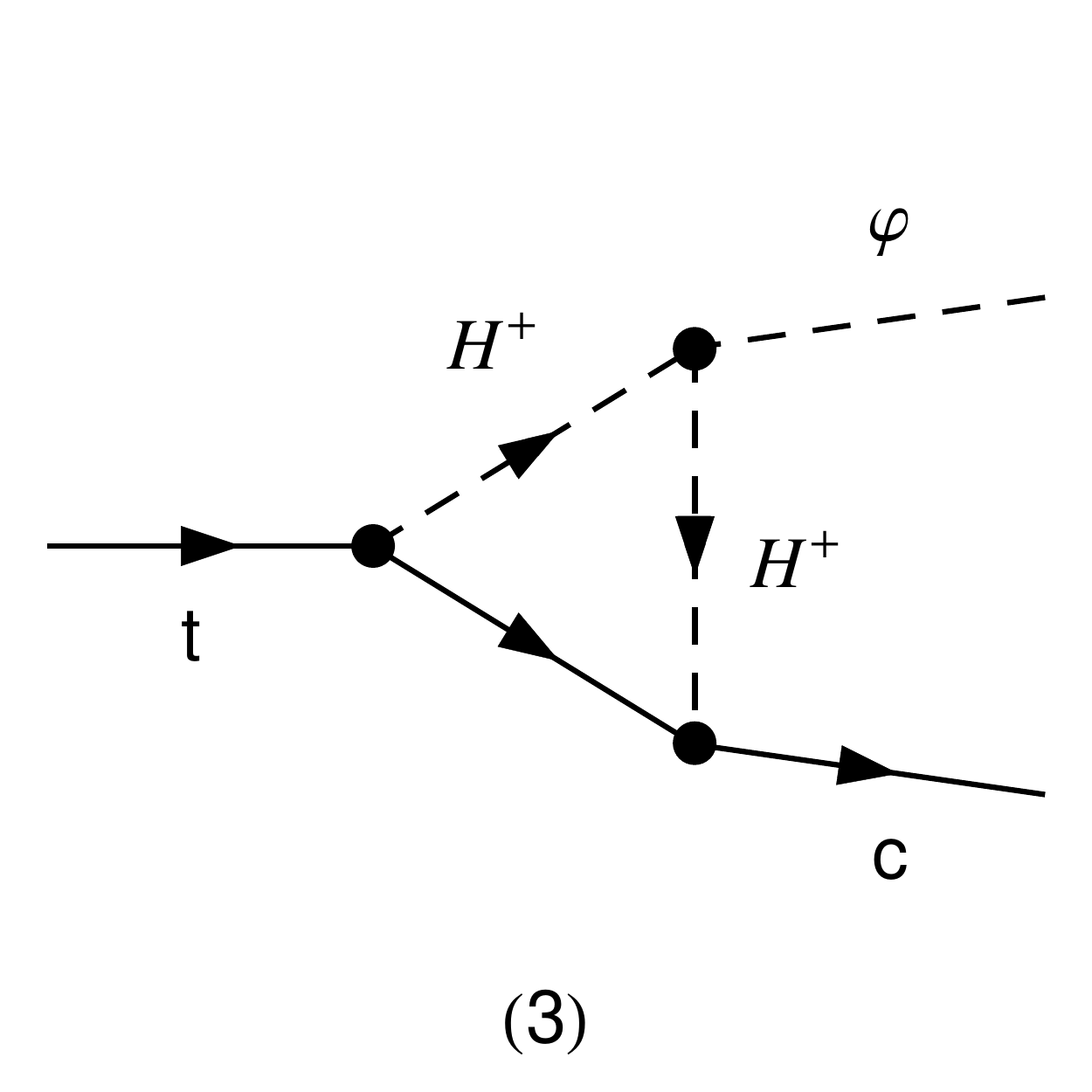}
~
\includegraphics[width=0.2\textwidth]{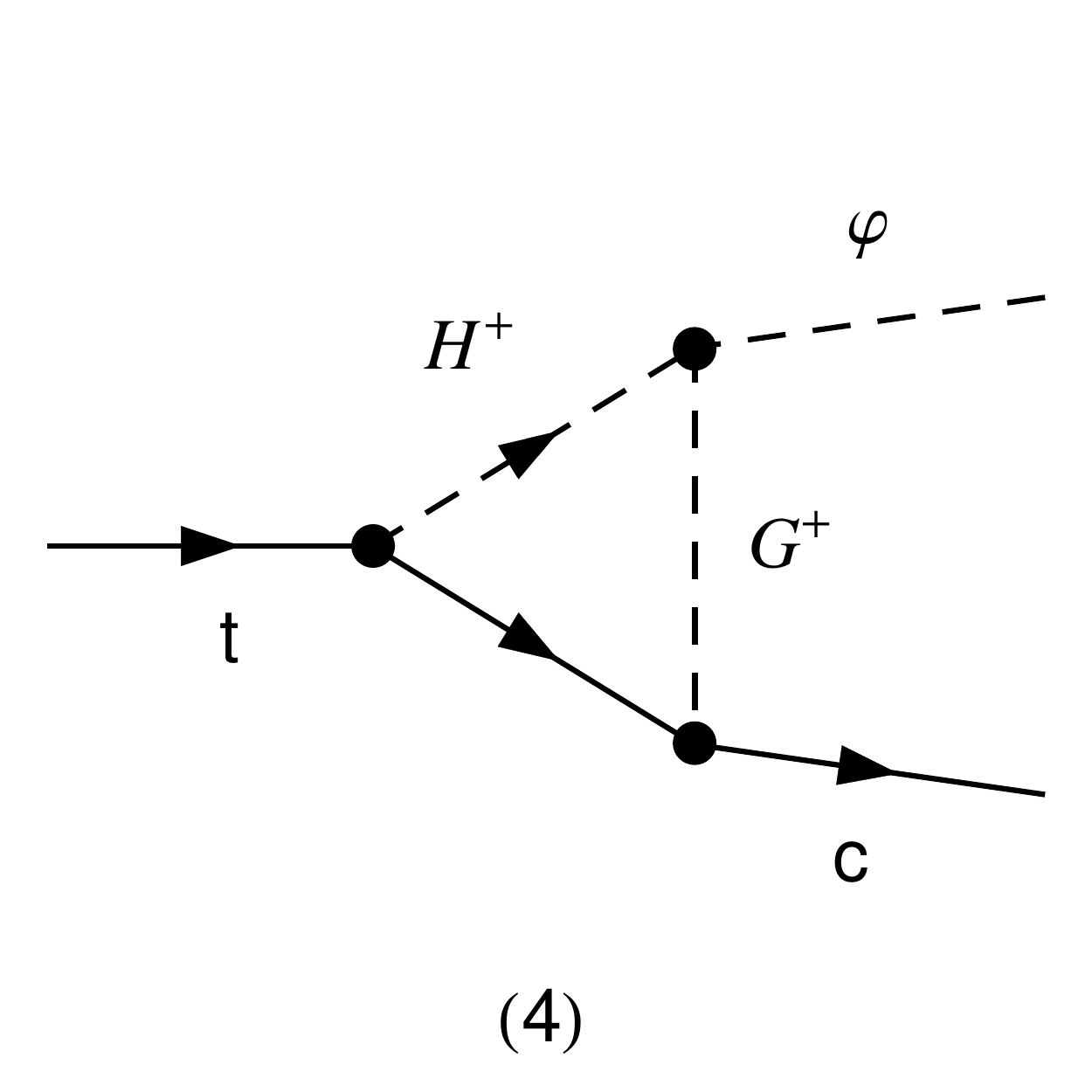}
~
\includegraphics[width=0.2\textwidth]{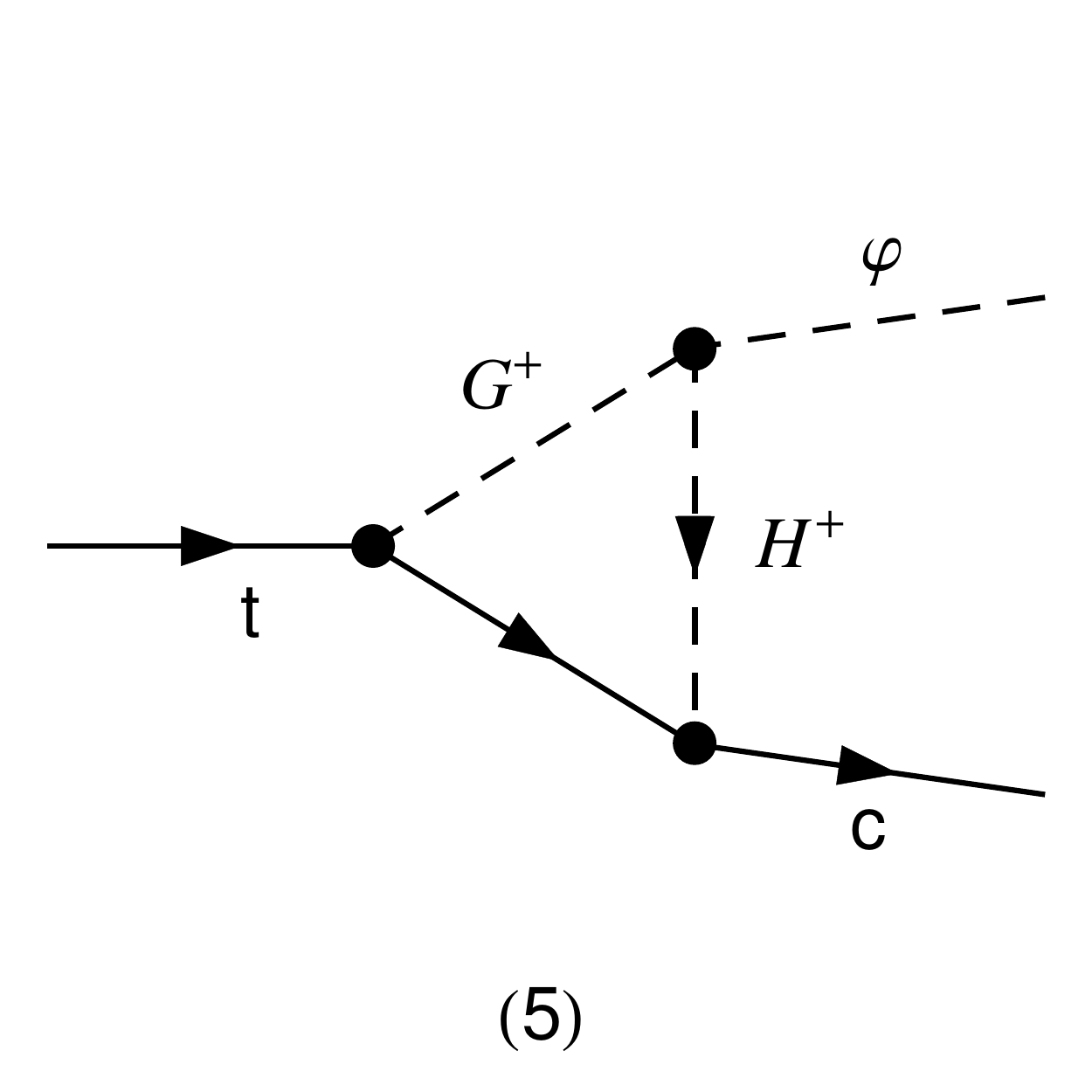}
~
\includegraphics[width=0.2\textwidth]{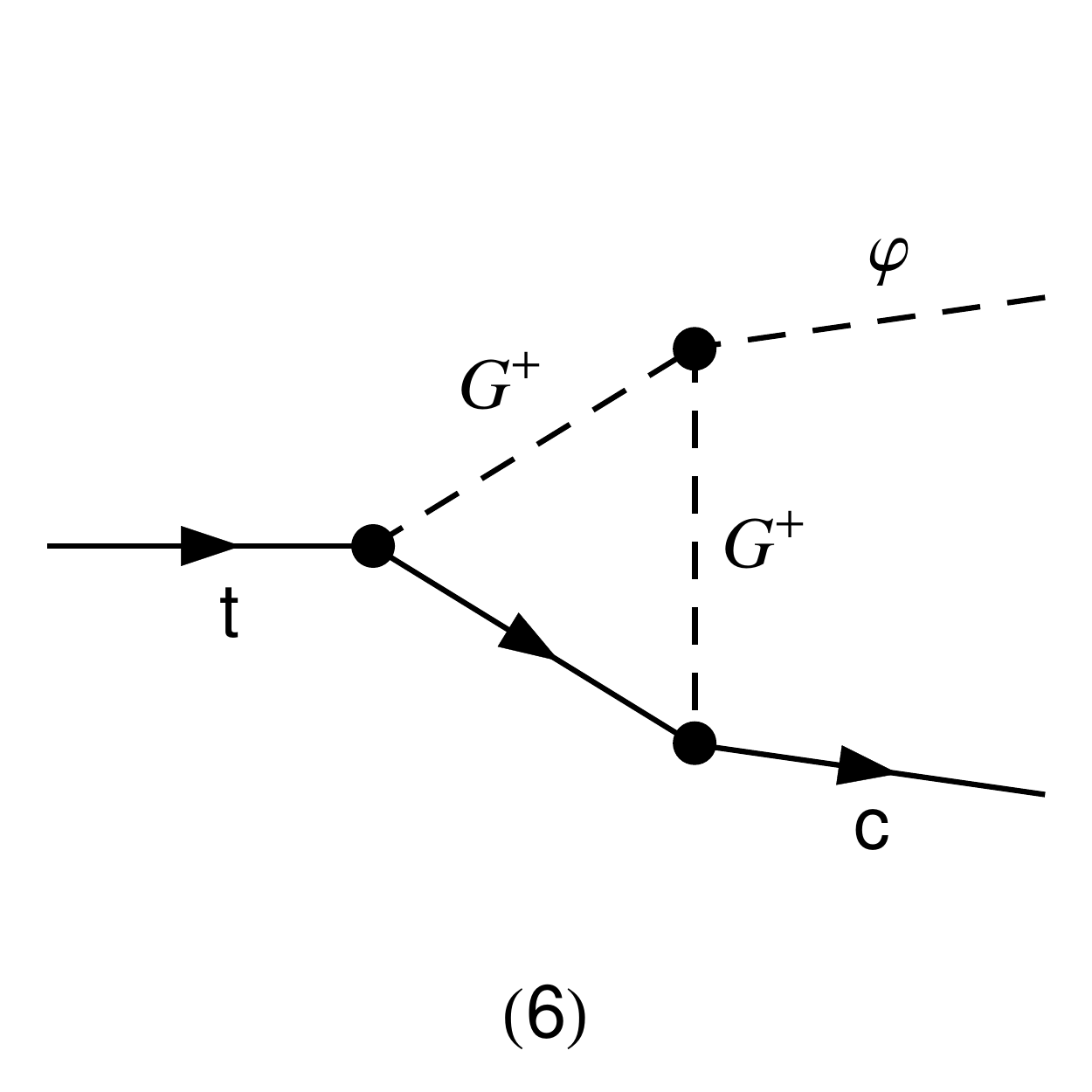}
~
\includegraphics[width=0.2\textwidth]{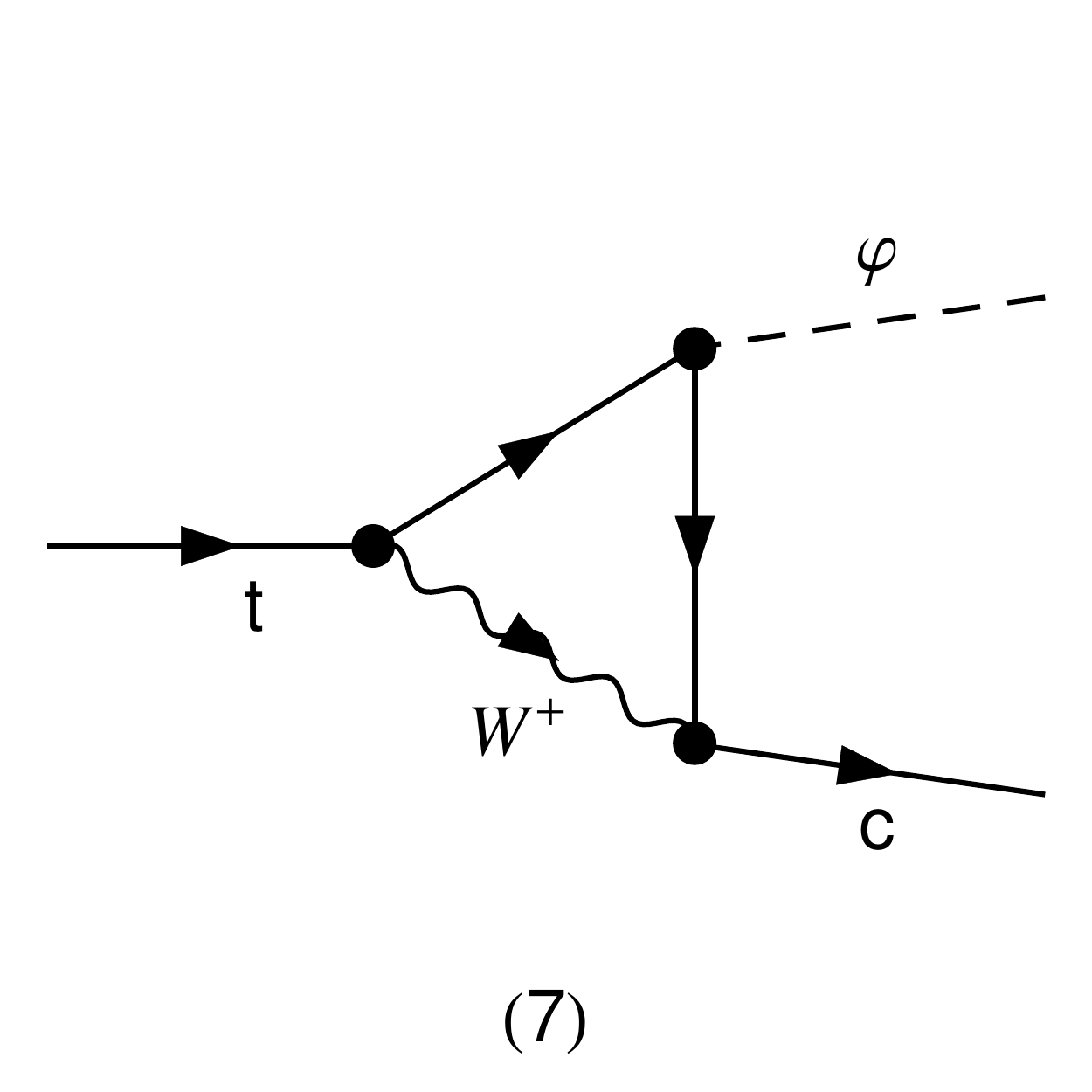}
~
\includegraphics[width=0.2\textwidth]{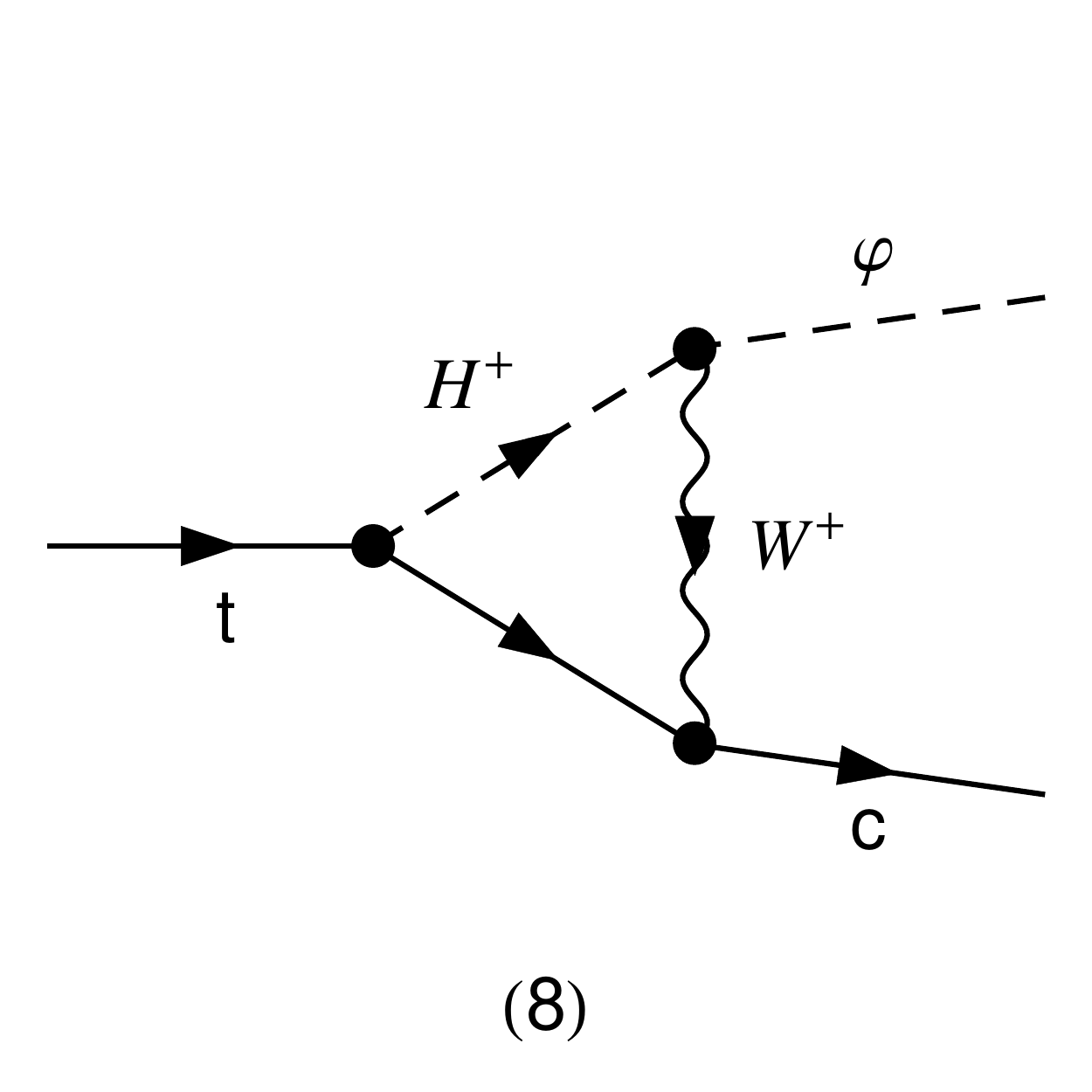}
~
\includegraphics[width=0.2\textwidth]{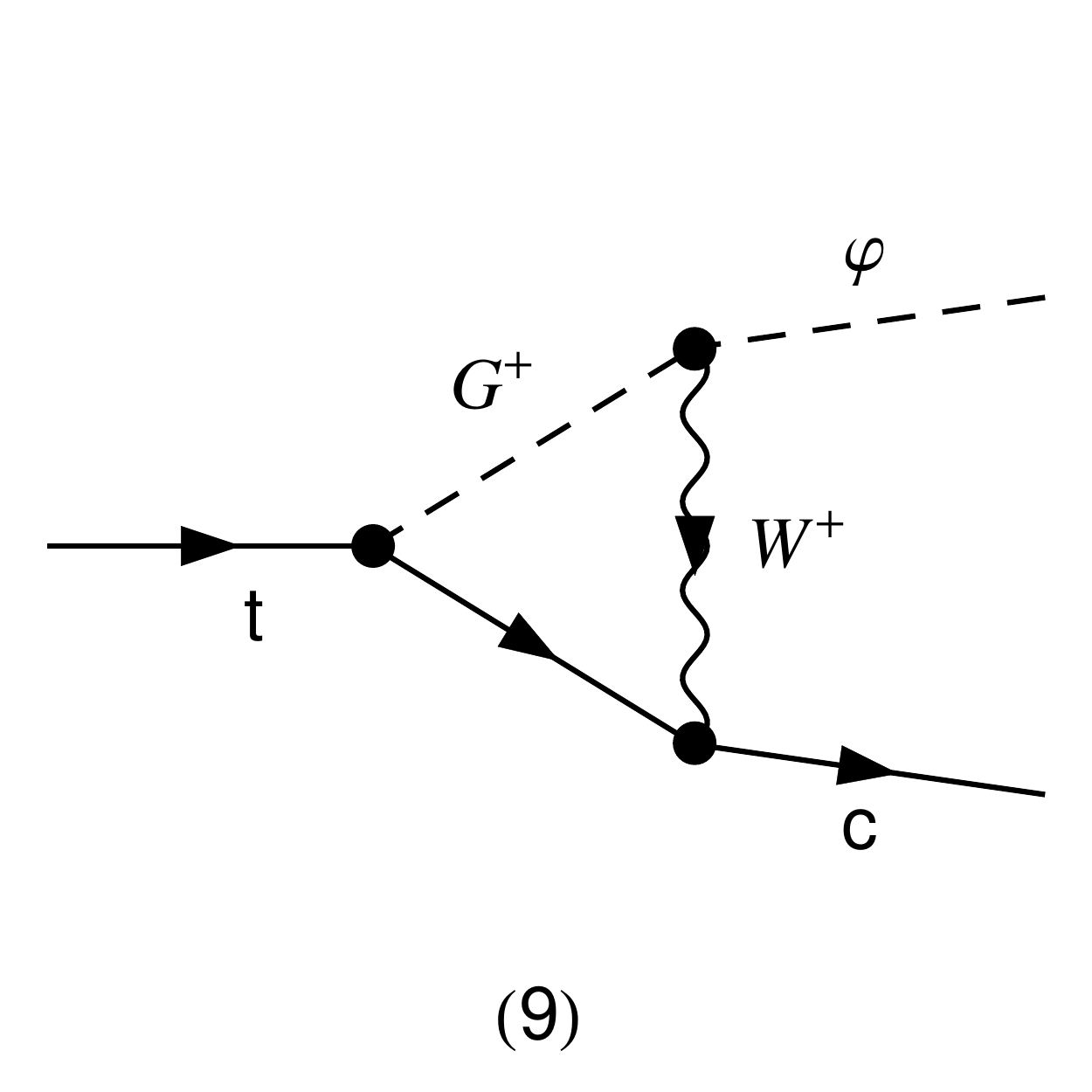}
~
\includegraphics[width=0.2\textwidth]{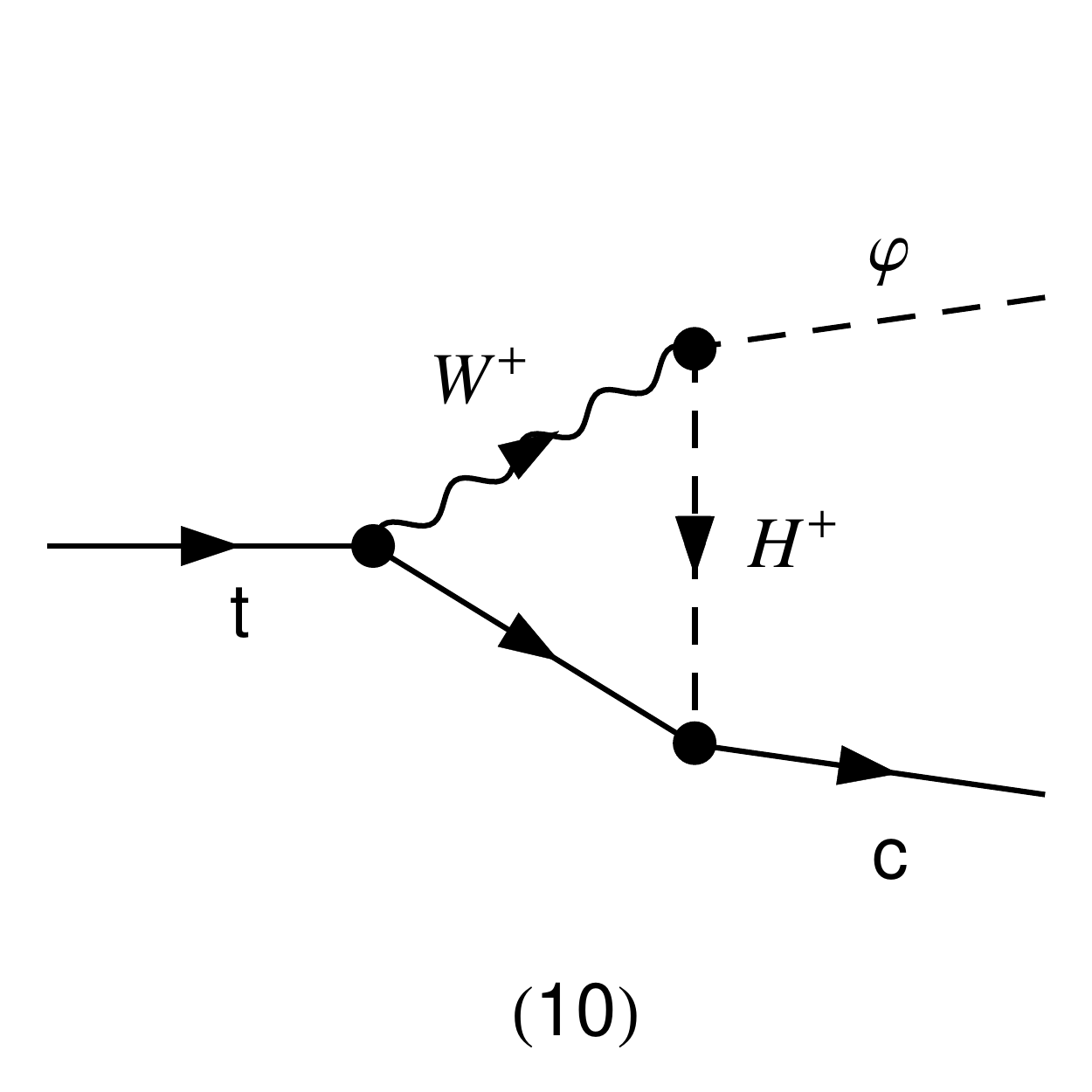}
~
\includegraphics[width=0.2\textwidth]{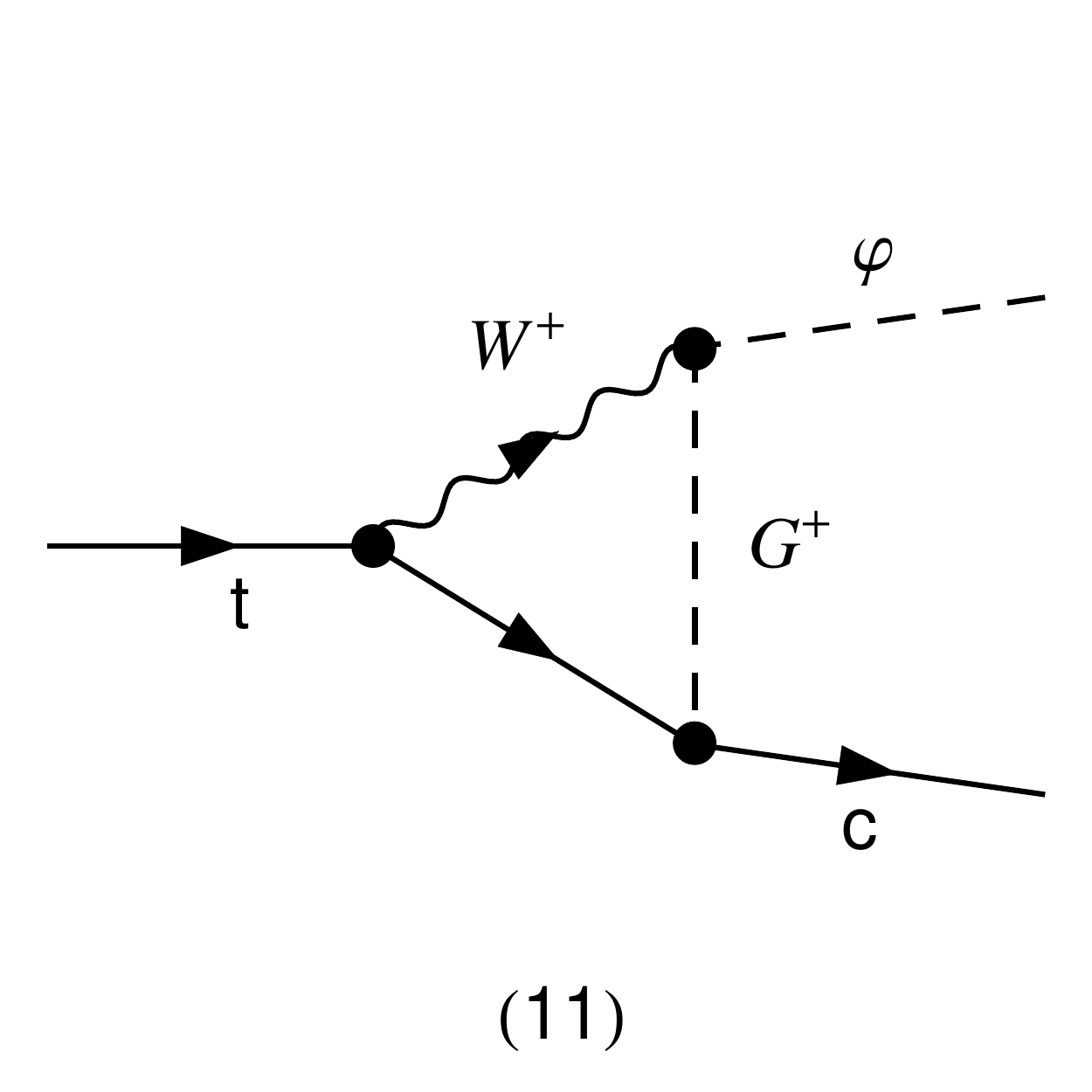}
~
\includegraphics[width=0.2\textwidth]{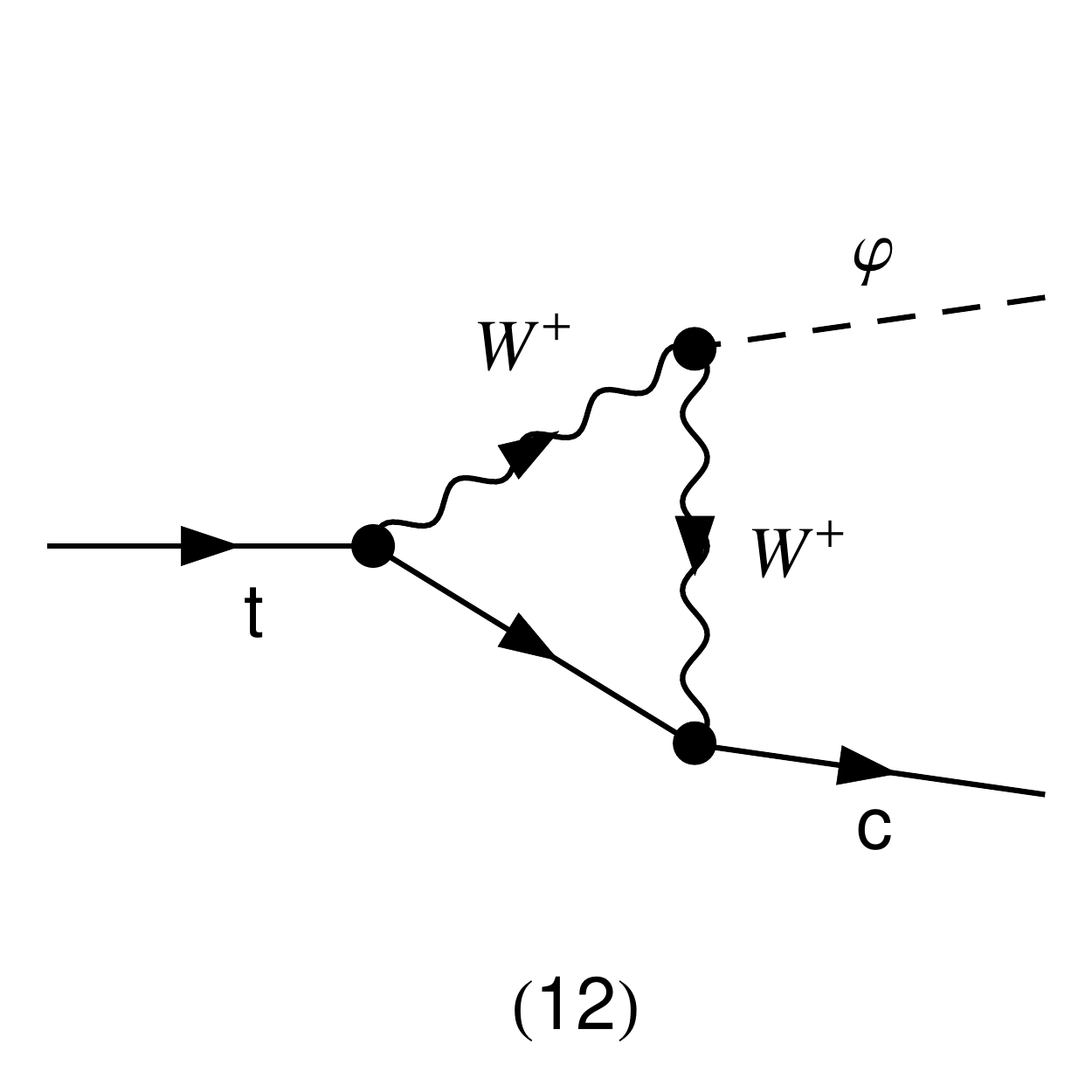}
\caption{\label{fig:diagsi} \it \small  Penguin diagrams contributing to $t \rightarrow c \varphi_j^0$ in the Feynman gauge. }
\end{figure}

\begin{figure}[ht!]
\centering
\includegraphics[width=0.2\textwidth]{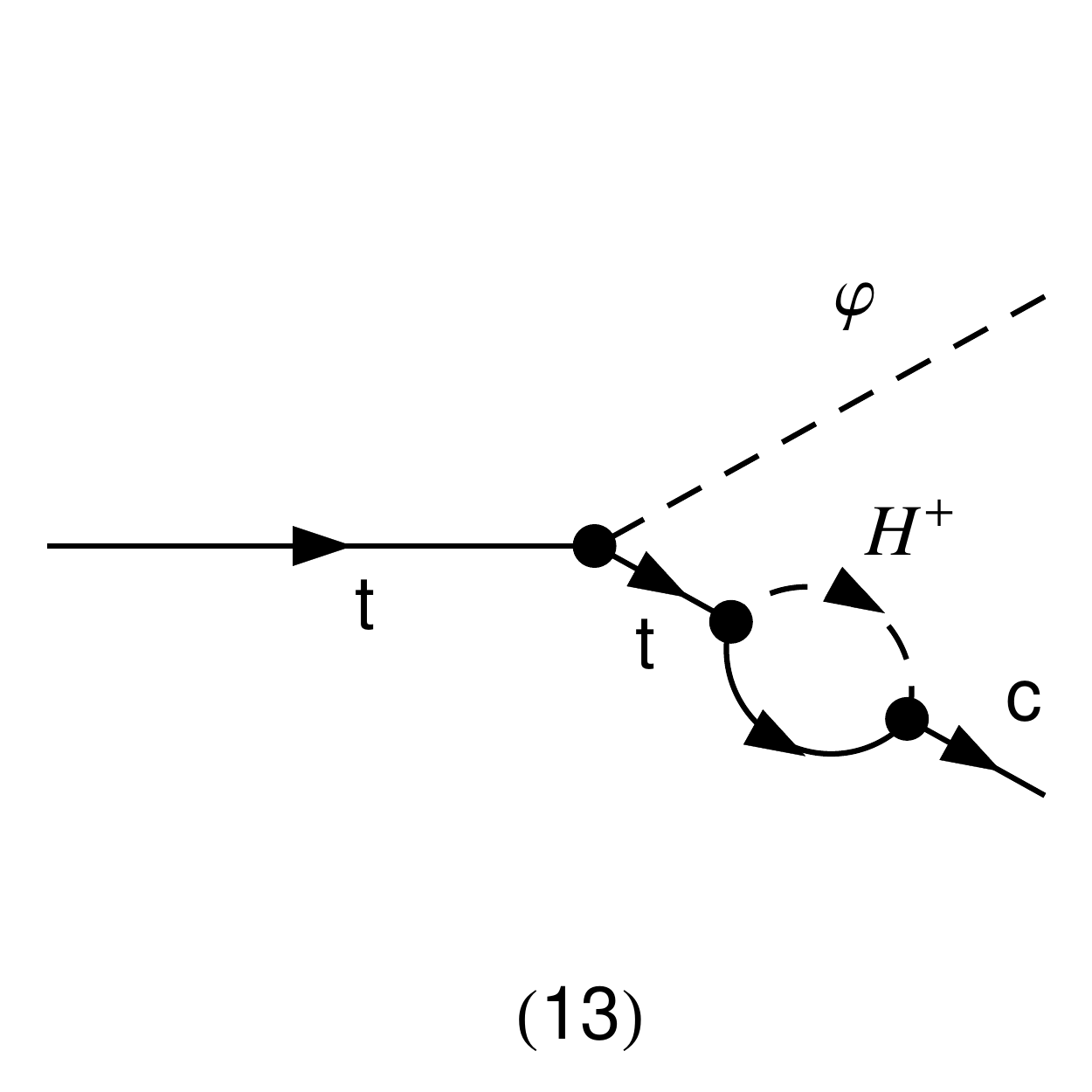}
~
\includegraphics[width=0.2\textwidth]{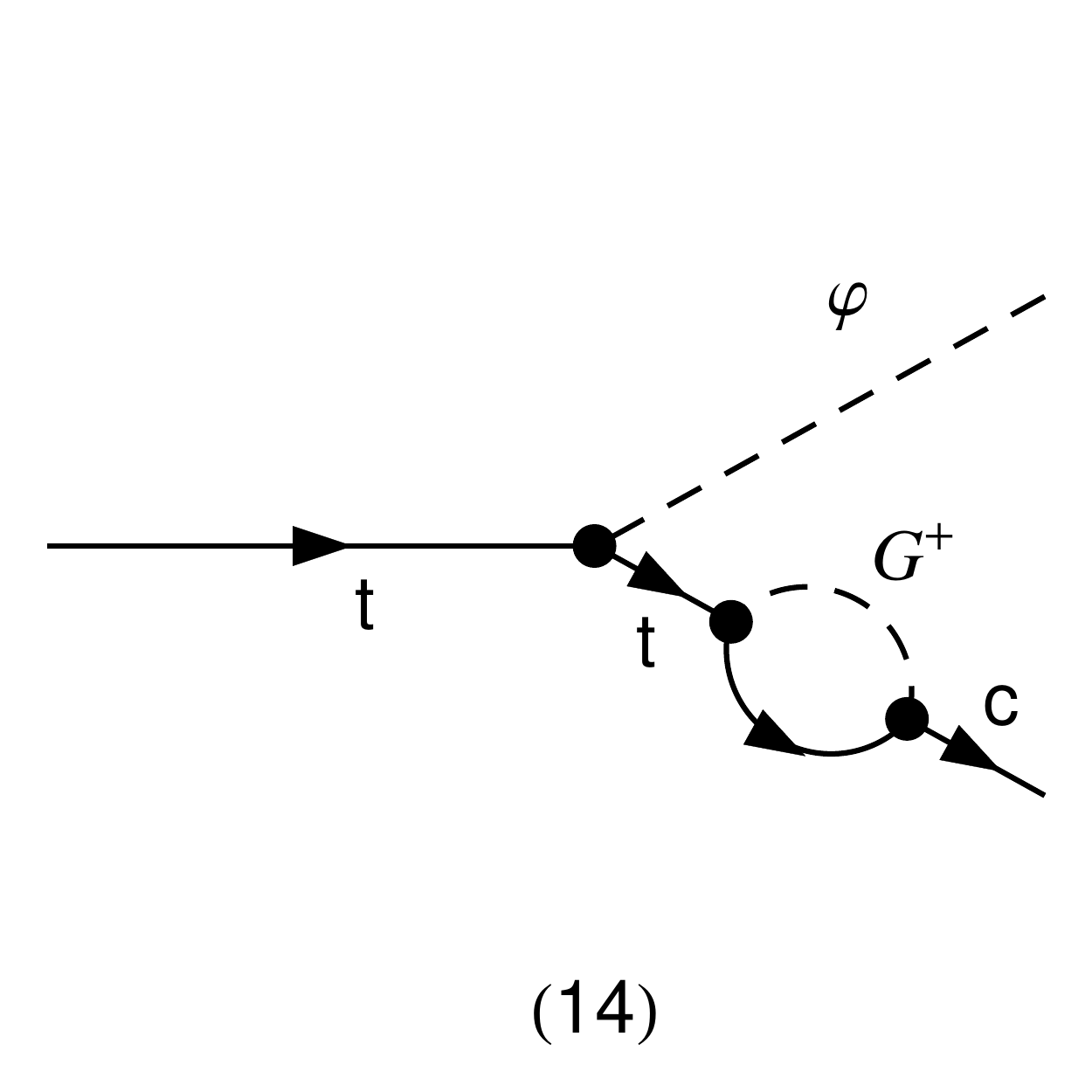}
~
\includegraphics[width=0.2\textwidth]{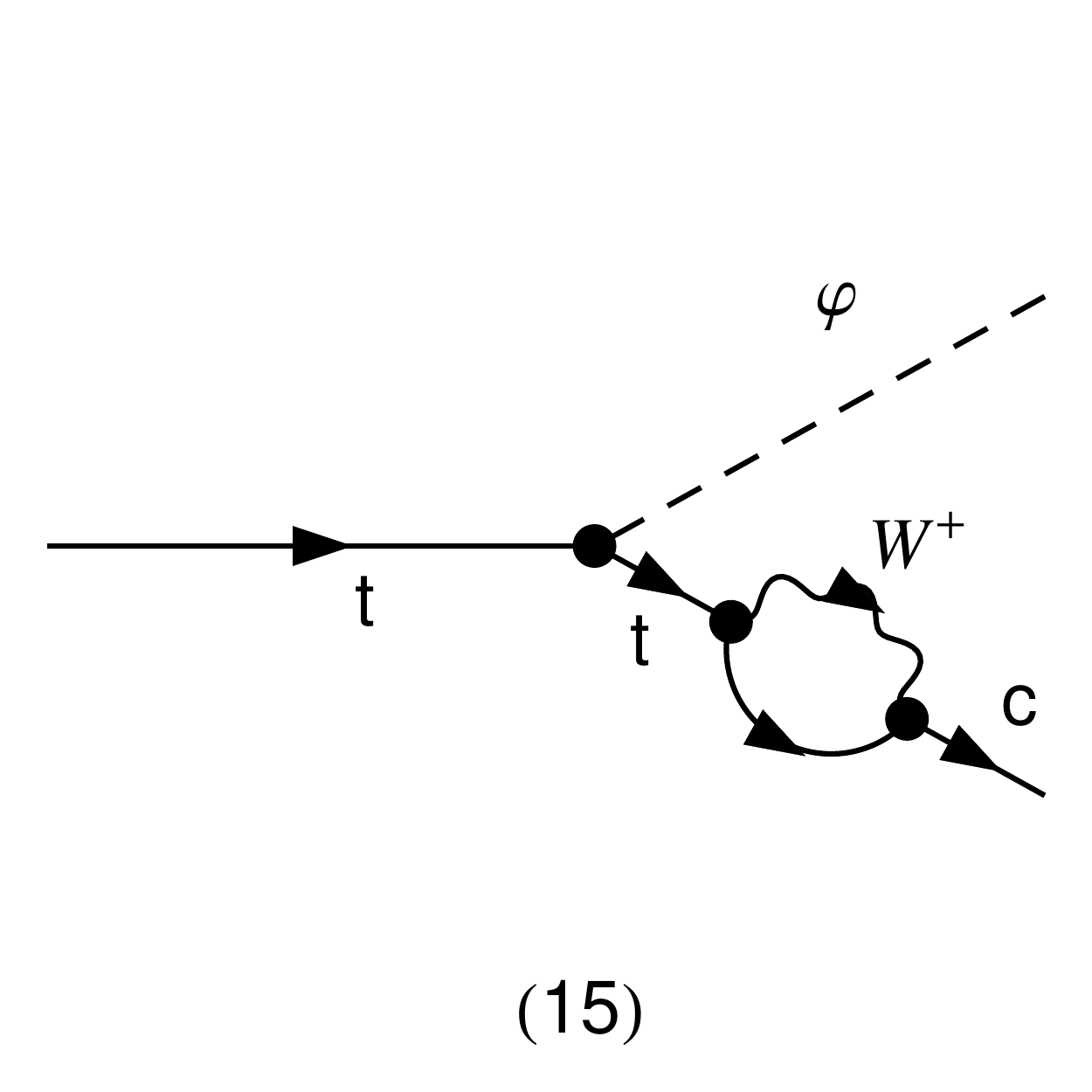}\\
\includegraphics[width=0.2\textwidth]{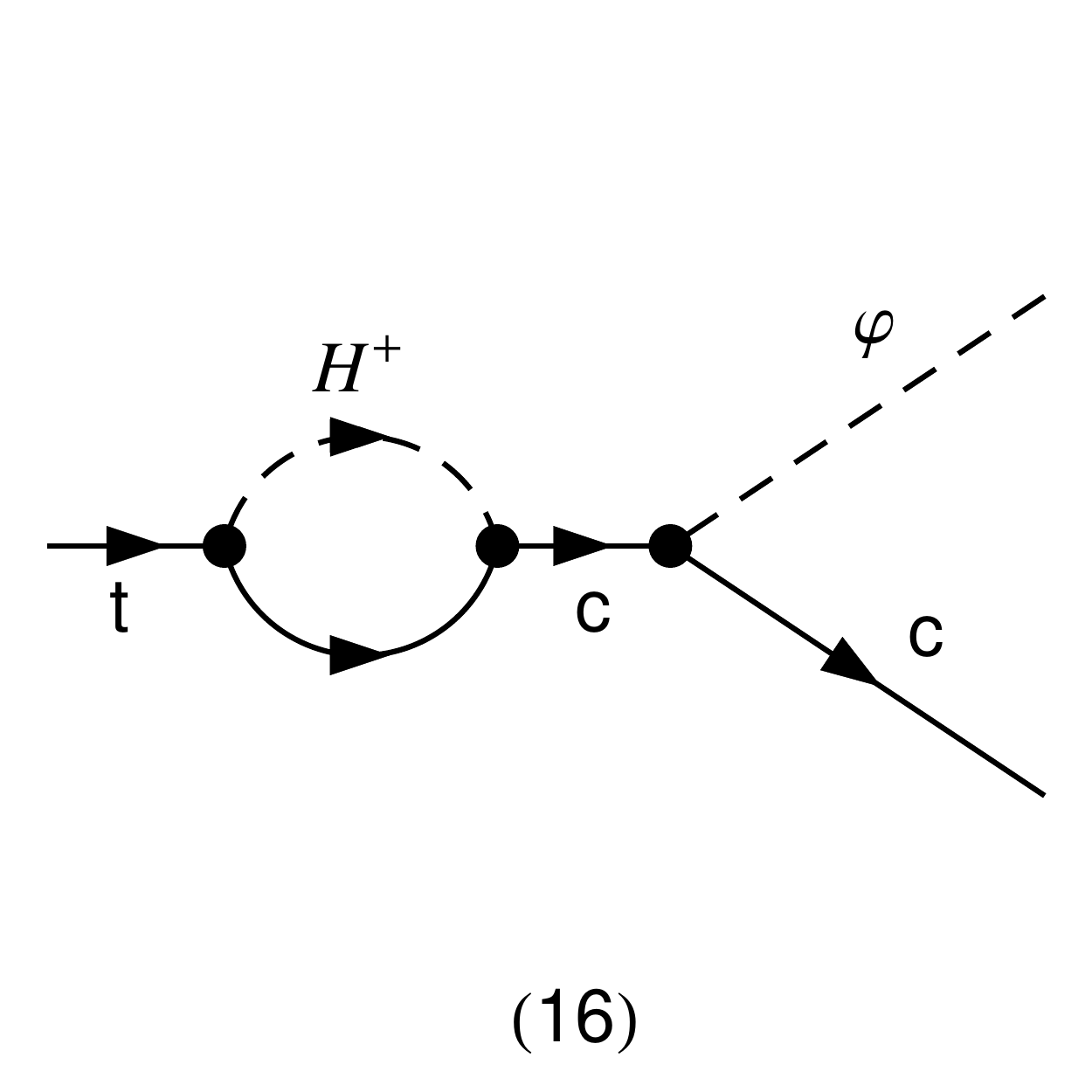}
~
\includegraphics[width=0.2\textwidth]{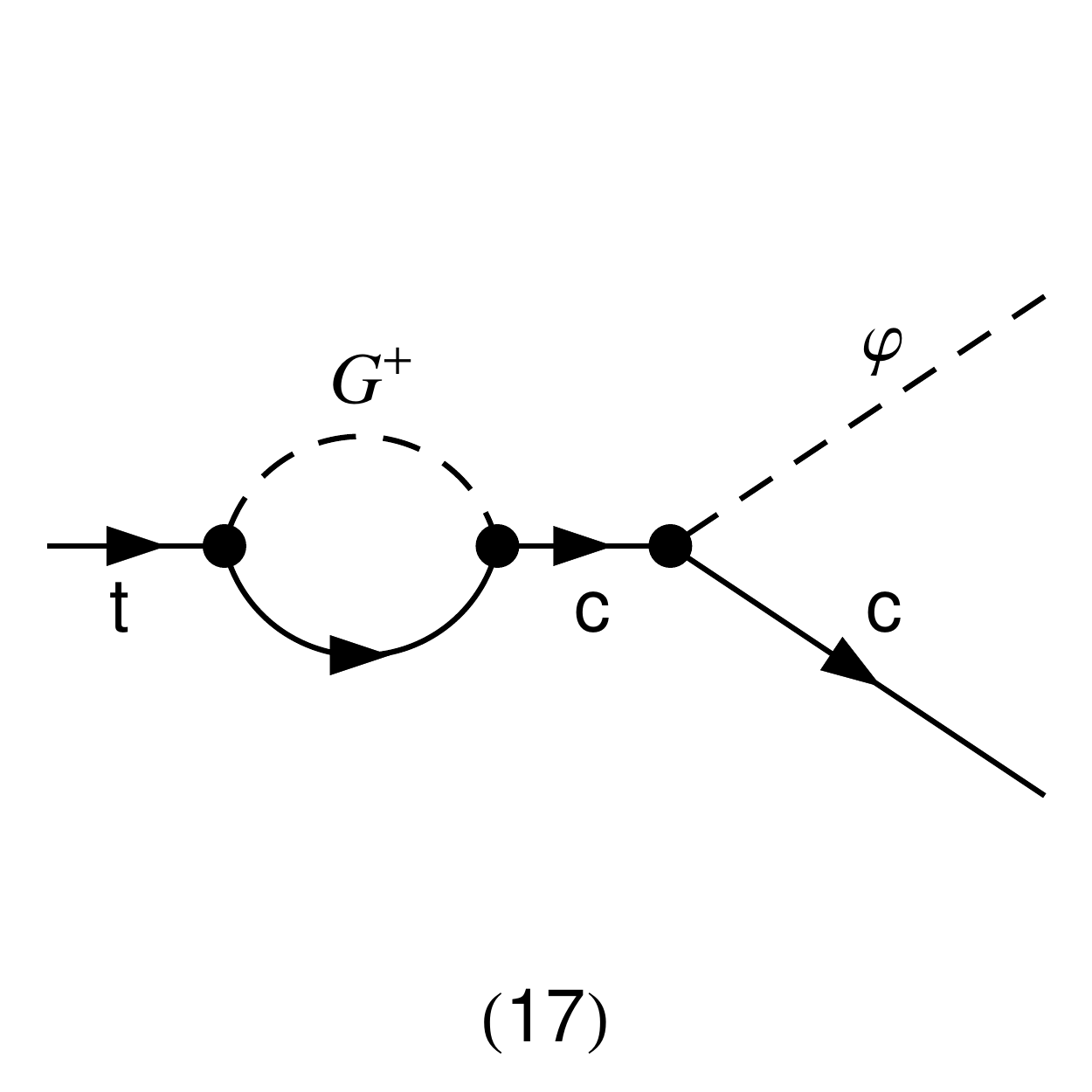}
~
\includegraphics[width=0.2\textwidth]{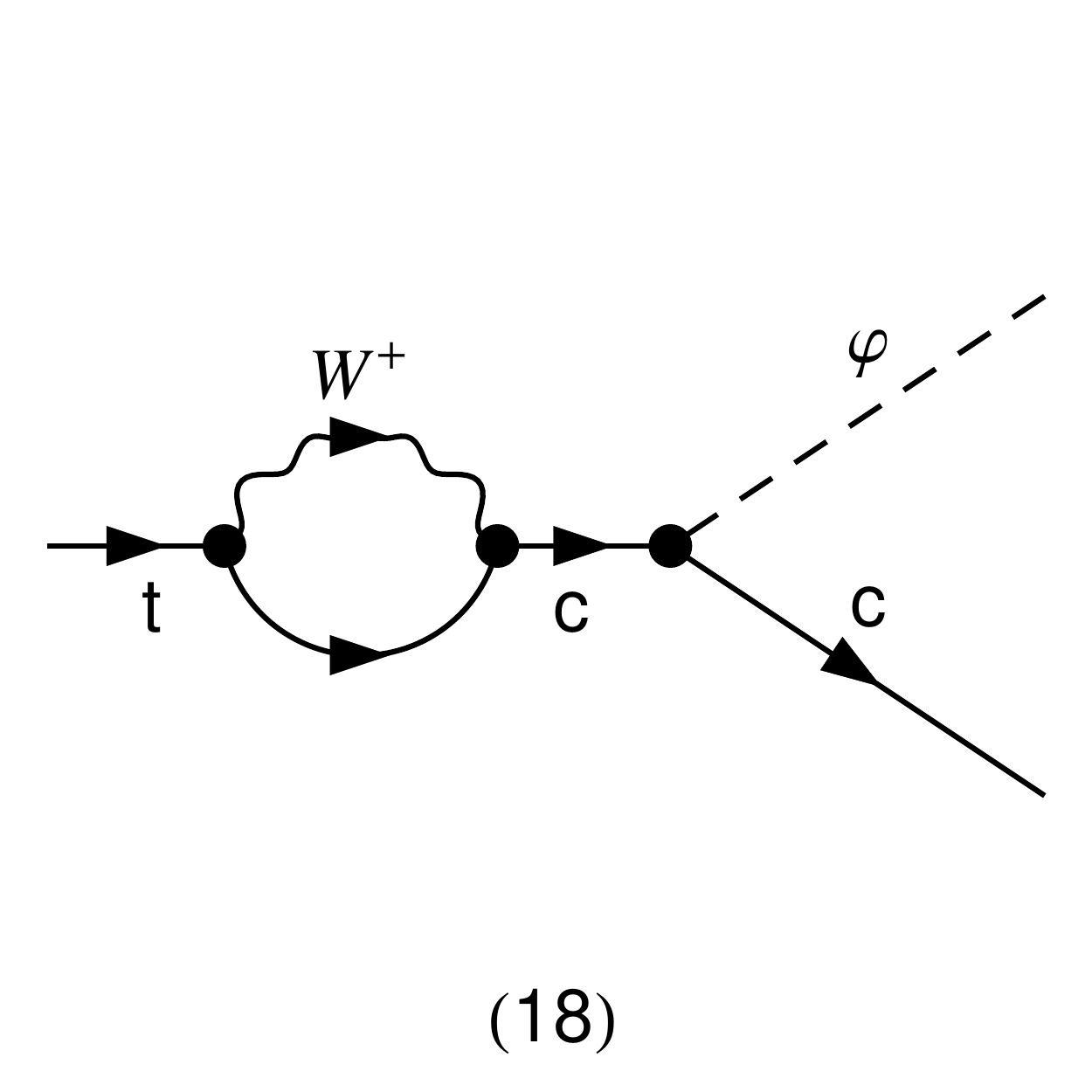}
\caption{\label{fig:diagsii} \it \small Self-energy diagrams contributing to $t \rightarrow c \varphi_j^0$ in the Feynman gauge. }
\end{figure}

The total amplitude for $t(p_t) \rightarrow c(p_c) \, \varphi_j^0(p_{\varphi}) $ decays can be written as $A_{\mbox{\scriptsize{tot}}} = A + A_{\mbox{\scriptsize{ct}}}$. Here $A$ collects all the contributions arising from one-loop diagrams and can be parametrized as~\cite{Eilam:1989zm,Eilam:1990zc}
\be \label{amp:ttoch}
A \; =\; \sum_{n =1}^{18}\, A_{n}\; =\; \sum_{n =1}^{18}  \sum_{q=d,s,b} \, V_{cq} V_{tq}^* \;\,  \bar u(p_c)  \Bigl[ \alpha^{(n)}  \, P_R + \beta^{(n)} \, P_L \Bigr] u(p_t) \,.
\ee
Diagrams contributing to this process in the Feynman gauge are shown in figures~\ref{fig:diagsi} and~\ref{fig:diagsii}. The total loop amplitude is ultraviolet divergent, except in the particular cases of 2HDMs with NFC where a finite result is obtained. As expected, the divergence gets reabsorbed into the renormalization of the counter-term coupling $\mathcal{C}$ in eq.~\eqref{fcnc:ct}. The tree-level contribution of the counter-term Lagrangian takes the form
\be  \label{ctpiece}
A_{\mbox{\scriptsize{ct}}} \;=\; \sum_{q=d,s,b}  V_{cq} V_{tq}^*  \;\,  \bar u(p_c)  \Bigl[ \alpha_{\mbox{\scriptsize{ct}}} \, P_R + \beta_{\mbox{\scriptsize{ct}}} \, P_L \Bigr]  u(p_t) \,,
\ee
with
\begin{align}  \label{ct:con}
\alpha_{\mbox{\scriptsize{ct}}} \;=\;&    -   \mathcal{C}\,  \frac{  i g^3 m_t }{  32 \pi^2 M_W^3  }\,    (1 + \varsigma_u^* \varsigma_d)  ( \varsigma_d - \varsigma_u  )^*   (   \mathcal{R}_{j2} - i \mathcal{R}_{j3} ) \,  m_q^2           \,, \nonumber \\[0.2cm]
\beta_{\mbox{\scriptsize{ct}}} \;=\;&    -   \mathcal{C}\,  \frac{  i g^3  m_c}{  32 \pi^2 M_W^3  } \,   (1 + \varsigma_d^* \varsigma_u)  ( \varsigma_d - \varsigma_u  )   (   \mathcal{R}_{j2}  + i \mathcal{R}_{j3} )  \, m_q^2    \,.
\end{align}
%

The partial decay width can be written as
\begin{align}  \label{eq:decayI}
\Gamma(t \rightarrow c \, \varphi_j^0) \;=\; \dfrac{\lambda^{1/2}(m_t^2,m_c^2,M_{\varphi_j^0}^2)}{32 \, \pi \, m_t^3}\, \left[ (  |\alpha|^2 + |\beta|^2 ) (  m_c^2 + m_t^2 -M_{\varphi_j^0}^2 ) + 2 m_c m_t\, ( \alpha^* \beta + \beta^* \alpha )  \right] .
\end{align}
Here
\begin{align}  \label{eq:deftot}
\alpha \;=\; \sum_{q=d,s,b} V_{cq}  V_{tq}^*\; \left( \sum_{n=1}^{18}  \alpha^{(n)} + \alpha_{\mbox{\scriptsize{ct}}}  \right) \,,
\qquad\quad
\beta \;=\; \sum_{q=d,s,b} V_{cq}  V_{tq}^*\; \left( \sum_{n=1}^{18}  \beta^{(n)} + \beta_{\mbox{\scriptsize{ct}}}  \right) \,.
\end{align}
The contributions to the amplitude from each diagram, encoded in the coefficients $\alpha^{(n)}$ and $\beta^{(n)}$, are collected in appendix~\ref{app:I}.

\subsection{\texorpdfstring{$\mathbf{\boldsymbol{t \rightarrow c V~(V = \gamma, Z)}}$}{Lg} decays}
\label{sec:t2}

\begin{figure}[ht!]
\centering
\includegraphics[width=0.2\textwidth]{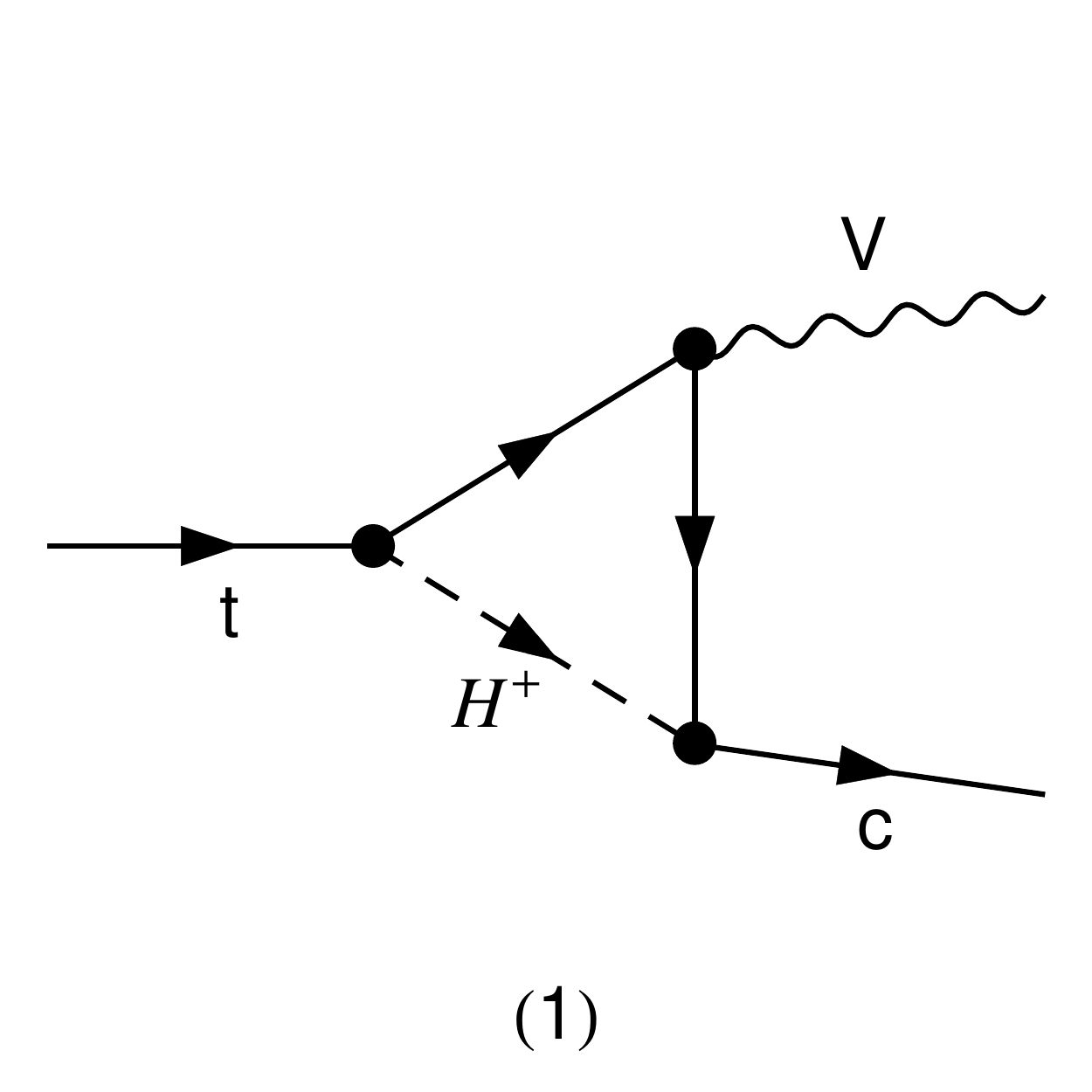}
~
\includegraphics[width=0.2\textwidth]{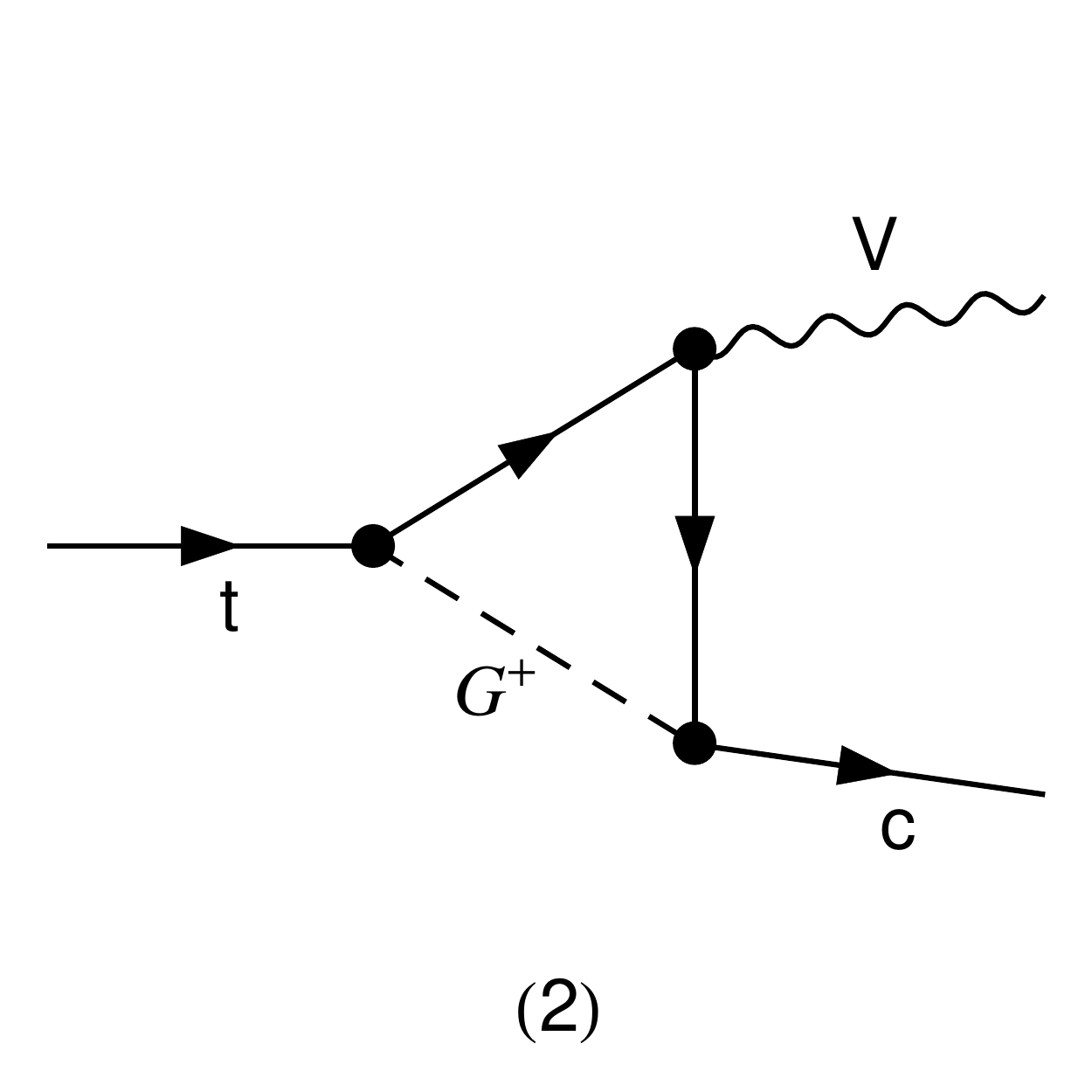}
~
\includegraphics[width=0.2\textwidth]{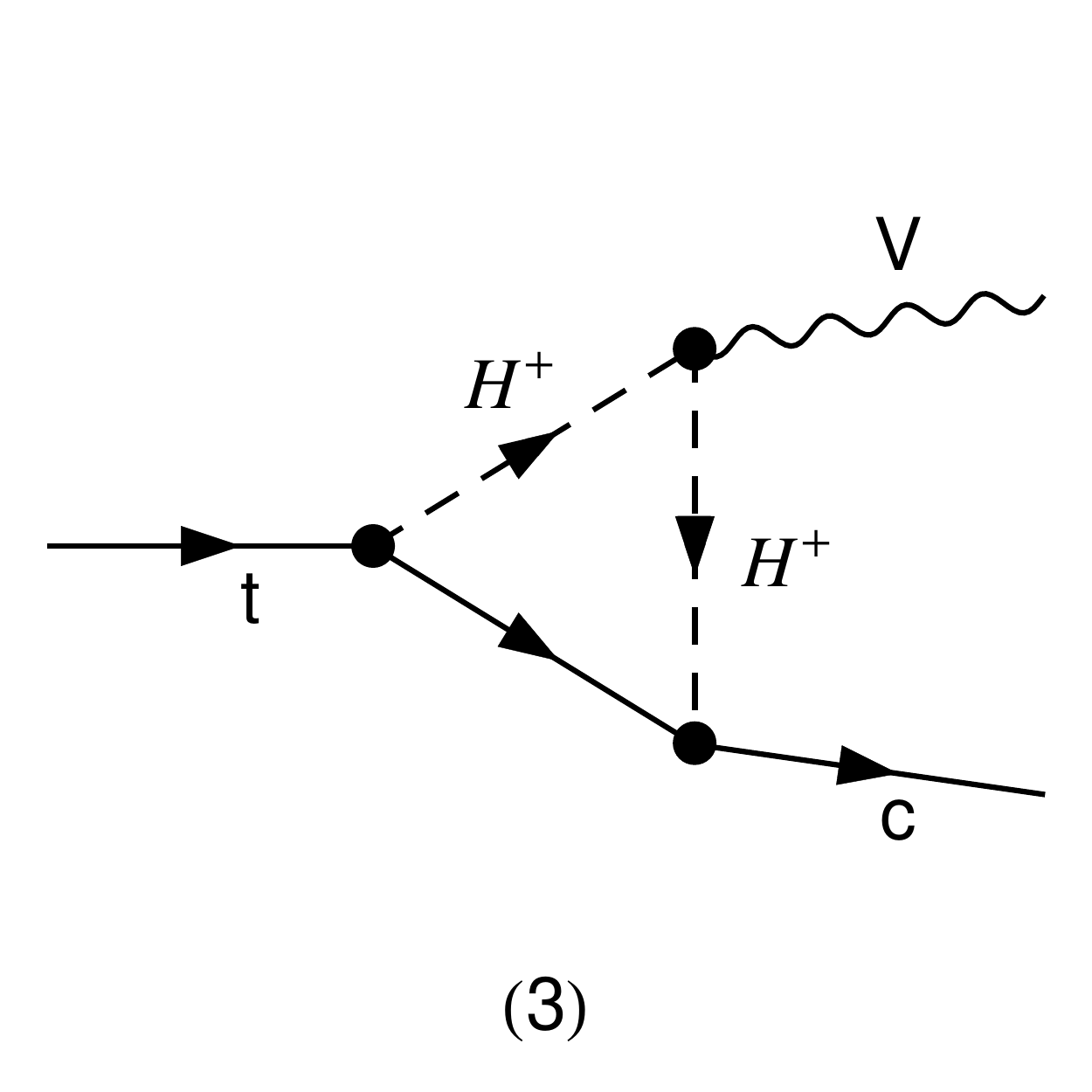}
~
\includegraphics[width=0.2\textwidth]{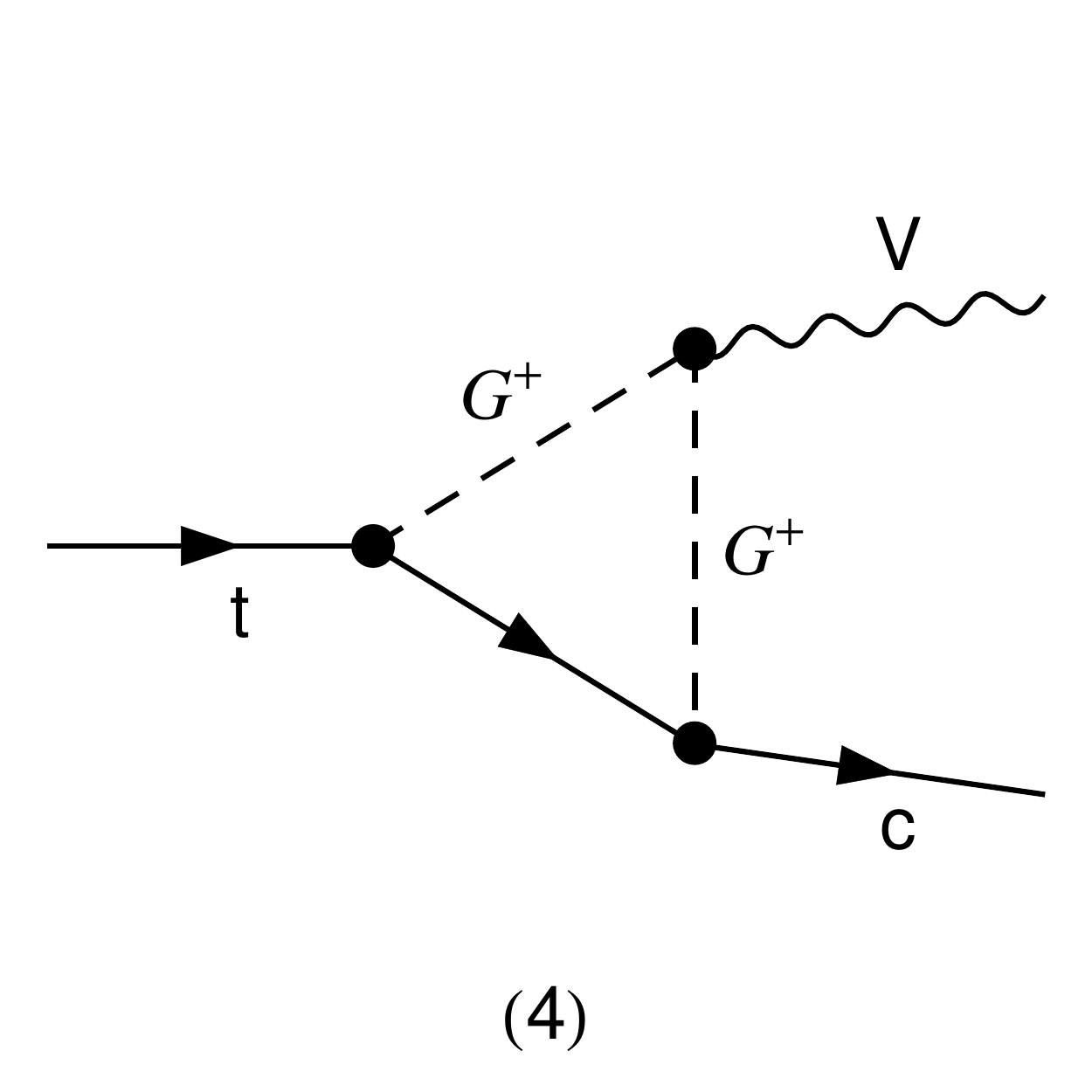}
~
\includegraphics[width=0.2\textwidth]{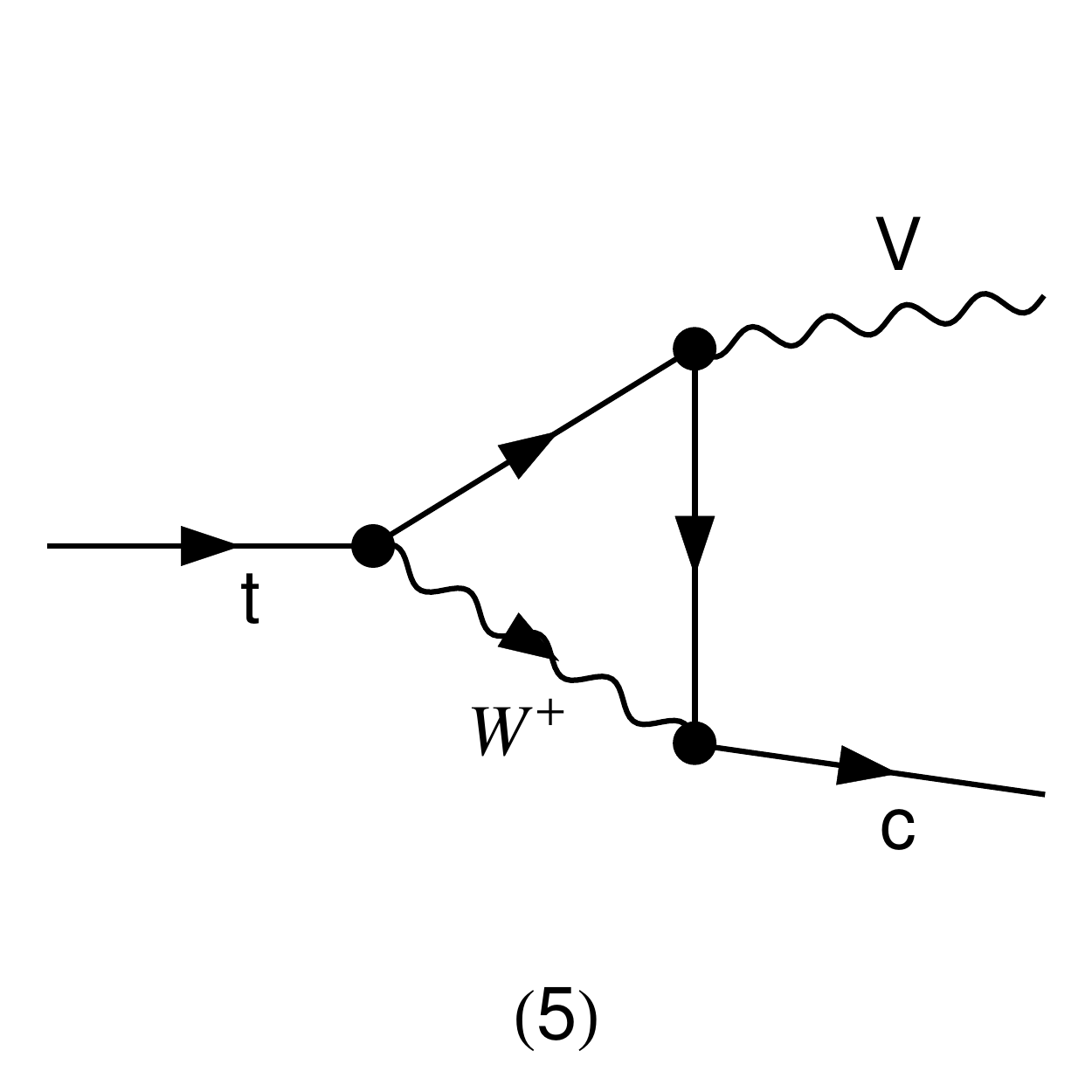}
~
\includegraphics[width=0.2\textwidth]{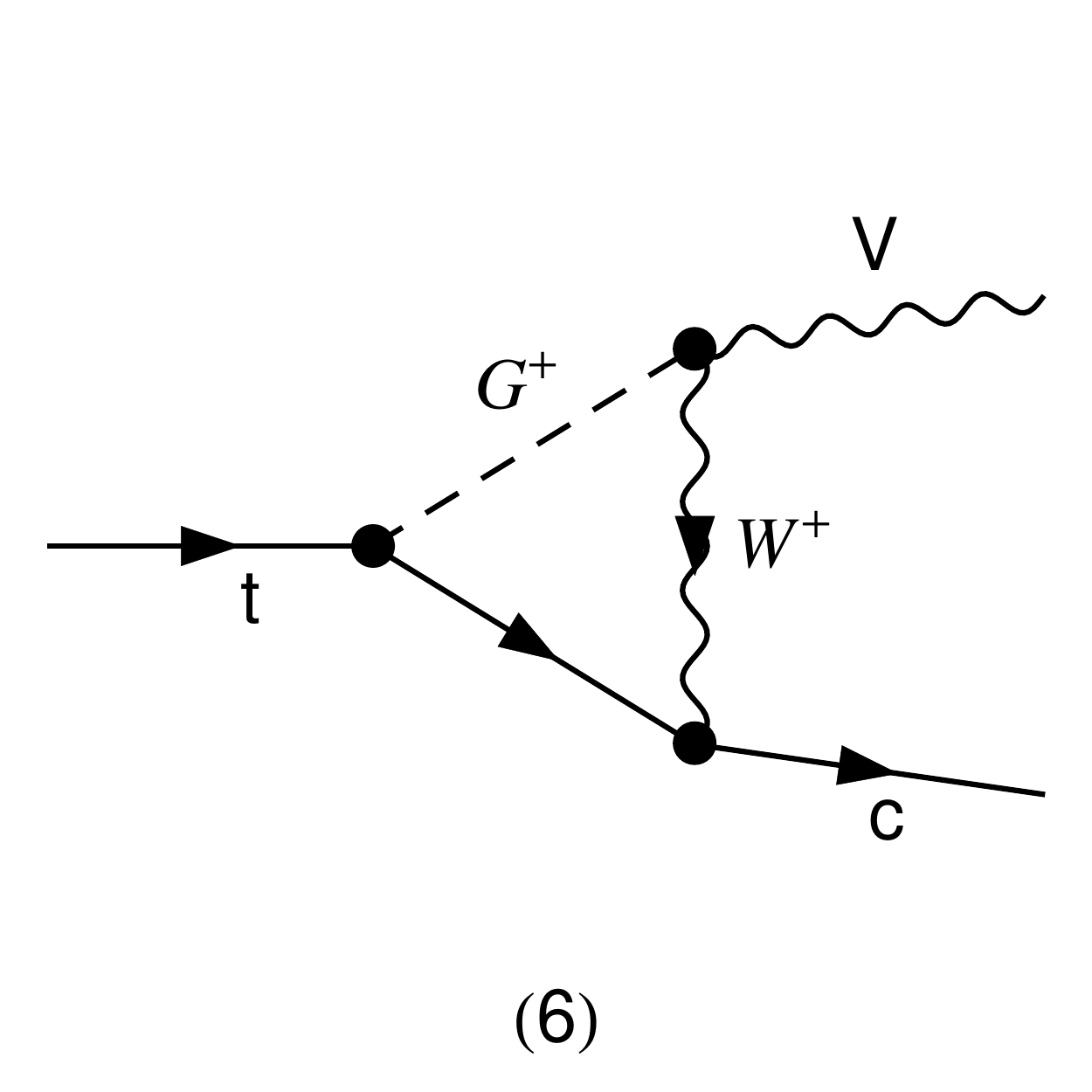}
~
\includegraphics[width=0.2\textwidth]{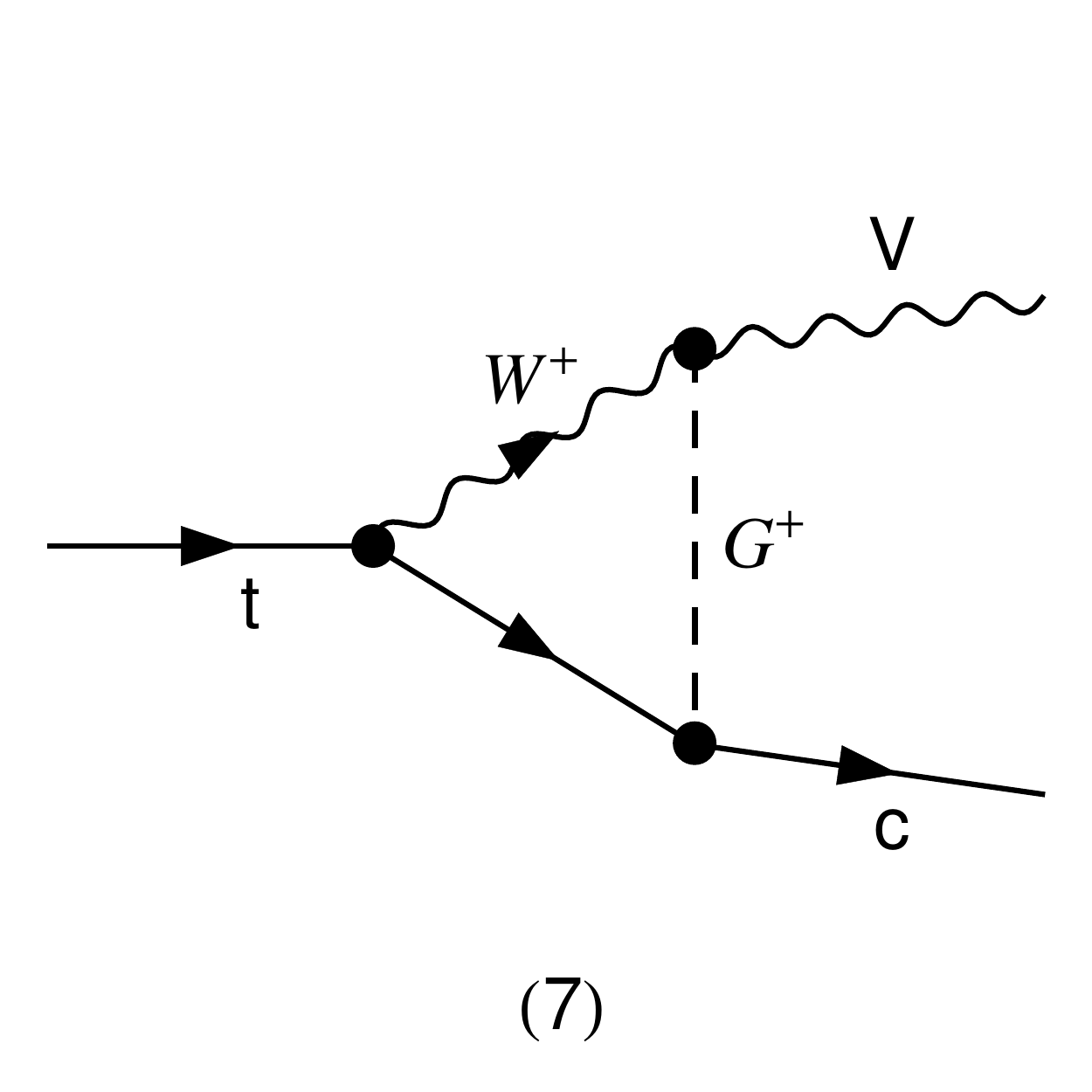}
~
\includegraphics[width=0.2\textwidth]{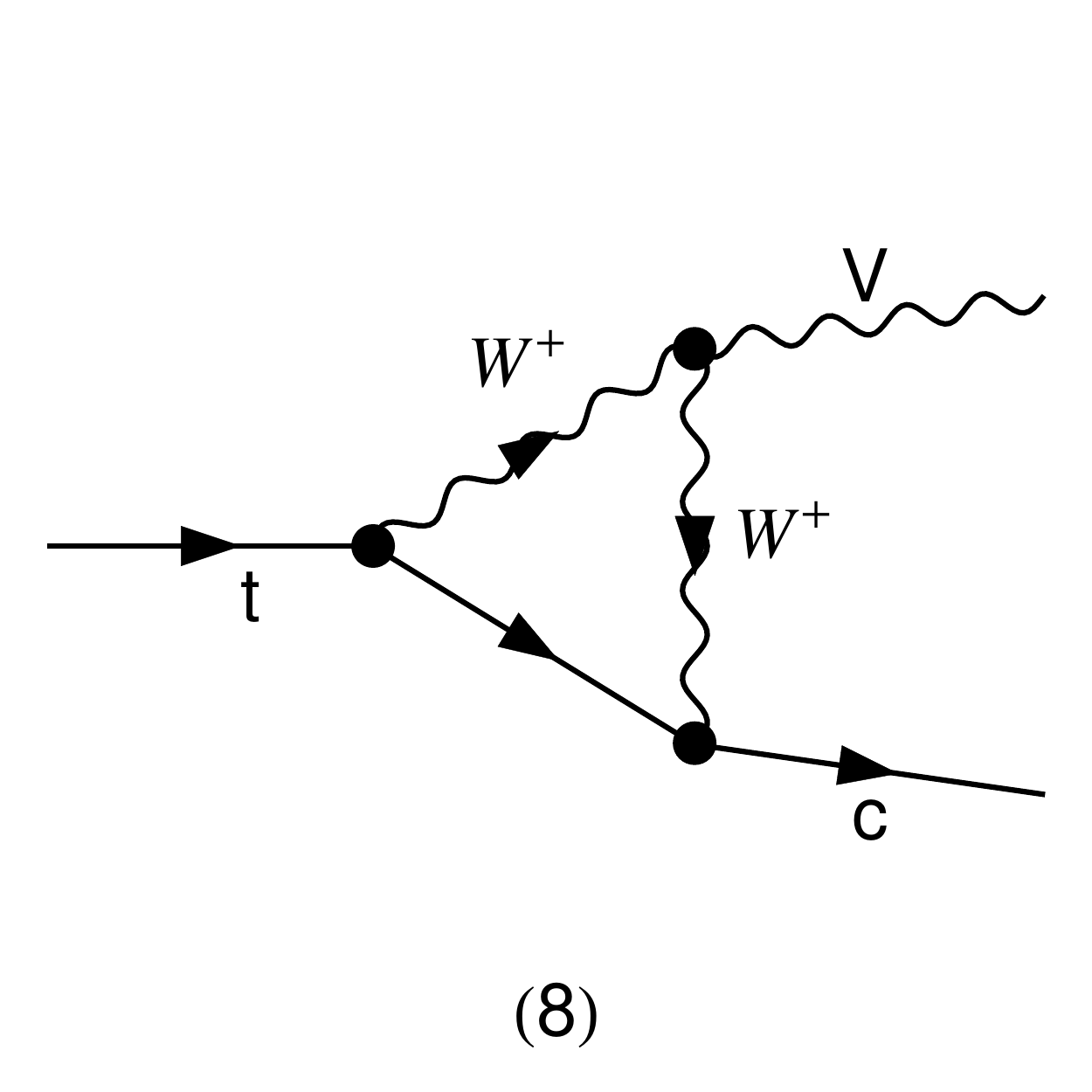}
\caption{\label{fig:diagsV} \it \small Penguin diagrams contributing to $t \rightarrow c V$ $(V=\gamma,Z)$ in the Feynman gauge.  }
\end{figure}

\begin{figure}[ht!]
\centering
\includegraphics[width=0.2\textwidth]{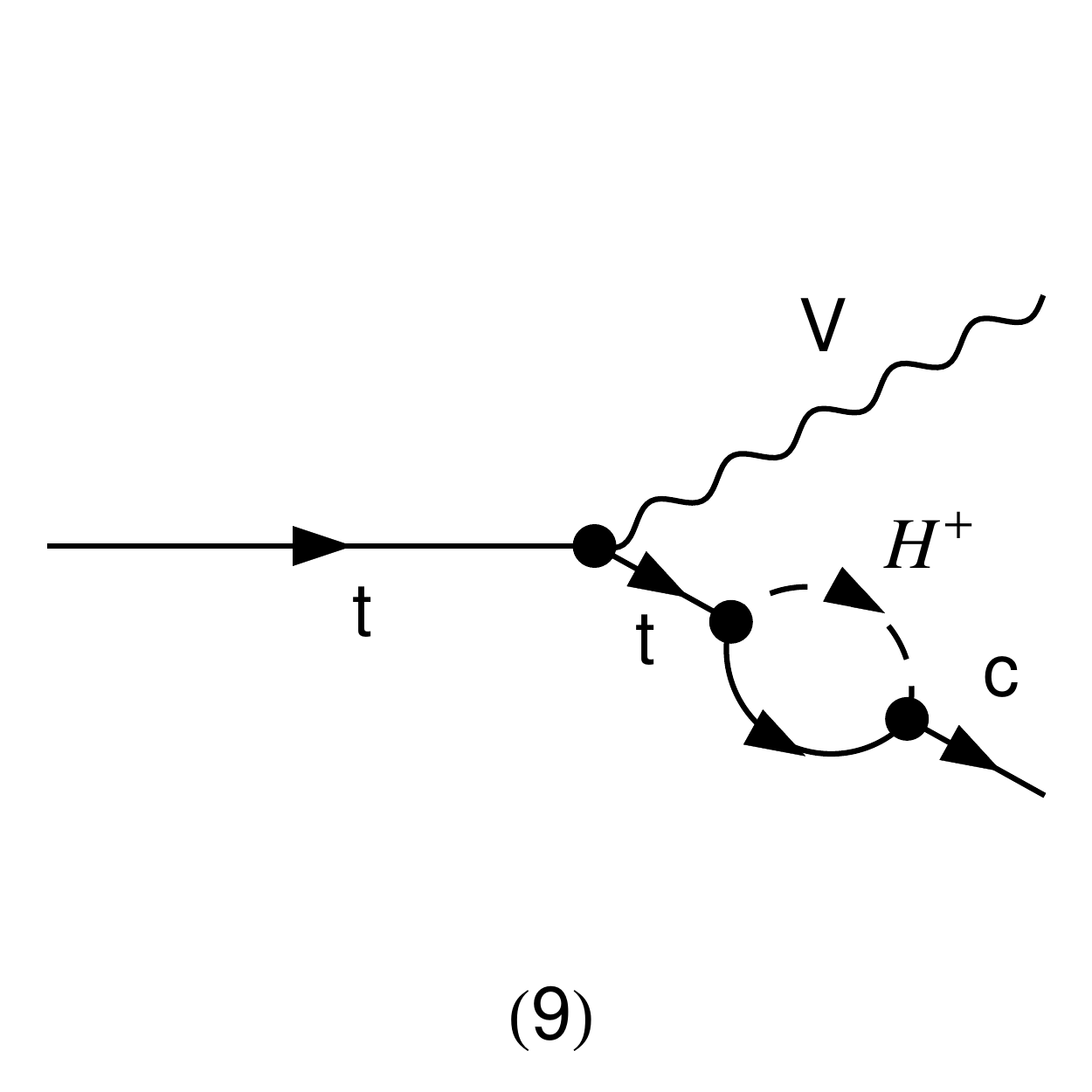}
\includegraphics[width=0.2\textwidth]{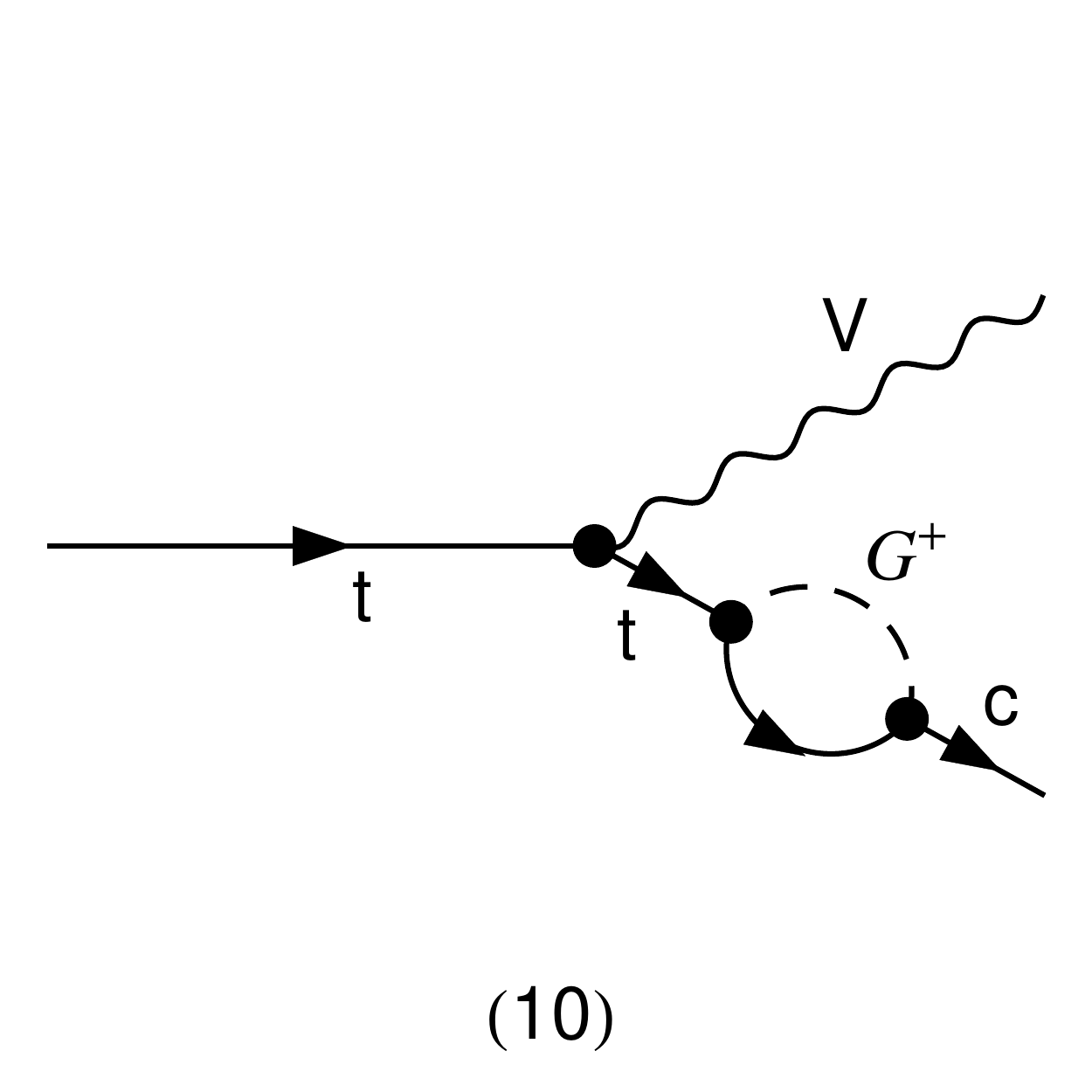}
~
\includegraphics[width=0.2\textwidth]{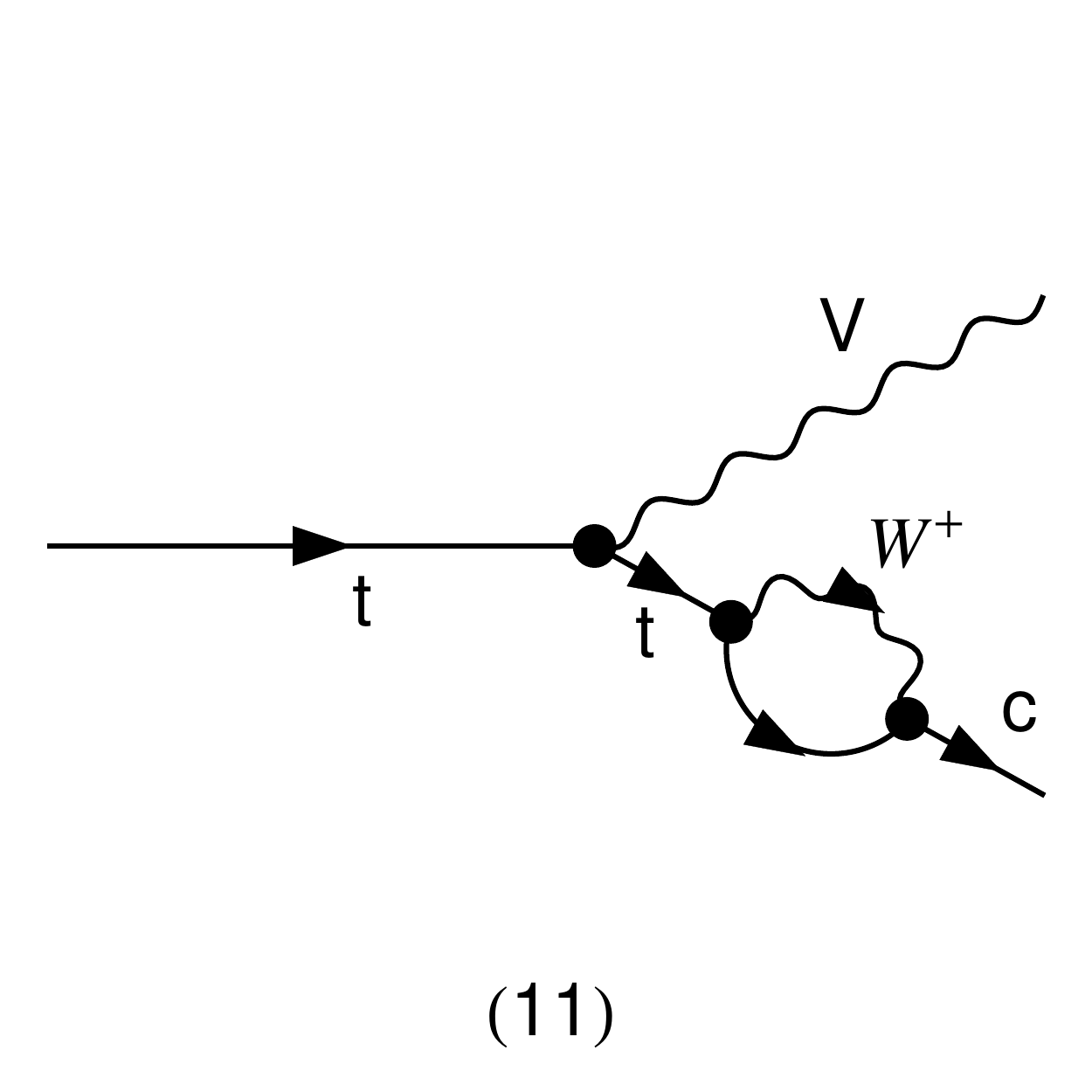}\\
\includegraphics[width=0.2\textwidth]{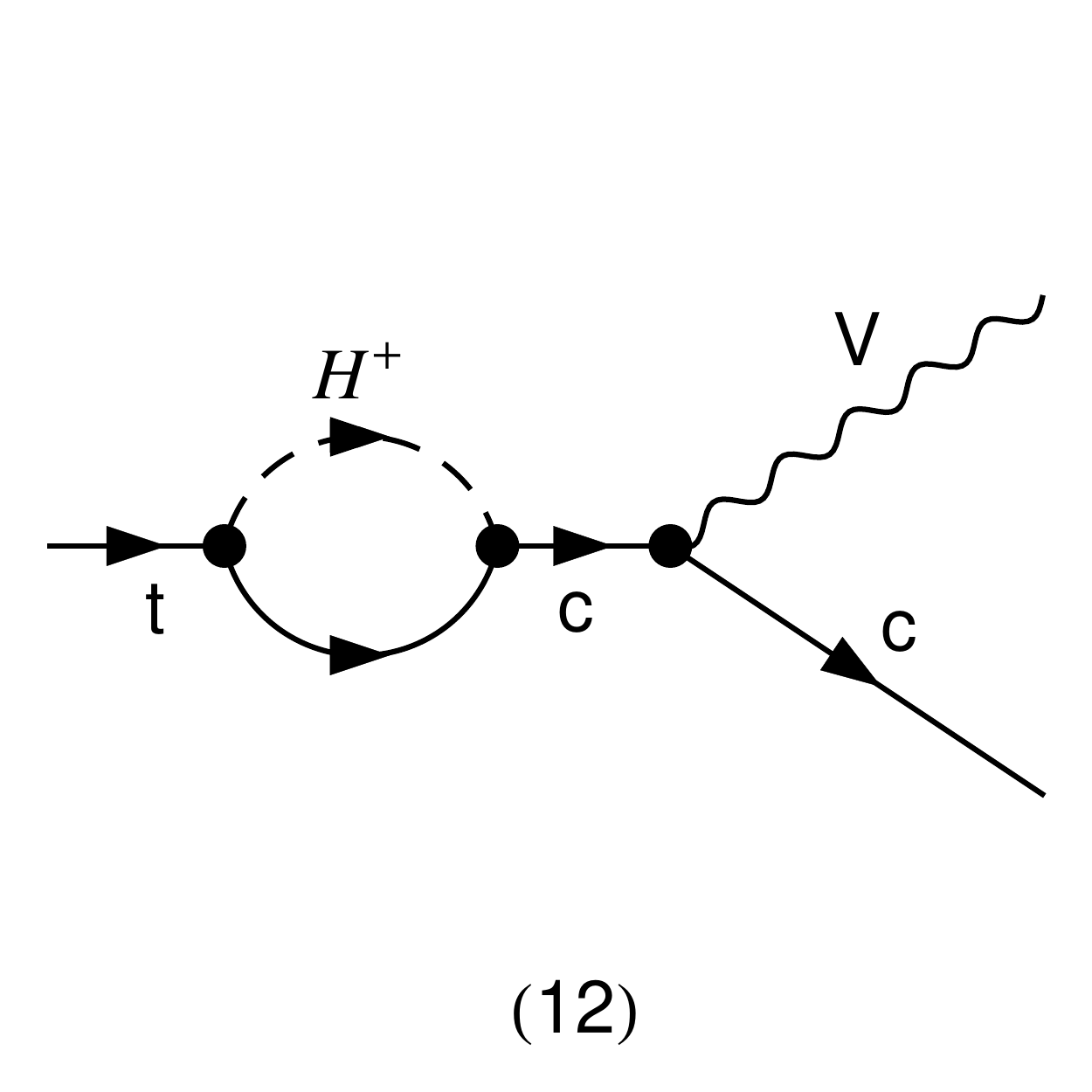}
~
\includegraphics[width=0.2\textwidth]{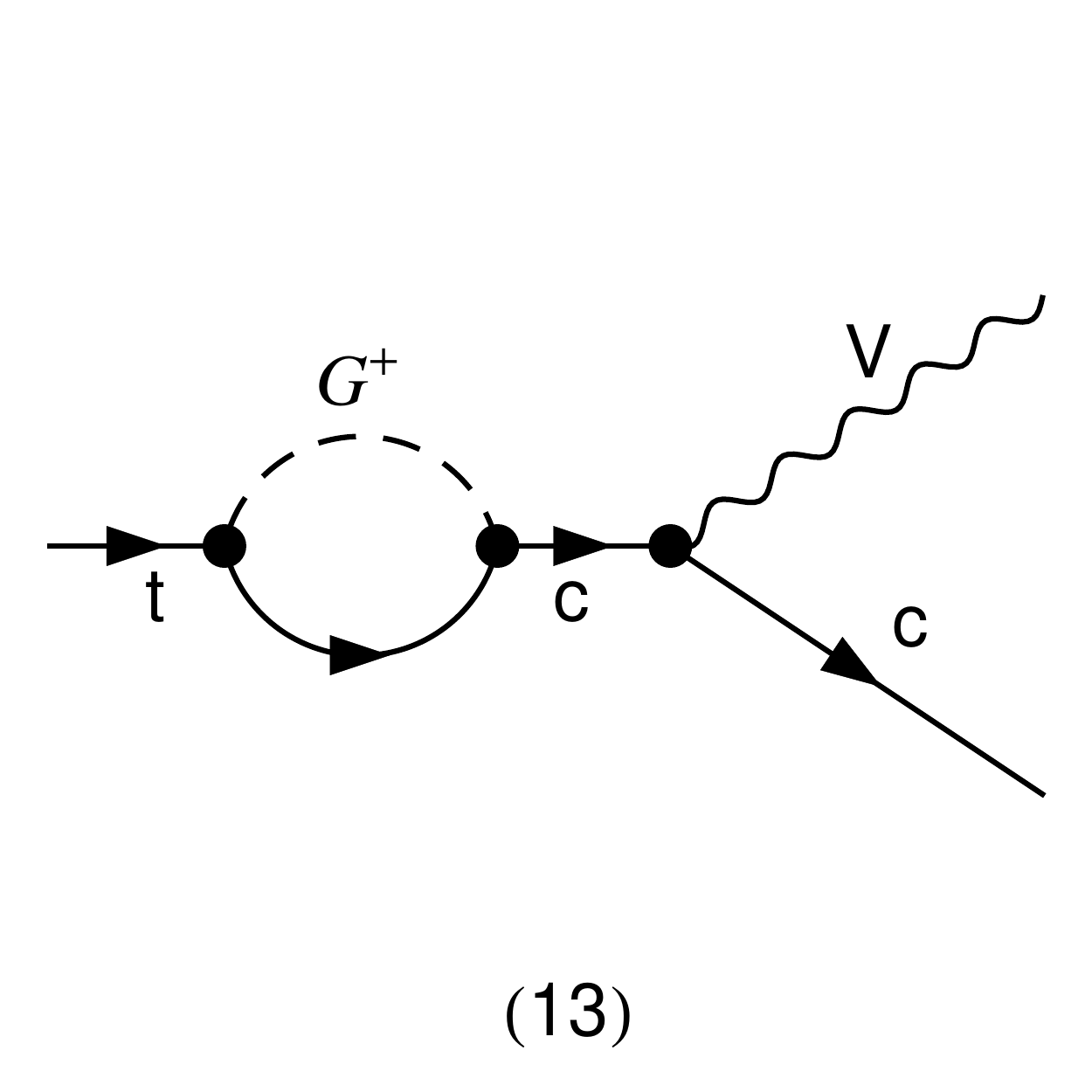}
~
\includegraphics[width=0.2\textwidth]{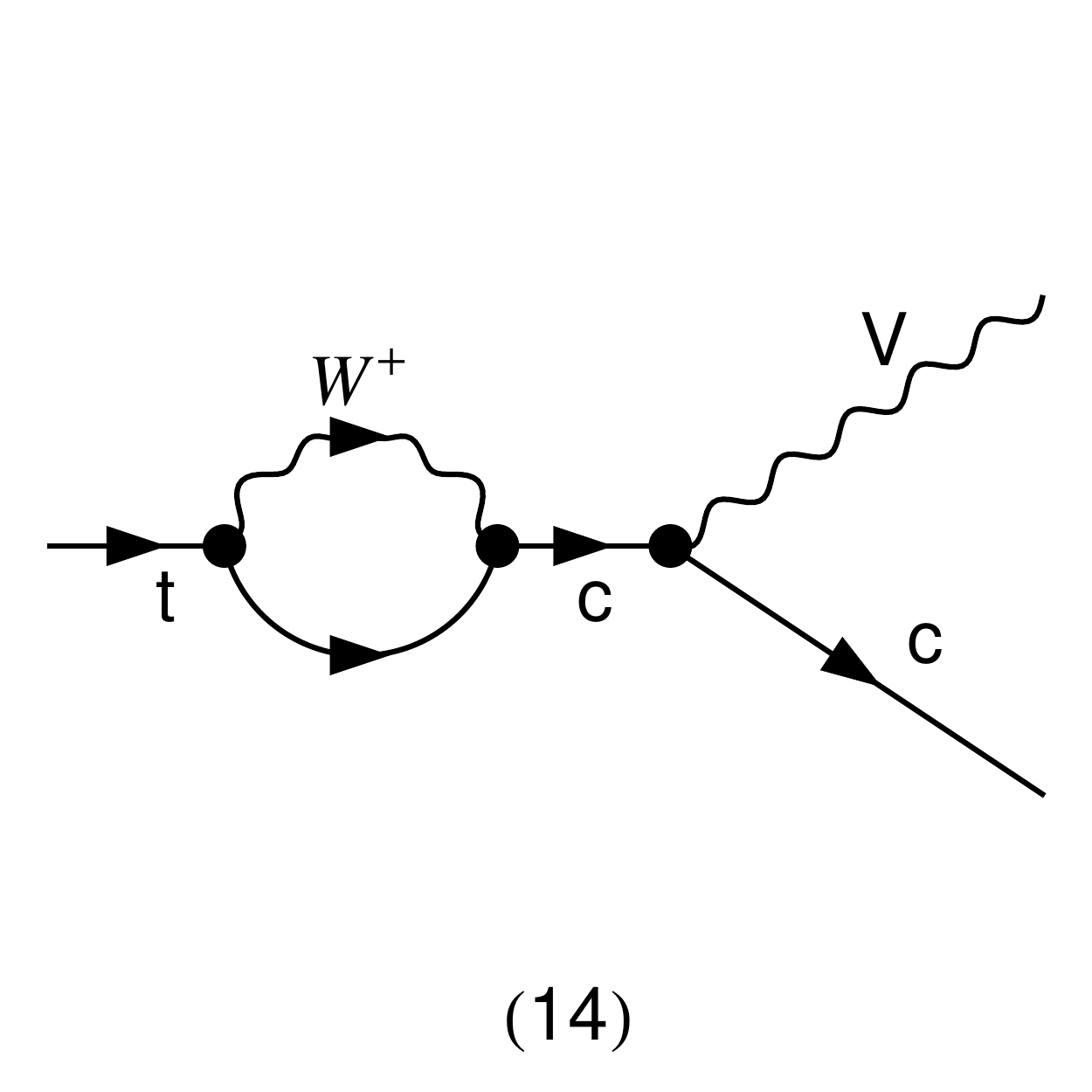}
\caption{\label{fig:diagsVII} \it \small Self-energy diagrams contributing to $t \rightarrow c V$  $(V=\gamma,Z)$ in the Feynman gauge.  }
\end{figure}

The decays $t(p_t) \rightarrow c(p_c) \, V(p_{V}) $ can only proceed at the loop level (there are no tree-level counter-terms in this case). The decay amplitude can be parametrized as~\cite{Eilam:1990zc}
\begin{align}  \label{amp:ttocV}
A \;=\; \sum_{n =1}^{14} A^{\mu}_{n} \, \epsilon_{\mu}^*(p_{V}) \;=\; & \sum_{n=1}^{14} \sum_{q=d,s,b} \, V_{cq} V_{tq}^* \; \bar u(p_c)  \biggl\{ \left[ a_{1}^{(n)} \, p_V^{\mu}  + a_{2}^{(n)}\, p_t^{\mu} + a_{3}^{(n)} \, \gamma^{\mu} \right]   \, P_L  \nonumber \\
& + \left[ b_{1}^{(n)} \,p_V^{\mu}  + b_{2}^{(n)} \,p_t^{\mu} + b_{3}^{(n)} \,\gamma^{\mu} \right] \, P_R \biggr\} u(p_t) \; \epsilon_{\mu}^*(p_{V}) \,,
\end{align}
where $\epsilon_{\mu}$ is the polarization vector of the gauge boson $V$. The one-loop diagrams contributing to this process in the Feynman gauge are shown in figures~\ref{fig:diagsV} and~\ref{fig:diagsVII}.  The total amplitude is of course ultraviolet finite.

The partial decay widths for $t \rightarrow c \, V$ ($V=\gamma, Z$) decays are given, in the limit $m_c=0$, by
\begin{align}
\Gamma(t \rightarrow c \gamma ) \;=\;& - \dfrac{m_t}{32 \, \pi }\,  \biggl\{ m_t^2\, \bigl[ |a_2|^2 + |b_2|^2  \bigr] - 2\, \bigl[ |a_3|^2 + |b_3|^2 \bigr] + m_t^2\, \mathrm{Re}(  a_1^* a_2 )  + m_t^2\, \mathrm{Re}( b_1^* b_2 ) \nonumber \\
& \qquad \qquad + 2 m_t\, \mathrm{Re}\bigl[ a_3^* (b_1+b_2)\bigr]+ 2 m_t \,\mathrm{Re}\bigl[ b_3^* (a_1 +a_2)\bigr] \biggr\} \,,
\end{align}
and
\begin{align}
\Gamma(t \rightarrow c Z ) \;=\;& \dfrac{(m_t^2-M_Z^2)^2}{128 \, \pi \, m_t^3 \, M_Z^2 } \, \biggl\{ ( m_t^2 - M_Z^2 )^2 \, \bigl[ |a_2|^2 + |b_2|^2 \bigr] + 4\, ( m_t^2 +2 M_Z^2) \, \bigl[ |a_3|^2 + |b_3|^2 \bigr] \nonumber \\
& \qquad \qquad \qquad \quad \; + 4 m_t\, (m_t^2 -M_Z^2) \, \bigl[ \mathrm{Re}(a_2 b_3^*) + \mathrm{Re}(a_3 b_2^*) \bigr] \biggr\} \,.
\end{align}
Here we have defined
\begin{align}
a_i \; =\; \sum_{q=d,s,b}  V_{cq} V_{tq}^* \, \left( \sum_{n=1}^{14} a_{i}^{(n)} \right) \,,
\qquad \quad
b_i\; =\;  \sum_{q=d,s,b} V_{cq} V_{tq}^*\, \left( \sum_{n=1}^{14} b_{i}^{(n)} \right) \,,
\end{align}
with $i=1,2,3$. For the numerical analysis we always keep finite charm mass effects into account. The contributions to the amplitude from each diagram, encoded in the coefficients $a_i^{(n)}$ and $b_i^{(n)}$, are collected in appendix~\ref{app:II}.

\section{Discussion}
\label{sec:pheno}

Relevant inputs for the evaluation of the flavour-changing top decay rates are listed in table~\ref{tab:inputs}. We assume that the combined measurement of the top-quark mass by the Tevatron and the LHC corresponds to the pole mass, but we increase its systematic error by $1$~GeV to account for the intrinsic ambiguity in the top-quark mass definition. The bottom- and charm-quark masses quoted in the table are $\overline{\rm MS}$ running masses at the quark-mass scale, {\it i.e.}, $\overline{m}_q(\overline{m}_q)$,
while the light-quark ones are $\overline{\rm MS}$ running masses renormalized at a scale of $2$~GeV. To fix the needed entries of the CKM matrix we use inputs with a minimal sensitivity to new physics contributions. Our SM predictions for the processes considered are presented in table~\ref{tab:SMpredictions}.

\begin{table}
\begin{center}
\caption{\it \small Relevant inputs for the evaluation of the decay rates.}
\vspace{0.2cm}
\doublerulesep 0.7pt \tabcolsep 0.025in
\begin{tabular}{|c|c|c||c|c|c|}
\hline \rowcolor{RGray}
Input & Value &  Comment  &   Input & Value  &  Comment   \\[0.1cm]
\hline
$G_F$  & $1.1663787(6)\times 10^{-5}$~GeV$^{-2}$    & Ref.~\cite{Agashe:2014kda}    &     $M_h$  & $125.14(24)$~GeV  &  Refs.~\cite{Aad:2014aba,Khachatryan:2014jba} \\[0.1cm]   \rowcolor{Gray}
$M_W$ & $80.385(15)$~GeV    &Ref.~\cite{Agashe:2014kda}&   $m_t$    &   $173.34(1.76)$~GeV &  Refs.~\cite{ATLAS:2014wva,Degrassi:2014sxa,Hoang:2014oea} \\[0.1cm]
$M_Z$ & $91.1876(21)$~GeV    &  Ref.~\cite{Agashe:2014kda}  &   $m_b$  &   $4.18(3)$~GeV&Ref.~\cite{Agashe:2014kda}  \\[0.1cm] \rowcolor{Gray}
$\gamma_{\text{\scriptsize{CKM}}}$ & $(73.2_{-7.0}^{+6.3})^\circ$ &  Ref.~\cite{Charles:2015gya}& $m_c$  & $1.275(25)$~GeV & Ref.~\cite{Agashe:2014kda}  \\[0.1cm]
$|V_{us}|$ & $0.2247(7)$ & Ref.~\cite{Aoki:2013ldr} &   $m_s$   &  $93.8(2.4)  \times10^{-3}$~GeV &Ref.~\cite{Aoki:2013ldr}  \\[0.1cm] \rowcolor{Gray}
$|V_{ub}|$ & $3.42(15)\times10^{-3}$ &Ref.~\cite{Aoki:2013ldr}  &   $m_d$ &  $4.68(16) \times10^{-3}$~GeV &Ref.~\cite{Aoki:2013ldr}  \\[0.1cm]
$|V_{cb}|$ & $42.21(78)\times10^{-3}$ &Refs.~\cite{Alberti:2014yda,Gambino:2013rza,Ricciardi:2014aya}  &   $\alpha_s(M_Z)$ & $0.1185(6)$ & Ref.~\cite{Agashe:2014kda}\\[0.1cm]
\hline
\end{tabular}
\label{tab:inputs}
\end{center}
\end{table}

\begin{table}
\begin{center}
\caption{\it \small  SM predictions for flavour-changing top decays.}
\vspace{0.2cm}  \tabcolsep 0.25in
\begin{tabular}{|c|c|c|c|}
\hline \rowcolor{RGray}
Observable & SM prediction   \\[0.1cm]
\hline
$\mathrm{Br}(t\rightarrow c\gamma)$  & $(4.31 \pm 0.24) \times10^{-14}$    \\[0.1cm]   \rowcolor{Gray}
$\mathrm{Br}(t\rightarrow c Z)$ & $(1.03\pm 0.06)\times10^{-14}$    \\[0.1cm]
$\mathrm{Br}(t\rightarrow c h)$ & $(3.00 \pm   0.17 )\times10^{-15}$     \\[0.1cm]
\hline
\end{tabular}
\label{tab:SMpredictions}
\end{center}
\end{table}

The ATLAS and CMS collaborations have searched for flavour-changing decays of the top quark. The ATLAS collaboration sets the bound $\mathrm{Br}(t \rightarrow q Z) < 0.73\%$ at the $95\%$ confidence level (CL), with $2.1$~fb$^{-1}$ of data at $\sqrt{s}=7$~TeV~\cite{Aad:2012ij}, where the $q$ in the final state denotes a sum over $q = u, c$. The CMS collaboration has set a better limit, $\mathrm{Br}(t \rightarrow q Z) < 0.05\%$, with $24.7$~fb$^{-1}$ of data at $\sqrt{s}=7\, \&\, 8$~TeV~\cite{Chatrchyan:2013nwa}. The strongest current bound on $t \rightarrow c \gamma$ decay has been obtained by the CMS collaboration, $\mathrm{Br}(t \rightarrow c \gamma) < 0.182\%$, using $19.1$~fb$^{-1}$ of data at $\sqrt{s}=8$~TeV~\cite{CMS:2014hwa}.\footnote{The CMS limit quoted here on $\mathrm{Br}(t \rightarrow c \gamma)$ is actually derived from a search for the anomalous single top quark production in association with a photon in proton-proton collisions, following an effective Lagrangian approach with the assumption of vanishing contributions from both $tqg$ and $tu\gamma$ interactions~\cite{CMS:2014hwa}.} The ATLAS collaboration sets the limit $\mathrm{Br}(t \rightarrow q h) < 0.79\%$, with $25$~fb$^{-1}$ of data at $\sqrt{s}=7 \,\&\,8$~TeV~\cite{Aad:2014dya}. A slightly stronger limit, $\mathrm{Br}(t \rightarrow q h) < 0.56\%$, has been obtained by the CMS collaboration, using 19.5~fb$^{-1}$ of data at $\sqrt{s}=8$~TeV~\cite{CMS:2014qxa}. Future prospects for these processes at the high luminosity LHC have been discussed in refs.~\cite{AguilarSaavedra:2000aj,Agashe:2013hma}. One expects to improve the limits to the $10^{-5}$ level for $\mathrm{Br}(t\rightarrow c V)$ ($V=\gamma, Z$), while for $\mathrm{Br}(t \rightarrow c h)$ it would be possible to reach the $10^{-4}-10^{-5}$ level.

For the phenomenological discussion we shall focus on the CP-conserving A2HDM, which contains $12$ free real parameters: $\mu_2$, $\lambda_{k}$ $(k=1,\ldots,7)$, the three alignment constants $\varsigma_{f}$ $(f=u,d,l)$ and the counter-term coupling $\mathcal{C}_R(\mu)$. Physical amplitudes are independent of the renormalization scale $\mu$, due to eq.~\eqref{eq:reno}; in the following, we choose $\mu= M_W$. Some of the parameters of the scalar potential can be traded by the physical scalar masses and the mixing angle $\tilde \alpha$.   The following relations
\begin{align}  \label{recon}
\lambda_1 &=   \frac{1}{2 v^2}  \left[ M_h^2 \, \cos^2 \tilde \alpha + M_H^2 \, \sin^2 \tilde \alpha    \right]\,,
\quad \quad
\lambda_4 =   \frac{1}{v^2} \left[  M_h^2 \, \sin^2 \tilde \alpha + M_H^2 \, \cos^2 \tilde \alpha + M_A^2 - 2 M_{H^{\pm}}^2  \right] \,,\nonumber \\[0.2cm]
\lambda_5 &=   \frac{1}{2 v^2} \left[  M_h^2 \, \sin^2 \tilde \alpha + M_H^2 \, \cos^2 \tilde \alpha - M_A^2     \right]  \,,
\quad \quad
\lambda_6 = - \frac{1}{v^2} (  M_H^2 - M_h^2  )  \cos \tilde \alpha \, \sin \tilde \alpha \,,
\end{align}
together with eq.~\eqref{eq:MHp}, allow us to work with a set of parameters more closely related to physical quantities:
\begin{align} \label{scanparam}
&\text{Scalar sector: \ } \hskip .25cm M_h, M_H, M_A, M_{H^{\pm}}, \cos \tilde \alpha ,\lambda_2,  \lambda_3, \lambda_7     \,, \nonumber \\[0.2cm]
&\text{Yukawa sector: \ } \varsigma_u, \varsigma_d, \varsigma_l, \mathcal{C}_R(M_W)\,.
\end{align}

Not all the parameters are relevant for the processes we are concerned about. The decays $t \rightarrow c \, \varphi_j^0$ are only sensitive to \{$M_{\varphi_j^0}$, $M_{H^{\pm}}$, $\cos \tilde \alpha$, $\lambda_3$, $\lambda_7$, $\varsigma_u$, $\varsigma_d$, $\mathcal{C}_{R}(M_W)$\}.  The transition amplitude does not depend on the other neutral scalar masses $M_{\varphi_i^0}$ ($i\neq j$), as can be seen explicitly from eq.~\eqref{expres_sa}. There is also no dependence on the coupling $\lambda_2$; the associated term in the scalar potential $(\Phi_2^{\dag} \Phi_2)^2$ does not generate the needed cubic vertices $H^+H^-\varphi_j^0$ because $\Phi_2$ has no vacuum expectation value (see eq.~\eqref{Higgsbasisintro}). The decays $t \rightarrow c V$ ($V= \gamma, Z$), on the other hand, depend only on $M_{H^{\pm}}$ and the alignment parameters $\varsigma_{u,d}$.  All the relevant cubic vertices are fixed in this case by the gauge symmetry and do not depend on free parameters of the scalar potential.

We assume in the following that the $125$~GeV Higgs boson corresponds to the lightest CP-even state $h$; {\it i.e.}, we fix $M_h \simeq 125$~GeV. The LHC data imply that it couples to the massive gauge vector bosons with a SM-like strength so that $\cos \tilde \alpha \simeq 1$.  We are interested in how large the enhancements of the flavour-changing top decay rates can be, compared with the SM predictions, focusing on the $125$~GeV Higgs boson in the case of $t \rightarrow c \, \varphi_j^0$ transitions. To address this question, we analyze the parameter space of the A2HDM, subject to the following assumptions and constraints:

\begin{itemize}
\item The LHC and Tevatron Higgs data imply that $\cos \tilde \alpha > 0.9$ ($68\%$ CL) and $|y^{h}_f| \sim 1$ ($f=u,d,l$)~\cite{Celis:2013rcs,Celis:2013ixa}. We work in the limit $\cos \tilde \alpha =1$ so that no constraints on the alignment parameters are obtained from the $125$~GeV Higgs data~\cite{Celis:2013rcs,Celis:2013ixa}.

\item We take into account constraints in the $\varsigma_u - \varsigma_d$ plane derived from the measurement of $\mathrm{Br}(\bar B \rightarrow X_s \gamma)$~\cite{Jung:2010ik,Jung:2012vu}.

\item We restrict the alignment parameter $|\varsigma_u|\leq 2$, in order to satisfy the constraints from $Z \rightarrow \bar b b$ decay and $B_{s,d}^0 - \bar B_{s,d}^0$ mixings~\cite{Jung:2010ik}.  The parameters $\varsigma_{d,l}$ are much less constrained phenomenologically; we take $|\varsigma_{d,\ell}|\leq 50$ as in ref.~\cite{Jung:2012vu}.

\item The four LEP collaborations, ALEPH, DELPHI, L3 and OPAL, have searched for pair-produced charged Higgs bosons in the framework of 2HDMs, excluding $M_{H^{\pm}} \lesssim 80$~GeV ($95\%$ CL) under the assumption that $H^{\pm}$ decays dominantly into fermions~\cite{Abbiendi:2013hk}.

\item Searches for a light charged Higgs via the decay $t \rightarrow H^+ b$  performed by the ATLAS~\cite{Aad:2013hla,Aad:2014kga} and CMS~\cite{CMS:2014cdp,CMS:2014kga} collaborations, together with the limits on a charged Higgs from the Tevatron~\cite{Gutierrez:2010zz}, are taken into account. These direct searches give an upper bound on the Yukawa combination $|\varsigma_u\varsigma_d|$, which, although being weaker than the one from $\mathrm{Br}(\bar{B} \rightarrow X_s \gamma)$, basically exclude one of the two possible strips allowed by the latter~\cite{Celis:2013ixa}.

\item We consider the perturbativity bound on the quartic scalar couplings $|\lambda_{3,7}| \leq 4\pi$~\cite{Celis:2013rcs}. Additionally, the loop-induced decay $h\rightarrow\gamma\gamma$ is sensitive to $\lambda_{3}$ and $\lambda_7$ through the charged Higgs contribution to this process~\cite{Celis:2013rcs,Celis:2013ixa}. We take into account the latest measurements of the Higgs signal strengths in the $h\rightarrow\gamma\gamma$ channel by ATLAS~\cite{ATLASr} and CMS~\cite{Khachatryan:2014jba}.
\end{itemize}

In the limit $\cos \tilde \alpha=1 $, the decay rate for $t \rightarrow c h $ does not depend on $\mathcal{C}_R(M_W)$ and $\lambda_7$. Explicit expressions for all the relevant cubic Higgs couplings are provided in appendix~\ref{app:I}. In particular, for $\cos \tilde \alpha =1$ we have $\lambda_{H^+ H^-}^{h} = \lambda_3$. The measured Higgs signal strengths by ATLAS and CMS in the di-photon channel are then only sensitive to $\lambda_3$ and $M_{H^{\pm}}$. Since in this case the Higgs production cross-section is the same as in the SM, one can write the Higgs signal strength in the di-photon channel as~\cite{Celis:2013rcs,Celis:2013ixa}:
\be \label{eqmu}
\mu_{\gamma\gamma}^{h}\;  =\;    \frac{ \sigma(pp \rightarrow h) \times \mathrm{Br}(h\rightarrow 2 \gamma)  }{  \sigma(pp\rightarrow h)_{\mbox{\scriptsize{SM}}} \times \mathrm{Br}(h\rightarrow 2 \gamma)_{\mbox{\scriptsize{SM}}}   } \;\simeq\; \left(1- 0.15 \,C_{H^{\pm}}^{h} \right)^2\,,
\ee
where $C_{H^{\pm}}^h$ encodes the charged Higgs contribution to $h\rightarrow 2 \gamma$ and is given by
\be
C_{H^{\pm}}^{h}\; =\; \frac{v^2}{2 M_{H^{\pm}}^2}\,  \lambda_{H^+ H^-}^{h}\,  \mathcal{A}(x_{H^{\pm}}) \,.
\ee
Here
\be
\mathcal{A}(x)\; =\; -x -\frac{x^2}{4}\, f(x)  \,,
\qquad \qquad
f(x)\; =\; -4 \arcsin^2(1/\sqrt{x}) \,,
\ee
with $x_{H^{\pm}} = 4 M_{H^{\pm}}^2/M_{h}^2$.  We require that the Higgs signal strength in eq.~\eqref{eqmu} lies within the $2\sigma$ range of the experimental measurements.  The latest results by ATLAS~\cite{ATLASr}: $\mu_{\gamma\gamma}^{h} = 1.17^{+0.28}_{-0.26}$, and by CMS~\cite{Khachatryan:2014jba}: $\mu_{\gamma\gamma}^{h} = 1.12 \pm 0.24 $, are consistent with the SM.

Performing a scan over $\{\varsigma_u$, $\varsigma_d$, $\varsigma_l$, $\lambda_3\}$, subject to the restrictions specified above, while fixing the charged Higgs mass to benchmark values, we obtain the upper bounds on $\mathrm{Br}(t \rightarrow c V)$ ($V= \gamma, Z$) and $\mathrm{Br}(t \rightarrow c h)$ shown in table~\ref{tab:scan:tcV}. In the window $90~\text{GeV}< M_{H^{\pm}}<150~\text{GeV}$ the alignment parameter $\varsigma_{d}$ is constrained to be small by the direct charged Higgs searches at the LHC via top decays, $|\varsigma_d| \lesssim 10$, implying a very strong suppression on the decay rates. For $M_{H^{\pm}} < 90~\text{GeV}$ a weaker bound on $|\varsigma_d|$ is obtained by a combination of LHC and Tevatron limits, $|\varsigma_d| \lesssim 25$. For $M_{H^{\pm}} > 150$~GeV the largest decay rates for these processes are obtained for $|\varsigma_u| < 1$ and $|\varsigma_d| \simeq 50$. The upper bounds obtained for $\mathrm{Br}(t \rightarrow c V)$ put these processes well beyond the reach of the high luminosity LHC, within the A2HDM~\cite{Agashe:2013hma}. Similar conclusions were obtained in refs.~\cite{Bejar:2000ub,Arhrib:2005nx} within the framework of 2HDMs with NFC.

The decay rate for $t \rightarrow c h$ can receive on the other hand much larger enhancements, due to the intermediate charged Higgs contribution involving the cubic Higgs coupling $\lambda_{H^+ H^-}^{h}$.  The maximum values for $\mathrm{Br}(t\rightarrow c h)$ are obtained when the cubic scalar coupling $\lambda_{H^+H^-}^{h}$ saturates either the $h\rightarrow 2\gamma$ limits or the perturbativity bound. Diagram 3 in figure~\ref{fig:diagsi} dominates the corresponding decay amplitude in this case. The contribution from this diagram to the decay amplitude is proportional to $\varsigma_u \varsigma_d \lambda_{H^+H^-}^{h}$ and $\varsigma_d^2 \lambda_{H^+H^-}^{h}$, see table~\ref{tab::topdecay1}. While the product $\varsigma_u \varsigma_d$ is constrained to be small in magnitude by $\mathrm{Br}(\bar B \rightarrow X_s \gamma)$, the term proportional to $\varsigma_d^2$ becomes greatly enhanced for large $|\varsigma_d|$ values. Such large values of $|\varsigma_d|$ can be obtained outside the window $90~\text{GeV} < M_{H^{\pm}} < 160$~GeV since the limits from direct charged Higgs searches via top decays at the LHC are avoided.

Analyses of $t \rightarrow c h$ decay within the type II 2HDM, prior to the Higgs discovery, have found that a light charged Higgs can enhance considerably the associated decay rate in this model, for large values of $\tan \beta$ and the cubic Higgs coupling $\lambda_{H^+H^-}^{h}$, and even reach the level of expected sensitivity at the high luminosity LHC: $\mathrm{Br}(t \rightarrow c h) \sim 10^{-5}$~\cite{Bejar:2000ub,Arhrib:2005nx}. Such behavior is compatible with our findings, given that in the limit $\varsigma_d = - \varsigma_u^{-1} = -\tan \beta$ we recover the Yukawa couplings of the type II 2HDM.  However, we find that current measurements of the $125$~GeV Higgs properties play an important role when evaluating possible enhancements of $\mathrm{Br}(t \rightarrow c h)$. In particular, measurements of the Higgs signal strengths in the di-photon channel restrict the allowed size of the cubic Higgs coupling $\lambda_{H^+H^-}^{h}$ for a light charged Higgs.  This in turn implies that the allowed enhancements of  $\mathrm{Br}(t \rightarrow c h)$ cannot be as large as previously speculated. Taking into account the measurements of the $125$~GeV Higgs properties, searches for a light charged Higgs via top decays, and the flavour constraints specified earlier, we find that the decay rate for $t \rightarrow c h$ lies beyond the reach of the high luminosity LHC in 2HDMs without tree-level FCNCs. Under the constraints considered the largest decay rate is obtained for $M_{H^{\pm}}$ being slightly below $90$~GeV, $\mathrm{Br}(t \rightarrow c h ) \lesssim 2 \times 10^{-7}$.

It is necessary to discuss the robustness of the previous statement. If small deviations from the limit $\cos \tilde \alpha =1$ are considered, the LHC Higgs data gives rise to strong bounds on the magnitude of the alignment parameters.  Since $|y_{f}^{h}|=  |\cos \tilde \alpha + \varsigma_f \sin \tilde \alpha|$ ($f=u,d,l$) is constrained to be close to one, one obtains $|\varsigma_f| \lesssim \mathcal{O}(1)$ when $\cos \tilde \alpha < 1$~\cite{Celis:2013rcs,Celis:2013ixa}.  This implies in particular that $|\varsigma_d|$ should be small and large enhancements of $\mathrm{Br}(t \rightarrow c h )$ are not possible. Allowing for CP violation would not led to any significant enhancement either, given the strong constraints on CP-violating couplings derived from electric dipole moment experiments~\cite{Jung:2013hka}.

\begin{table}[t]
\begin{center}
\caption{\it \small  Upper bounds for $\mathrm{Br}(t \rightarrow c V)$ ($V=\gamma,Z$) and $\mathrm{Br}(t\rightarrow c h)$ in the CP-conserving A2HDM.}
\vspace{0.2cm}  \tabcolsep 0.25in
\begin{tabular}{|c|c|c|c|c|}
\hline \rowcolor{RGray}
$M_{H^{\pm}}$~[GeV]  & $\mathrm{Br}(t\rightarrow c \gamma)$  & $\mathrm{Br}(t\rightarrow c Z)$   &  $\mathrm{Br}(t\rightarrow c h)$    \\[0.1cm]
\hline
 100 &    $ \lesssim  2 \times 10^{-12}$    &   $ \lesssim 2 \times 10^{-13}$  &    $\lesssim 6 \times 10^{-9}$  \\[0.1cm]   \rowcolor{Gray}
   200  & $ \lesssim 10^{-10}$   & $ \lesssim 3 \times 10^{-11}$  &   $\lesssim 3 \times 10^{-8}$  \\[0.1cm]
    300 &  $ \lesssim 10^{-11}$    & $ \lesssim5 \times 10^{-12}$   &  $\lesssim 2 \times 10^{-8}$ \\[0.1cm]     \rowcolor{Gray}
        400 &    $ \lesssim 2 \times 10^{-12}$  &  $ \lesssim2\times 10^{-12}$   &   $\lesssim 5 \times 10^{-9}$ \\[0.2cm]
           500 &   $ \lesssim 10^{-12}$   &  $ \lesssim 10^{-12}$ &  $\lesssim 2 \times 10^{-9}$ \\[0.1cm]   \rowcolor{Gray}
\hline  \hline
        Exp. limit &   $ <1.8\times 10^{-3}$~\cite{CMS:2014hwa}   &  $ <5\times 10^{-4}$~\cite{Chatrchyan:2013nwa}   &  $ <5.6\times10^{-3}$~\cite{CMS:2014qxa}  \\[0.1cm] \hline
\end{tabular}
\label{tab:scan:tcV}
\end{center}
\end{table}

We turn now to discuss the role of the direct counter-term contribution to $t \rightarrow c h$ decay, which is not present in 2HDMs with NFC. In the limit $\cos \tilde \alpha = 1$ this contribution vanishes because of the orthogonality of $\mathcal{R}$. The LHC data imply that $\cos \tilde \alpha$ is very close to one so that the counter-term contribution to the flavour-changing $t \rightarrow c h$ decay will be suppressed by a small factor $ \sin \tilde \alpha$ at the amplitude level. Furthermore, the characteristic flavour structure of the A2HDM counter-term~\eqref{fcnc:ct} implies a strong suppression of its effects, due to the explicit powers of quark masses and the unitarity of the quark mixing matrix~\cite{Pich:2009sp}. Neglecting the loop contribution (at $\mu=M_W$),
\begin{align}  \label{eq:tree-contrib}
\mathrm{Br}(t\to c h)_{\mathrm{tree}}\,\approx\;\, &
\frac{\alpha^2\,\pi^2\,|V_{cb}|^2\,m_b^4}{2\,\sin^4{\theta_W}\,M_W^4}
\;\frac{(1-M_h^2/m_t^2)^2}{(1-M_W^2/m_t^2)^2\,(1+2 M_W^2/m_t^2)}
\, \sin^2{\tilde\alpha}\, |E_d|^2
\nonumber \\
\;\approx\;\, & 2\times 10^{-11} \, \sin^2{\tilde\alpha}\, |E_d|^2\, ,
\end{align}
where
\be
E_d\; =\; \frac{1}{4 \pi^2}\, \mathcal{C}_R(M_W) \, \left(1 + \varsigma_u \varsigma_d\right) \left(   \varsigma_d - \varsigma_u \right)   \,.
\ee
The size of $E_d$ is constrained experimentally by the measured amount of mixing between the neutral $B_s^{0}$ meson and its antiparticle, which receives also contributions from the Lagrangian~\eqref{fcnc:ct}, mediated by the three neutral scalars $\varphi^0_i = \{h,H,A\}$. One finds that this process allows for $|E_d| \sim \mathcal{O}(1)$, even when the masses of the neutral scalars are of $\mathcal{O}(100~\text{GeV})$~\cite{Braeuninger:2010td}, but this is far too small to generate any observable signal in $t\to ch$.\footnote{
Assuming Yukawa alignment to hold at the high-energy scale $\Lambda_A$, we have $\mathcal{C}_R(M_W) = \ln(\Lambda_A/M_W)$. Therefore, $\mathcal{C}_R(M_W)/(4\pi^2) \lesssim 1$ for $\Lambda_A \lesssim 10^{19}$~GeV.}

\begin{figure}[ht!]
\centering
\includegraphics[width=0.6\textwidth]{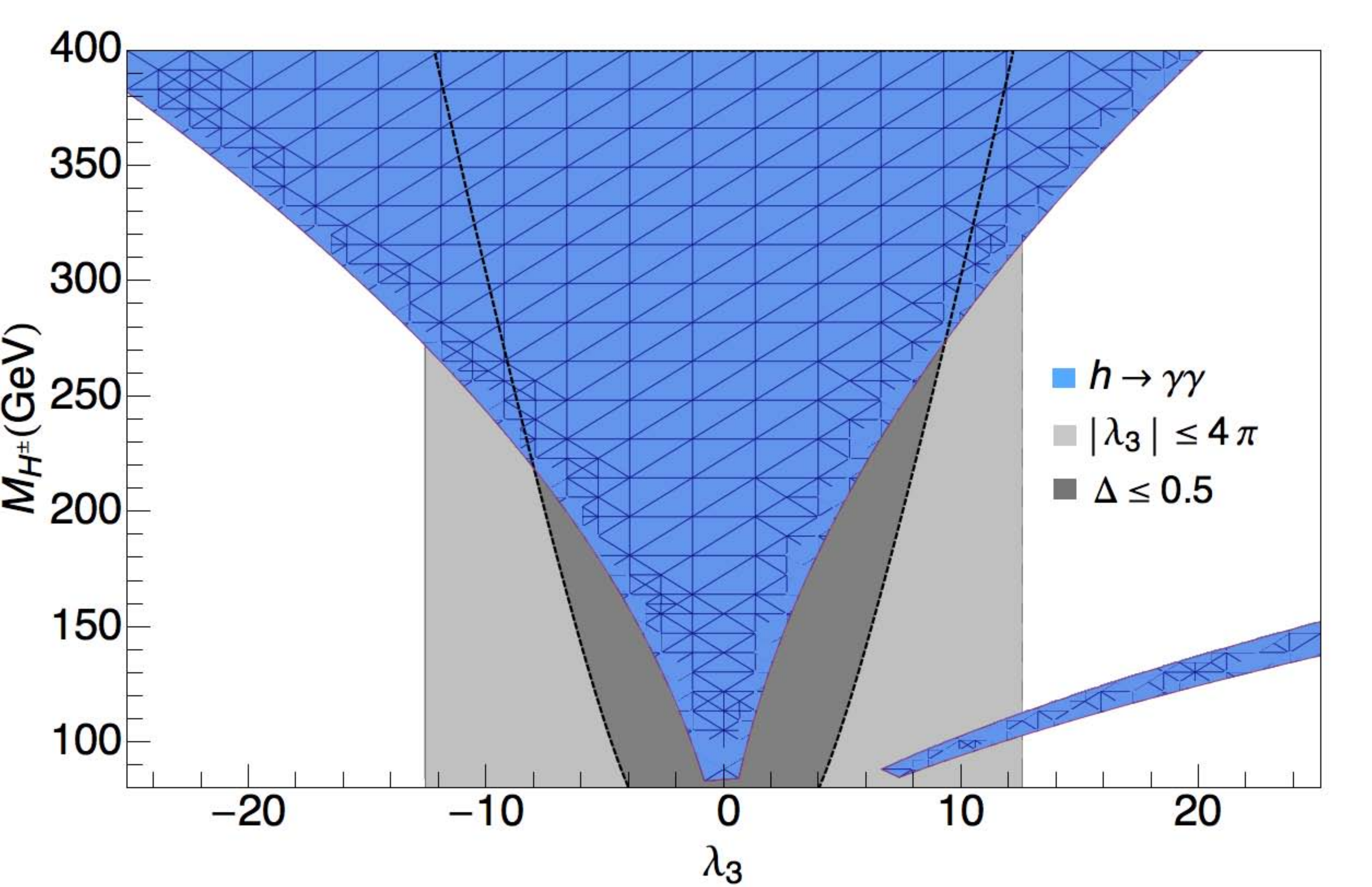}
\caption{\label{fig:final} \it \small  Allowed region by measurements of the Higgs signal strengths in the di-photon channel (blue-meshed) together with the perturbativity limits $|\lambda_3| \leqslant  4 \pi$ (light gray) or $\Delta \leqslant  0.5$ (dark gray). See text for details. }
\label{fig:allwval}
\end{figure}

It is important to analyze also the impact of the perturbativity bound on the results obtained. A different upper limit on the relevant cubic coupling $|\lambda_{H^+H^-}^{h}|$ has been considered in ref.~\cite{Celis:2013rcs}. The charged Higgs gives the following finite correction to the $h H^+H^-$ vertex, at the one-loop level:
\begin{align}
(\lambda_{H^+H^-}^{h})_{\mbox{\scriptsize{eff}}}\; =\;
\lambda_{H^+H^-}^{h}\, \left[ 1 + \frac{v^2 (\lambda_{H^+H^-}^{h})^2}{16 \pi^2 M_{H^\pm}^2}\, \mathcal{Z}\left(\frac{M_{h}^2}{M_{H^\pm}^2}\right) \right]\; \equiv\;\lambda_{H^+H^-}^{h} \; \left(1+\Delta\right) \,,
\end{align}
where
\begin{equation}
\mathcal{Z}(X)\; =\; \int_0^1 dy\; \int_0^{1 - y}  dz  \;  \left[ (y+z)^2+ X \, (1- y- z- y z)\right]^{-1} \, .
\end{equation}
Large values of $|\lambda_{H^+H^-}^{h}|$ could make this loop correction comparable to the leading-order result, which would cast doubts on the perturbative expansion. We therefore allow this correction to be at most $50\%$ ($\Delta \leqslant 0.5$). In figure~\ref{fig:allwval} we show the region allowed at $2\sigma$ by the measurement of the Higgs signal strengths in the di-photon channel, together with the bounds extracted with the perturbativity limits $|\lambda_3| \leqslant  4 \pi$ and $\Delta \leqslant 0.5$.

The constraints from $h \rightarrow \gamma \gamma$ give rise to a large allowed region, centered  around $\lambda_3=0$, whose width increases for higher values of $M_{H^{\pm}}$. In this area the $h \rightarrow \gamma \gamma$ decay amplitude is dominated by the $W$-boson and top-quark loop contributions, as in the SM; the charged Higgs contribution remains subdominant. For light charged Higgs masses a small disjoint allowed region appears with $\lambda_3 \gtrsim 6$. In this small area the charged Higgs contribution dominates over the $W$-boson and top-quark loops and flips the sign of the amplitude, $(1- 0.15 \,C_{H^{\pm}}^{h} ) \sim -1$, giving a SM-like Higgs signal strength (see eq.~\eqref{eqmu}). In principle there is no reason to expect such an accidental tuning of the charged Higgs contribution to occur. This separate small region is therefore not to be seen as very realistic. It is possible to argue that such region brings problems to the perturbative expansion. The perturbativity limit $\Delta \leqslant 0.5$, being more stringent for light charged Higgs masses, excludes this small region.   For a light charged Higgs the maximum values of $\mathrm{Br}(t\rightarrow c h)$ are obtained precisely in this separate region, where the value for $|\lambda_{H^+H^-}^{h}|$ reaches its maximum allowed value. The limits on $\mathrm{Br}(t \rightarrow c h)$ would therefore be even stronger for a light charged Higgs if the perturbativity limit $\Delta \leqslant 0.5$ is taken into account. Once this perturbativity limit is considered, we get the limit $\mathrm{Br}(t \rightarrow c h) \lesssim 6 \times 10^{-8}$.

Recent works have studied the possibility to look for flavour-changing top-quark anomalous interactions via production processes. In ref.~\cite{Durieux:2014xla} a fully gauge-invariant effective-field-theory approach was adopted for parametrizing the top-quark FCNC interactions, while in ref.~\cite{Greljo:2014dka} it was assumed that the Higgs boson posses tree-level flavour-changing couplings with the top quark. It was pointed out in these works that these top-quark flavour-changing effects can often be probed with better sensitivities in production processes than via the top-quark decays. Whether this is also the case within the A2HDM with a light charged scalar deserves a detailed analysis but lies beyond the scope of the present work.

\section{Conclusions}
\label{concl}

We have performed a complete one-loop calculation of flavour-changing top decays ($t \rightarrow c \gamma$, $t \rightarrow c Z$, $t \rightarrow c \varphi_j^0 $), within the A2HDM. Here $\varphi_j^0 = \{h, H, A \}$ represents any of the neutral scalar mass eigenstates. Our results agree with the available SM results in the literature when the corresponding limit is taken~\cite{Eilam:1989zm,Eilam:1990zc}. We have also checked the gauge independence of our results by carrying out the calculation in the Feynman and unitary gauges. Explicit analytical expressions in the Feynman gauge are provided in appendices \ref{app:I} and \ref{app:II}. The results are presented in the limit $m_c=0$, in order to avoid lengthy expressions; however, in our numerical analyses we have always used the exact expressions.

The SM predictions for these transitions are given in table~\ref{tab:SMpredictions}. They are orders of magnitude too small to be accessible even at the high-luminosity phase of the LHC. We have investigated whether significant enhancements of the branching ratios could be possible within the A2HDM. Assuming that the $125$~GeV Higgs-like boson corresponds to the lightest CP-even state $h$ of the CP-conserving A2HDM, we have discussed the impact of the relevant model parameters on the decay rates. We have taken into account the constraints from flavour experiments as well as the measurements of the Higgs-boson properties at the LHC. Upper bounds obtained for these rare top decays within the A2HDM are listed in table~\ref{tab:scan:tcV} for benchmark values of the charged Higgs mass.

While sizeable enhancements are indeed possible, compared with the SM predictions, we find that the decay rates for $t \rightarrow c V$ ($V= \gamma, Z$) remain well below the expected sensitivity levels at the high luminosity LHC, across all of the parameter space considered. As long as the charged Higgs is relatively light, the decay rate for $t \rightarrow c h$ receives much larger enhancements for large values of $\varsigma_d$ and the cubic Higgs coupling $\lambda_{H^+H^-}^{h}$. The LHC measurements of the Higgs signal strengths in the di-photon channel are found to play a very important role in estimating the maximum allowed values for $\mathrm{Br}(t \rightarrow c h)$. The charged Higgs also contributes at the loop level to the decay $h \rightarrow \gamma\gamma$ and, for a light charged Higgs, a large cubic coupling $\lambda_{H^+H^-}^{h}$ would led to large deviations of the Higgs signal strengths in the di-photon channel. We find that $\mathrm{Br}(t \rightarrow c h)$ lies also beyond the reach of the high luminosity LHC, once the constraints from the Higgs data are taken into account.

\section*{Acknowledgments}

This work has been supported in part by the Spanish Government and the European Commission [grants FPA2011-23778 and CSD2007-00042 (Consolider Project CPAN)], and the Generalitat Valenciana [PrometeoII/2013/007]. The work of A.C. is supported by the Alexander von Humboldt Foundation. The work of X.L. is supported by the National Natural Science Foundation of China~(NSFC) under contract Nos.~11005032 and 11435003. X.L. is also supported in part by the Scientific Research Foundation for the Returned Overseas Chinese Scholars, State Education Ministry, by the Open Project Program of State Key Laboratory of Theoretical Physics, Institute of Theoretical Physics, Chinese Academy of Sciences, China~(No.Y4KF081CJ1), and by the self-determined research funds of CCNU from the colleges' basic research and operation of MOE~(CCNU15A02037).

\begin{appendix}

\section{Loop functions}
\label{app:III}

Dimensional regularization is used in our calculations. The scalar loop functions appearing are given by~\cite{'tHooft:1978xw}
\begin{align}
A_0(M_1) &\;=\; \int d^{D} \tilde k \;\dfrac{1}{  k^2 -M_1^2  }\,, \nonumber \\[0.2cm]
B_0(l, M_1, M_2) &\;=\; \int d^{D} \tilde k \;\dfrac{1}{    (k^2 -M_1^2)   [    (k +l)^2 - M_2^2 ]  } \,, \nonumber \\[0.2cm]
C_0(l, s, M_1, M_2, M_3 ) &\;=\;   \int d^{D} \tilde k \;\dfrac{1}{    (k^2 - M_1^2)    [   (k+l)^2  -M_2^2 ]   [   (k+l +s)^2 -M_3^2     ]}  \,.
\end{align}
Here
\be
d^{D} \tilde k\; =\; \mu^{3 \epsilon/2}\,   \dfrac{d^{D} k}{   (2 \pi)^{D} }\,,
\ee
is the integration measure with $\epsilon = 4 - D$, and $g \mu^{\epsilon/2}$ is the $\mathrm{SU(2)}_L$ gauge coupling constant in $D$ dimensions. Vector and tensor integrals are reduced to the scalar loop integrals via the Passarino--Veltman method~\cite{Passarino:1978jh}. Following the notation of refs.~\cite{Eilam:1989zm,Axelrod:1982yc}, we have
\begin{align}
B_1 &\;=\; \frac{1}{2 l^2}   \bigl[ A_0(M_1) - A_0(M_2)  - s_1 B_0  \bigr] \,, \nonumber \\[0.2cm]
\widetilde C_0 &\;=\;   B_0(s,M_2,M_3)  + M_1^2 C_0 \,,
\end{align}
where $s_1= l^2 + M_1^2 -M_2^2$. The other relevant loop functions are given by
\begin{align}\label{eq:CsII}
\left(\ba C_{11}\\  C_{12}\ea\right)\; &= \;  Y
\left[\ba B_0(l +s,M_1, M_3) -B_0(s,M_2,M_3)  - s_1 C_0     \\[0.2cm]  B_0(l,M_1, M_2) -B_0(l+s,M_1,M_3)  - s_2 C_0 \ea\right] \,, \nonumber \\[0.3cm]
\left(\ba C_{21} \\  C_{23}\ea\right)\; &= \; Y \left[\ba B_1(l +s,M_1, M_3) +B_0(s,M_2,M_3)  - s_1 C_{11} -2 C_{24}  \\[0.2cm]  B_1(l,M_1, M_2) -B_1(l+s,M_1,M_3)  - s_2 C_{11} \ea\right] \,,
\end{align}
and
\begin{align}
C_{22} \;=\; & \dfrac{1}{  2 \bigl[  l^2 s^2 - (l \cdot s)^2 \bigr] }\; \biggl\{  - l \cdot s \Bigl[ B_1(l+s,M_1,M_3)    -B_1(s,M_2,M_3)    - s_1 C_{12}   \Bigr] \nonumber \\
&+ l^2 \Bigl[  -B_1(l+s,M_1,M_3) -s_2 C_{12}  -2 C_{24}      \Bigr]
\biggr\}    \,, \nonumber \\[0.3cm]
C_{24} \;=\; & \frac{1}{   2 ( D - 2) }   \Bigl[     B_0(s,M_2,M_3)   + 2 M_1^2 C_0   + s_1 C_{11}  + s_2 C_{12}   \Bigr]   \,.
\end{align}
Here we have defined $s_2= s^2 + 2 l \cdot s + M_2^2 - M_3^2 $ and
\bel{eq:factor}
Y = \; \dfrac{1}{  2 \bigl[ l^2 s^2 - (l \cdot s)^2 \bigr] }\,
\left[\bat   s^2   &   - l \cdot s  \\[0.2cm]
- l \cdot s  &  l^2 \ea\right]\, .
\ee

\section{Decay amplitude for \texorpdfstring{$\mathbf{\boldsymbol{t \rightarrow c \, \varphi_j^0}}$}{Lg}}
\label{app:I}

We parametrize the one-loop contribution to the $t \rightarrow c \, \varphi_j^0$ decay amplitude as indicated in eq.~\eqref{amp:ttoch}. In table~\ref{tab::topdecay1} we give the analytical expressions for the coefficients $\alpha^{(n)}$ and $\beta^{(n)}$, obtained from the $18$ Feynman diagrams in figures~\ref{fig:diagsi} and~\ref{fig:diagsii}. For simplicity, we only give the results in the limit $m_c=0$, although we have used the exact expressions, including finite charm masses, in our numerical results. In this limit, all coefficients $\beta^{(n)} =0$ ($n=1,\ldots,18$), while $\alpha^{(n)} = 0$ for $n =15, 16, 17, 18$.

We have defined the combination
\be
\gamma_{d}^{\varphi_j^0} \;\equiv\;   \mathcal{R}_{j1}  + \mathcal{R}_{j2}\;
\mathrm{Re}(\varsigma_d) -  \mathcal{R}_{j3}\;\mathrm{Im}(\varsigma_d)\, .
\ee
The matrix $\cR$ determines the neutral Higgs boson states in terms of the neutral components of the scalar doublets in the Higgs basis (see sec.~\ref{sec:frame}). The parameters $y_{u,d}^{\varphi_j^0}$ appearing in table~\ref{tab::topdecay1} have been defined in eq.~\eqref{yukascal}. The relevant cubic couplings in this case read~\cite{Li:2014fea}
\begin{align}   \label{expres_sa}
\lambda_{W^+W^-}^{\varphi_j^0} \;=\;\; & \lambda_{G^+ W^-}^{\varphi_j^0}\; =\; \cR_{j1} \,, \nonumber \\[0.2cm]
\lambda_{H^+W^-}^{\varphi_j^0}\;=\;\; & \cR_{j2} - i \,\cR_{j3} \,, \nonumber \\[0.2cm]
\lambda_{H^+H^-}^{\varphi_j^0} \;=\;\; & \lambda_3 \,\cR_{j1} + \lambda_7^R \,\cR_{j2} - \lambda_7^{I} \,\cR_{j3} \,, \nonumber \\[0.2cm]
\lambda_{G^+G^-}^{\varphi_j^0} \;=\;\; & 2 \lambda_1 \,\cR_{j1} + \lambda_6^R \,\cR_{j2} - \lambda_6^{I} \,\cR_{j3} \; = \; \frac{M_{\varphi_j^0}^2}{v^2}\; \cR_{j1} \,, \nonumber \\[0.2cm]
\lambda_{H^+ G^-}^{\varphi_j^0}\;=\;\; & \lambda_6 \,\cR_{j1}  + \frac{1}{2}\, ( \lambda_4 + 2 \lambda_5 ) \,\cR_{j2}  - \frac{i}{2}\, (\lambda_4 - 2 \lambda_5) \,\cR_{j3} \; = \; \frac{ M_{\varphi_j^0}^2 - M_{H^{\pm}}^2 }{v^2}\; ( \cR_{j2} - i\, \cR_{j3} ) \,.
\end{align}
Here $\lambda_k^{R,I}$ denote the real and imaginary parts of $\lambda_k$ respectively.

\begin{table}[t]
\setlength{\extrarowheight}{3pt}
\begin{center}
\caption{\label{tab::topdecay1} \it \small Amplitude for $t \rightarrow c\, \varphi_j^0$ in the limit $m_c =0$.  }
\vspace{0.1cm}
\doublerulesep 0.6pt \tabcolsep 0.035in
\small{
\begin{tabular}{||c|c|c||}\hline\hline
\rowcolor{RGray}
$n$    &   $\qquad \qquad  \alpha^{(n)}$                    &     argument \\[0.1cm] \hline
  1                                         &    $  - \dfrac{  \rule{0cm}{0.4cm}     g^3 m_q^2 m_t  }{  4 M_W^3 }   \varsigma_d   \Bigl\{    \varsigma_u^* \,  (y_d^{\varphi_j^0})^* \,  \tilde C_0  +   m_q^2 (   \varsigma_u - \varsigma_d   )^*  \, y_{d}^{\varphi_j^0}\,  C_0$        &  $(p_{\varphi_j^0},-p_t, m_q, m_q, M_{H^{\pm}})$   \\
          &         $-  \bigl[ 2 m_q^2 \varsigma_d^*  \gamma_{d}^{\varphi_j^0}    - M_{\varphi_j^0}^2  \varsigma_u^*  \, (y_d^{\varphi_j^0})^*  \bigr]    (C_{11} - C_{12})      \Bigr\}   $  &  \\[0.5cm]
\rowcolor{Gray}
2                               &$  \dfrac{   \rule{0cm}{0.4cm}    g^3 m_q^2 m_t  }{  4 M_W^3 }   \Bigl\{  \bigl[  2 m_q^2  \gamma_{d}^{\varphi_j^0}     - M_{\varphi_j^0}^2   (y_d^{\varphi_j^0})^*      \bigr]       (  C_{11} - C_{12} )       -  (y_d^{\varphi_j^0})^*   \tilde C_{0}      \Bigr\}     $            &   $(p_{\varphi_j^0},-p_t, m_q, m_q, M_W)$    \\[0.5cm]
3                              &$  - \dfrac{ \rule{0cm}{0.4cm}      g m_q^2 m_t  }{  M_W }  \lambda_{H^+ H^-}^{\varphi_j^0}   \varsigma_d     \Bigl\{    \varsigma_u^*  C_0 + \varsigma_d^* (   C_{11} - C_{12} )      \Bigr\}   $            &  $(p_{\varphi_j^0},-p_t, M_{H^{\pm}}, M_{H^{\pm}}, m_q)$ \\[0.5cm]
\rowcolor{Gray}
      4   &   $  - \dfrac{   \rule{0cm}{0.4cm}    g m_q^2 m_t  }{  M_W }  \lambda_{H^+ G^-}^{\varphi_j^0}     \Bigl\{    \varsigma_u^*  C_0 + \varsigma_d^* (   C_{11} - C_{12} )      \Bigr\}    $              & $(p_{\varphi_j^0},-p_t, M_{W}, M_{H^{\pm}}, m_q)$  \\[0.5cm]
         5      &  $ - \dfrac{   \rule{0cm}{0.4cm}    g m_q^2 m_t  }{  M_W }  (\lambda_{H^+ G^-}^{\varphi_j^0})^*   \varsigma_d     \Bigl(     C_0 +    C_{11} - C_{12}       \Bigr)  $       & $ (p_{\varphi_j^0},-p_t, M_{H^{\pm}}, M_{W},m_q) $ \\[0.5cm]
          \rowcolor{Gray}
         6                                     &$ - \dfrac{ \rule{0cm}{0.4cm}   g m_q^2 m_t  }{ M_W }  \lambda_{G^+ G^-}^{\varphi_j^0}  \Bigl(   C_0 + C_{11} - C_{12}           \Bigr)     $                  & $ (p_{\varphi_j^0},-p_t, M_{W}, M_{W},m_q)  $  \\[0.5cm]
                 7                                   &  $  ( D - 2 )   \dfrac{ \rule{0cm}{0.4cm}      g^3 m_q^2 m_t  }{  4 M_W }   \Bigl\{     (y_d^{\varphi_j^0})^* C_0    +  2 \gamma_{d}^{\varphi_j^0}   ( C_{11}   - C_{12})         \Bigr\}    $                &  $(p_{\varphi_j^0},-p_t, m_q, m_q, M_W) $ \\[0.5cm]
                 \rowcolor{Gray}
          8   &  $-   \dfrac{  \rule{0cm}{0.4cm}     g^3  m_t  }{  4 M_W }   \lambda_{H^+ W^-  }^{ \varphi_j^0 }       \Bigl\{  \varsigma_u^*   \tilde  C_0 + 2 \bigl[  m_q^2 \varsigma_d^* + \varsigma_u^* (  M_{\varphi_j^0}^2 - m_t^2 )  \bigr] C_0 $ &  $(p_{\varphi_j^0},-p_t, M_{W}, M_{H^{\pm}}, m_q) $  \\   \rowcolor{Gray}
          & $+ \bigl[  m_q^2  \varsigma_d^* +( 3 M_{\varphi_j^0}^2 - m_t^2 ) \varsigma_u^*     \bigr]     C_{11}  - \bigl[   m_q^2 \varsigma_d^* +  \varsigma_u^*  (    M_{\varphi_j^0}^2  + m_t^2  )  \bigr] C_{12}    \Bigr\}  $  &      \\[0.5cm]
                    9   & $ \dfrac{ \rule{0cm}{0.4cm}   g^3 m_t  }  {  4 M_W }   \lambda_{ G^+ W^- }^{ \varphi_j^0 }  \Bigl\{    - \tilde C_{0}  - 2 (    m_q^2 + M_{\varphi_j^0}^2   - m_t^2 )  C_0   $ &   $(p_{\varphi_j^0},-p_t, M_{W}, M_{W},m_q) $ \\
                      &    $+ (  m_t^2 - m_q^2 - 3  M_{\varphi_j^0}^2  )  C_{11}  + (   m_q^2 +  M_{\varphi_j^0}^2 + m_t^2 ) C_{12}     \Bigr\} $ &    \\[0.5cm]     \rowcolor{Gray}
                        10   & $  \dfrac{ \rule{0cm}{0.4cm}      g^3   m_q^2 m_t  }{  4 M_W }    \varsigma_d   (  \lambda_{H^+ W^-}^{ \varphi_j^0 }  )^*          \Bigl\{     C_0 - C_{11}   + C_{12}    \Bigr\}    $ & $(p_{\varphi_j^0},-p_t, M_{H^{\pm}}, M_{W}, m_q) $  \\[0.5cm]
                        11   & $ \dfrac{ \rule{0cm}{0.4cm}   g^3 m_q^2 m_t   }   {   4 M_W }    (\lambda_{ G^+ W^- }^{ \varphi_j^0 } )^*    \Bigl\{   C_0 - C_{11} + C_{12}       \Bigr\}     $ & $(p_{\varphi_j^0},-p_t, M_{W}, M_{W},m_q) $  \\[0.5cm]    \rowcolor{Gray}
                                  12   & $- \dfrac{1\rule{0cm}{0.4cm}  }{2}  (  D - 2 )   g^3 m_t M_W   \, \lambda_{W^+W^-}^{\varphi_j^0} \,   \left(   C_{11} - C_{12} \right)    $& $(p_{\varphi_j^0},-p_t, M_{W}, M_{W},m_q) $   \\[0.5cm]
                                            13  & $ \dfrac{  \rule{0cm}{0.4cm}     g^3 m_q^2 m_t  }{  4 M_W^3 }   \varsigma_d \varsigma_u^*  \, y_{u}^{\varphi_j^0} \,   B_0 $& $(-p_c, M_{H^{\pm}}, m_q)$  \\[0.5cm]     \rowcolor{Gray}
                                                      14   & $   \dfrac{g^3\rule{0cm}{0.4cm}  }{    4 M_W^3}   m_q^2 m_t \, y_u^{\varphi_j^0} \, B_0   $ & $ (-p_c, m_q, M_W) $  \\[0.4cm]
\hline\hline
\end{tabular}}
\end{center}
\end{table}

\section{Decay amplitude for \texorpdfstring{$\mathbf{\boldsymbol{t \rightarrow c \, V}}$}{Lg}}
\label{app:II}

Following the notation of eq.~\eqref{amp:ttocV}, all non-vanishing contributions to the $t \rightarrow c V$ ($V=\gamma,Z$) decay amplitude have been given, for $m_c=0$, in tables~\ref{tab::topdecay2}, \ref{tab::topdecay3} and \ref{tab::topdecay4}. Here we have defined
\begin{align}  \label{legneq}
g_{Vq}\;=\;A_{Vq} + B_{Vq} \,  \gamma_5  \,,
\end{align}
and
\be \label{legf}
S_q \;\equiv\; A_{Vq}  + B_{Vq} \,,
\qquad \qquad
P_q \;\equiv\; A_{Vq}  - B_{Vq}  \,,
\ee
with $q=(u,d)$. Values for the relevant constants are given in tables~\ref{tab:legend} and \ref{tab:legendII}. The weak mixing angle is determined by $e = g \sin \theta_W$, $M_W =g v/2$ and $M_W = M_Z \cos \theta_W$. We use the notations: $s_W \equiv \sin \theta_W$ and $c_W \equiv \cos \theta_W$.

\begin{table}[t]
\setlength{\extrarowheight}{3pt}
\begin{center}
\caption{\label{tab::topdecay2} \it \small Amplitude for $t \rightarrow c\, V$ $(V=\gamma, Z)$ in the limit $m_c =0$: Coefficients $a_3^{(n)}$. }
\vspace{0.2cm}
\doublerulesep 0.7pt \tabcolsep 0.09in
\small{
\begin{tabular}{||c|c|c||}\hline\hline
\rowcolor{RGray}
$n$    &   $\qquad \qquad  a_3^{(n)}$                    &     argument \\[0.1cm] \hline
  1                                         &    $ \dfrac{\rule{0cm}{0.4cm}  i  g^2 m_q^2}{  2 M_W^2}    \varsigma_d   \Bigl\{     - \varsigma_d^* S_d \widetilde C_0   + P_d (   \varsigma_d^* m_q^2  -  \varsigma_u^* m_t^2  )   C_0           $  &  $(p_V,-p_t,m_q,m_q,M_{H^{\pm}})$       \\
          &         $    - \bigl[     \varsigma_d^*  M_V^2 S_d - 2  \varsigma_u^*  B_{Vd}  m_t^2  \bigr]   (    C_{11} - C_{22} )    + 2 \varsigma_d^*  S_d C_{24}  \Bigr\}  $  &       \\[0.5cm]
\rowcolor{Gray}
2                               &$  \dfrac{\rule{0cm}{0.4cm}   i  g^2 m_q^2 }{2 M_W^2}    \Bigl\{   - S_d \widetilde C_0 + P_d (m_b^2 -m_t^2) C_0     $            &   $(p_{V},-p_t, m_q, m_q, M_W)$      \\ \rowcolor{Gray}
& $+ \bigl[2 B_{Vd}    m_t^2 - M_V^2 S_d \bigr]  ( C_{11} - C_{12})    + 2 S_d C_{24}    \Bigr\}  $  &        \\[0.5cm]
3                              &$   - \dfrac{ \rule{0cm}{0.4cm}    g^2 m_q^2 |\varsigma_d|^2 }{     M_W^2 }   g_{VH^+H^-}  C_{24}   $            &  $(p_{V},-p_t, M_{H^{\pm}}, M_{H^{\pm}}, m_q)$ \\[0.5cm]
\rowcolor{Gray}
      4   &   $  - \dfrac{ \rule{0cm}{0.4cm}    g^2 m_q^2  }{     M_W^2 }   g_{VG^+G^-}  C_{24}    $              & $(p_{V},-p_t, M_{W}, M_{W}, m_q)$  \\[0.5cm]
         5      &  $   - \dfrac{i g^2\rule{0cm}{0.4cm}  }{2}  \Bigl\{   (D-2)  P_d \widetilde C_0 - (D-2)   m_q^2   S_d C_0     + P_d  \bigl[      (D-2)  M_V^2 C_{11}  $  & $ (p_{V},-p_t, m_q, m_q, M_W) $    \\
         & $\rule{0cm}{0.4cm}     -   \left(  (D-4)  M_V^2 + 2 m_t^2     \right)     C_{12}    - 2 (D-2)   C_{24}   \bigr]   \Bigr\}      $   &             \\[0.5cm]
          \rowcolor{Gray}
         6                                     &$     - \dfrac{g^2\rule{0cm}{0.4cm}  }{2 M_W}    g_{VG^+W^-}    \Bigl\{  m_q^2 C_0 + m_t^2 (  C_{11} - C_{12} )   \Bigr\}      $                  & $ (p_{V},-p_t, M_{W}, M_{W},m_q)  $  \\[0.5cm]
                 7                                   &  $  - \dfrac{ g^2  m_q^2\rule{0cm}{0.4cm}  }{  2 M_W } g_{VG^+W^-} C_0     $                &  $(p_{V},-p_t, M_W, M_W, m_q) $ \\[0.5cm]
                 \rowcolor{Gray}
          8   &  $\dfrac{g^2\rule{0cm}{0.4cm}  }{2}   g_{VW^+W^-}   \Bigl\{   -2 \tilde C_0   + 2 (  m_t^2 - M_V^2)   C_0        $ &  $(p_{V},-p_t, M_{W}, M_{W}, m_q) $  \\   \rowcolor{Gray}
          & $ + (m_t^2 -2 M_V^2)  C_{11}      + m_t^2 C_{12}   -2 (D-2)  C_{24}   \Bigr\}      $  &      \\[0.5cm]
                    9   & $  i \dfrac{g^2 m_q^2\rule{0cm}{0.4cm}   }{2 M_W^2}   \varsigma_d \varsigma_u^* P_u B_0     $ &   $(p_{V}-p_t, M_{H^{\pm}},m_q) $     \\[0.5cm]     \rowcolor{Gray}
                        10   & $ i  \dfrac{g^2 m_q^2\rule{0cm}{0.4cm}   }{  2 M_W^2 }  P_u B_0 $ & $(p_{V} -p_t, M_{W},  m_q) $  \\[0.5cm]
                                  12   & $ i \dfrac{ g^2 m_q^2\rule{0cm}{0.4cm}   }{  2 M_W^2 } \varsigma_d  P_u \Bigl\{   ( \varsigma_d^* - \varsigma_u^* )   B_0 + \varsigma_d^*  B_1     \Bigr\}  $  & $(-p_t, M_{H^{\pm}},m_q) $   \\[0.5cm] \rowcolor{Gray}
                                            13  & $i \dfrac{g^2 m_q^2\rule{0cm}{0.4cm}  }{  2 M_W^2}   P_u B_1 $& $(-p_t, M_{W}, m_q)$  \\[0.5cm]
                                                      14   & $ \dfrac{i\rule{0cm}{0.4cm}  }{2}  (D-2)  g^2 P_u \left(B_0 +B_1\right)  $ & $ (-p_t, M_W, m_q) $  \\[0.3cm]
\hline\hline
\end{tabular}}
\end{center}
\end{table}

\begin{table}[t]
\setlength{\extrarowheight}{3pt}
\begin{center}
\caption{\label{tab::topdecay3} \it \small Amplitude for $t \rightarrow c\, V$ $(V=\gamma, Z)$ in the limit $m_c =0$: Coefficients $b_1^{(n)}$. }
\vspace{0.2cm}
\doublerulesep 0.7pt \tabcolsep 0.035in
\small{
\begin{tabular}{||c|c|c||}\hline\hline
\rowcolor{RGray}
$n$    &   $\qquad \qquad  b_1^{(n)}$                    &     argument \\[0.1cm] \hline
  1                                         &    $ i \dfrac{   g^2 m_q^2 m_t  \rule{0cm}{0.4cm}  }{   M_W^2 }  \varsigma_d  \Bigl\{   \varsigma_u^*  P_d C_0  + (   \varsigma_d^* S_d      -   \varsigma_u^*  P_d  )   C_{11}       - \varsigma_d^*  S_d  (  C_{12}     - C_{21}  + C_{23} )  \Bigr\}        $  &  $(p_V,-p_t,m_q,m_q,M_{H^{\pm}})$       \\
\rowcolor{Gray}
2                               &$   -  i\dfrac{g^2 m_q^2 m_t \rule{0cm}{0.4cm}   }{M_W^2}   \Bigl\{   P_d C_0 + S_d   ( C_{12} - C_{21}   + C_{23} )   - 2 B_{Vd}   C_{11}        \Bigr\}        $            &   $(p_{V},-p_t, m_q, m_q, M_W)$       \\[0.5cm]
3                              &$ \dfrac{-g^2 m_q^2 m_t  \rule{0cm}{0.4cm}  }{  2 M_W^2  }   g_{VH^+H^-}  \varsigma_d     \Bigl\{     \varsigma_u^*   C_0  + (   \varsigma_d^* + 2 \varsigma_u^*  )        C_{11}   - \varsigma_d^*  (  C_{12} - 2 C_{21}     + 2 C_{23}  )     \Bigr\}  $            &  $(p_{V},-p_t, M_{H^{\pm}}, M_{H^{\pm}}, m_q)$ \\[0.5cm]
\rowcolor{Gray}
      4   &   $- \dfrac{g^2 m_q^2 m_t  \rule{0cm}{0.4cm}   }{ 2 M_W^2}   g_{VG^+G^-}     \Bigl(    C_0 + 3 C_{11} - C_{12}   + 2 C_{21}  - 2 C_{23}   \Bigr)  $              & $(p_{V},-p_t, M_{W}, M_{W}, m_q)$  \\[0.5cm]
         5      &  $\rule{0cm}{0.4cm}   i g^2 m_t P_d  \Bigl\{   (D-2)   C_{11}    - (D-4)   C_{12}  + (D-2)   \bigl[ C_{21} - C_{23} \bigr]    \Bigr\} $  &   $(p_V,-p_t,m_q,m_q,M_W)$        \\[0.5cm]
          \rowcolor{Gray}
         6                                     &$  - \dfrac{   g^2 m_t \rule{0cm}{0.4cm}   }{M_W}      g_{VG^+W^-}    \Bigl(    C_0 + C_{11}  \Bigr)    $                  & $ (p_{V},-p_t, M_{W}, M_{W},m_q)  $  \\[0.5cm]
                 8                                   &  $  \dfrac{g^2 m_t \rule{0cm}{0.4cm}   }{2}   g_{VW^+W^-}    \Bigl\{   2 C_0 - D \, C_{11}   + (D+2)  C_{12}  -2 (D-2) \bigl[    C_{21}     -C_{23} \bigr]     \Bigr\}    $                &  $(p_{V},-p_t, M_W, M_W, m_q) $     \\[0.3cm]
\hline\hline
\end{tabular}}
\end{center}
\end{table}

\begin{table}[t]
\setlength{\extrarowheight}{3pt}
\begin{center}
\caption{\label{tab::topdecay4} \it \small Amplitude for $t \rightarrow c\, V$ $(V=\gamma, Z)$ in the limit $m_c =0$:  Coefficients $b_2^{(n)}$.   }
\vspace{0.2cm}
\doublerulesep 0.7pt \tabcolsep 0.035in
\small{
\begin{tabular}{||c|c|c||}\hline\hline
\rowcolor{RGray}
$n$    &   $\qquad \qquad  b_2^{(n)}$                    &     argument \\[0.1cm]  \hline
  1                                         &    $\dfrac{  i g^2   m_q^2 m_t  \rule{0cm}{0.4cm}  }{  M_W^2}  \varsigma_d \Bigl\{   \varsigma_u^*   P_d C_0 + S_d   \bigl(      \varsigma_u^*   C_{12}  + \varsigma_d^* C_{22}   - \varsigma_{d}^*  C_{23}      \bigr)  - 2 B_{Vd}  \varsigma_u^* C_{11}       \Bigr\}              $  &  $(p_V,-p_t,m_q,m_q,M_{H^{\pm}})$           \\[0.5cm]
\rowcolor{Gray}
2                               &$\dfrac{   i g^2 m_q^2 m_t \rule{0cm}{0.4cm}   }{M_W^2}  \Bigl\{   P_d C_0 +  S_d (    C_{12}  + C_{22}     - C_{23}  )    - 2 B_{Vd}  C_{11}      \Bigr\}       $            &   $(p_{V},-p_t, m_q, m_q, M_W)$      \\[0.5cm]
3                              &$\dfrac{      m_q^2 m_t   \rule{0cm}{0.4cm}   }{M_W^2}  g^2 g_{VH^+H^-}     \varsigma_d \Bigl\{        \varsigma_u^*   C_{12}   + \varsigma_d^* (   C_{23} - C_{22} )             \Bigr\}  $            &  $(p_{V},-p_t, M_{H^{\pm}}, M_{H^{\pm}}, m_q)$ \\[0.5cm]
\rowcolor{Gray}
      4   &   $\dfrac{   g^2 m_q^2 m_t  \rule{0cm}{0.4cm}  }{M_W^2}   g_{VG^+G^-}    \Bigl(     C_{12}    - C_{22}   + C_{23}    \Bigr)    $               & $(p_{V},-p_t, M_{W}, M_{W}, m_q)$  \\[0.5cm]
         5      &  $\rule{0cm}{0.4cm}  -  i g^2 m_t P_d \Bigl\{    2 C_{12} - (D-2) \bigl[ C_{22} - C_{23}  \bigr]   \Bigr\}   $   &   $ (p_{V},-p_t, m_{q}, m_{q},M_W)  $      \\[0.5cm]      \rowcolor{Gray}
         6                                     &$\dfrac{ \rule{0cm}{0.4cm}   g^2 m_t  }{M_W}      g_{VG^+W^-}    \left(   C_0  + C_{11}  \right)   $                  & $ (p_{V},-p_t, M_{W}, M_{W},m_q)  $  \\[0.5cm]
                 8                                   &  $\rule{0cm}{0.4cm}   g^2  m_t  g_{VW^+W^-}   \Bigl\{  C_{11}  -C_0   -2 C_{12} - (D-2) \bigl[ C_{22} - C_{23} \bigr]        \Bigr\}     $                &  $(p_{V},-p_t, M_W, M_W, m_q) $   \\[0.3cm]
\hline\hline
\end{tabular}}
\end{center}
\end{table}

\begin{table}[t]
\begin{center}
\caption{\it \small Quark couplings with neutral vector bosons.}
\vspace{0.2cm}
\doublerulesep 0.7pt \tabcolsep 0.16in
\begin{tabular}{||c|c|c|c|c||}
\hline \hline  \rowcolor{RGray}
$V$ &   $A_{Vd}$ & $B_{Vd}$  &    $A_{Vu}$ & $B_{Vu}$     \\[0.1cm]
\hline
$\gamma$  &    $ i \dfrac{e}{3}$        &$0$ & $-i \dfrac{2 e \rule{0cm}{0.4cm}  }{3}$   &  $0$  \\[0.4cm]     \rowcolor{Gray}
$Z$ & $- i \dfrac{ g\rule{0cm}{0.4cm}    }{  c_W}    \Bigl(- \dfrac{1 }{4}  + \dfrac{1}{3} s_W^2   \Bigr) $ & $- i \dfrac{g}{4 c_W}     $ & $- i \dfrac{ g\rule{0cm}{0.4cm}    }{  c_W}    \Bigl( \dfrac{1}{4}  - \dfrac{2}{3} s_W^2   \Bigr) $ & $ i \dfrac{  g }{4 c_W}$   \\[0.4cm]
\hline \hline
\end{tabular}
\label{tab:legend}
\end{center}
\end{table}

\begin{table}[t]
\begin{center}
\caption{\it \small Cubic couplings of the neutral vector bosons.}
\vspace{0.2cm}
\doublerulesep 0.7pt \tabcolsep 0.25in
\begin{tabular}{||c|c|c|c|c||}
\hline  \hline\rowcolor{RGray}
$V$ &   $g_{VH^+H^-}$  &   $g_{VG^+G^-}$  &  $g_{VG^+W^-}$  &  $g_{VW^+W^-}$   \\[0.1cm]
\hline
$\gamma$  &    $e\rule{0cm}{0.4cm}  $   &  $e$ &   $-e M_W$  &   $e$ \\[0.2cm]     \rowcolor{Gray}
$Z$          &    $\rule{0cm}{0.4cm}  e \cot(2 \theta_W) $    &   $e \cot(2 \theta_W) $     &   $g s_W^2 M_Z $    &  $g c_W$ \\[0.2cm]
\hline \hline
\end{tabular}
\label{tab:legendII}
\end{center}
\end{table}

\end{appendix}


\begin{thebibliography}{99}

\bibitem{Aad:2012tfa}
  G.~Aad {\it et al.}  [ATLAS Collaboration],
  Phys.\ Lett.\ B {\bf 716} (2012) 1
  [arXiv:1207.7214 [hep-ex]].

\bibitem{Chatrchyan:2012ufa}
  S.~Chatrchyan {\it et al.}  [CMS Collaboration],
  Phys.\ Lett.\ B {\bf 716} (2012) 30
  [arXiv:1207.7235 [hep-ex]].

\bibitem{Glashow:1970gm}
  S.~L.~Glashow, J.~Iliopoulos and L.~Maiani,
  Phys.\ Rev.\ D {\bf 2} (1970) 1285.

\bibitem{Branco:2011iw}
  G.~C.~Branco, P.~M.~Ferreira, L.~Lavoura, M.~N.~Rebelo, M.~Sher and J.~P.~Silva,
  Phys.\ Rept.\  {\bf 516} (2012) 1
  [arXiv:1106.0034 [hep-ph]].

\bibitem{Glashow:1976nt}
  S.~L.~Glashow and S.~Weinberg,
  Phys.\ Rev.\ D {\bf 15} (1977) 1958.

\bibitem{Paschos:1976ay}
  E.~A.~Paschos,
  Phys.\ Rev.\ D {\bf 15} (1977) 1966.

\bibitem{Pich:2009sp}
  A.~Pich and P.~Tuzon,
  Phys.\ Rev.\ D {\bf 80} (2009) 091702
  [arXiv:0908.1554 [hep-ph]].

\bibitem{Serodio:2011hg}
  H.~Serodio,
  Phys.\ Lett.\ B {\bf 700} (2011) 133
  [arXiv:1104.2545 [hep-ph]].

\bibitem{Cree:2011uy}
  G.~Cree and H.~E.~Logan,
  Phys.\ Rev.\ D {\bf 84} (2011) 055021
  [arXiv:1106.4039 [hep-ph]].

\bibitem{Varzielas:2011jr}
  I.~de Medeiros Varzielas,
  Phys.\ Lett.\ B {\bf 701} (2011) 597
  [arXiv:1104.2601 [hep-ph]].

\bibitem{Celis:2014zaa}
  A.~Celis, J.~Fuentes-Martin and H.~Serodio,
  Phys.\ Lett.\ B {\bf 737} (2014) 185
  [arXiv:1407.0971 [hep-ph]].

\bibitem{Botella:2015yfa}
  F.~J.~Botella, G.~C.~Branco, A.~M.~Coutinho, M.~N.~Rebelo and J.~I.~Silva-Marcos,
  arXiv:1501.07435 [hep-ph].

\bibitem{Jung:2010ik}
  M.~Jung, A.~Pich and P.~Tuzon,
  JHEP {\bf 1011} (2010) 003
  [arXiv:1006.0470 [hep-ph]].

\bibitem{Jung:2012vu}
  M.~Jung, X.~-Q.~Li and A.~Pich,
  JHEP {\bf 1210} (2012) 063
  [arXiv:1208.1251 [hep-ph]].

\bibitem{Celis:2012dk}
  A.~Celis, M.~Jung, X.~-Q.~Li and A.~Pich,
  JHEP {\bf 1301} (2013) 054
  [arXiv:1210.8443 [hep-ph]].

\bibitem{Jung:2013hka}
  M.~Jung and A.~Pich,
  JHEP {\bf 1404} (2014) 076
  [arXiv:1308.6283 [hep-ph]].

\bibitem{Li:2014fea}
  X.~-Q.~Li, J.~Lu and A.~Pich,
  JHEP {\bf 1406} (2014) 022
  [arXiv:1404.5865 [hep-ph]].

\bibitem{Dekens:2014jka}
  W.~Dekens  {\it et al.},
  JHEP {\bf 1407} (2014) 069
  [arXiv:1404.6082 [hep-ph]].

\bibitem{Altmannshofer:2012ar}
  W.~Altmannshofer, S.~Gori and G.~D.~Kribs,
  Phys.\ Rev.\ D {\bf 86} (2012) 115009
  [arXiv:1210.2465 [hep-ph]].

\bibitem{Bai:2012ex}
  Y.~Bai, V.~Barger, L.~L.~Everett and G.~Shaughnessy,
  Phys.\ Rev.\ D {\bf 87} (2013) 115013
  [arXiv:1210.4922 [hep-ph]].

\bibitem{Celis:2013rcs}
  A.~Celis, V.~Ilisie and A.~Pich,
  JHEP {\bf 1307} (2013) 053
  [arXiv:1302.4022 [hep-ph]].

\bibitem{Barger:2013ofa}
  V.~Barger, L.~L.~Everett, H.~E.~Logan and G.~Shaughnessy,
  Phys.\ Rev.\ D {\bf 88} (2013) 11,  115003
  [arXiv:1308.0052 [hep-ph]].

\bibitem{Lopez-Val:2013yba}
  D.~Lopez-Val, T.~Plehn and M.~Rauch,
  JHEP {\bf 1310} (2013) 134
  [arXiv:1308.1979 [hep-ph]].

\bibitem{Duarte:2013zfa}
  L.~Duarte, G.~A.~Gonzalez-Sprinberg and J.~Vidal,
  JHEP {\bf 1311} (2013) 114
  [arXiv:1308.3652 [hep-ph]].

\bibitem{Celis:2013ixa}
  A.~Celis, V.~Ilisie and A.~Pich,
  JHEP {\bf 1312} (2013) 095
  [arXiv:1310.7941 [hep-ph]].

\bibitem{Wang:2013sha}
  L.~Wang and X.~F.~Han,
  JHEP {\bf 1404} (2014) 128
  [arXiv:1312.4759 [hep-ph]].

\bibitem{Eilam:1989zm}
  G.~Eilam, B.~Haeri and A.~Soni,
  Phys.\ Rev.\ D {\bf 41} (1990) 875.

\bibitem{DiazCruz:1989ub}
  J.~L.~Diaz-Cruz, R.~Martinez, M.~A.~Perez and A.~Rosado,
  Phys.\ Rev.\ D {\bf 41} (1990) 891.

\bibitem{Eilam:1990zc}
  G.~Eilam, J.~L.~Hewett and A.~Soni,
  Phys.\ Rev.\ D {\bf 44} (1991) 1473
  [Erratum-ibid.\ D {\bf 59} (1999) 039901].

\bibitem{Hou:1991un}
  W.~-S.~Hou,
  Phys.\ Lett.\ B {\bf 296} (1992) 179.

\bibitem{Atwood:1996vj}
  D.~Atwood, L.~Reina and A.~Soni,
  Phys.\ Rev.\ D {\bf 55} (1997) 3156
  [hep-ph/9609279].

\bibitem{Mele:1998ag}
  B.~Mele, S.~Petrarca and A.~Soddu,
  Phys.\ Lett.\ B {\bf 435} (1998) 401
  [hep-ph/9805498].

\bibitem{AguilarSaavedra:2004wm}
  J.~A.~Aguilar-Saavedra,
  Acta Phys.\ Polon.\ B {\bf 35} (2004) 2695
  [hep-ph/0409342].

\bibitem{Arhrib:2005nx}
  A.~Arhrib,
  Phys.\ Rev.\ D {\bf 72} (2005) 075016
  [hep-ph/0510107].

\bibitem{Bejar:2000ub}
  S.~Bejar, J.~Guasch and J.~Sola,
  Nucl.\ Phys.\ B {\bf 600} (2001) 21
  [hep-ph/0011091].

\bibitem{Aad:2012ij}
  G.~Aad {\it et al.}  [ATLAS Collaboration],
  JHEP {\bf 1209} (2012) 139
  [arXiv:1206.0257 [hep-ex]].

\bibitem{Aad:2014dya}
  G.~Aad {\it et al.}  [ATLAS Collaboration],
  JHEP {\bf 1406} (2014) 008
  [arXiv:1403.6293 [hep-ex]].

\bibitem{Chatrchyan:2013nwa}
  S.~Chatrchyan {\it et al.}  [CMS Collaboration],
  Phys.\ Rev.\ Lett.\  {\bf 112} (2014) 17,  171802
  [arXiv:1312.4194 [hep-ex]].

\bibitem{CMS:2014hwa}
  CMS Collaboration,
  CMS-PAS-TOP-14-003.

\bibitem{CMS:2014qxa}
  CMS Collaboration,
  CMS-PAS-HIG-13-034.

\bibitem{Goldouzian:2014xfa}
  R.~Goldouzian [CDF and D0 and ATLAS and CMS Collaborations],
  arXiv:1412.2524 [hep-ex].

\bibitem{AguilarSaavedra:2000aj}
  J.~A.~Aguilar-Saavedra and G.~C.~Branco,
  Phys.\ Lett.\ B {\bf 495} (2000) 347
  [hep-ph/0004190].

\bibitem{Agashe:2013hma}
  K.~Agashe {\it et al.}  [Top Quark Working Group Collaboration],
  arXiv:1311.2028 [hep-ph].

\bibitem{O'Neil:2009nr}
  D.~O'Neil,
  arXiv:0908.1363 [hep-ph].

\bibitem{Asner:2013psa}
  D.~M.~Asner {\it et al.},
  arXiv:1310.0763 [hep-ph].

\bibitem{Cabibbo:1963yz}
  N.~Cabibbo,
  Phys.\ Rev.\ Lett.\  {\bf 10} (1963) 531.

\bibitem{Kobayashi:1973fv}
  M.~Kobayashi and T.~Maskawa,
  Prog.\ Theor.\ Phys.\  {\bf 49} (1973) 652.

\bibitem{Ferreira:2010xe}
  P.~M.~Ferreira, L.~Lavoura and J.~P.~Silva,
  Phys.\ Lett.\ B {\bf 688} (2010) 341
  [arXiv:1001.2561 [hep-ph]].

\bibitem{Braeuninger:2010td}
  C.~B.~Braeuninger, A.~Ibarra and C.~Simonetto,
  Phys.\ Lett.\ B {\bf 692} (2010) 189
  [arXiv:1005.5706 [hep-ph]].

\bibitem{Bijnens:2011gd}
  J.~Bijnens, J.~Lu and J.~Rathsman,
  JHEP {\bf 1205} (2012) 118
  [arXiv:1111.5760 [hep-ph]].

\bibitem{Cvetic:1997zd}
  G.~Cvetic, S.~S.~Hwang and C.~S.~Kim,
  Int.\ J.\ Mod.\ Phys.\ A {\bf 14} (1999) 769
  [hep-ph/9706323].

\bibitem{Cvetic:1998uw}
  G.~Cvetic, C.~S.~Kim and S.~S.~Hwang,
  Phys.\ Rev.\ D {\bf 58} (1998) 116003
  [hep-ph/9806282].

\bibitem{Haeri:1988jt}
  B.~Haeri, A.~Soni and G.~Eilam,
  Phys.\ Rev.\ Lett.\  {\bf 62} (1989) 719.

\bibitem{Agashe:2014kda}
  K.~A.~Olive {\it et al.}  [Particle Data Group Collaboration],
  Chin.\ Phys.\ C {\bf 38} (2014) 090001.

\bibitem{Charles:2015gya}
  J.~Charles, O.~Deschamps, S.~Descotes-Genon, H.~Lacker, A.~Menzel, S.~Monteil, V.~Niess and J.~Ocariz {\it et al.},
  Phys.\ Rev.\ D {\bf 91} (2015) 7,  073007
  [arXiv:1501.05013 [hep-ph]].

\bibitem{Aoki:2013ldr}
  S.~Aoki {\it et al.},
  Eur.\ Phys.\ J.\ C {\bf 74} (2014) 9,  2890
  [arXiv:1310.8555 [hep-lat]].

\bibitem{Alberti:2014yda}
  A.~Alberti, P.~Gambino, K.~J.~Healey and S.~Nandi,
  Phys.\ Rev.\ Lett.\  {\bf 114} (2015) 6,  061802
  [arXiv:1411.6560 [hep-ph]].

\bibitem{Gambino:2013rza}
  P.~Gambino and C.~Schwanda,
  Phys.\ Rev.\ D {\bf 89} (2014) 1,  014022
  [arXiv:1307.4551 [hep-ph]].

\bibitem{Ricciardi:2014aya}
  G.~Ricciardi,
  arXiv:1412.4288 [hep-ph].

\bibitem{Aad:2014aba}
  G.~Aad {\it et al.}  [ATLAS Collaboration],
  Phys.\ Rev.\ D {\bf 90} (2014) 5,  052004
  [arXiv:1406.3827 [hep-ex]].

\bibitem{Khachatryan:2014jba}
  V.~Khachatryan {\it et al.}  [CMS Collaboration],
  arXiv:1412.8662 [hep-ex].

\bibitem{ATLAS:2014wva}
  ATLAS and CDF and CMS and D0 Collaborations,
  arXiv:1403.4427 [hep-ex].

\bibitem{Degrassi:2014sxa}
  G.~Degrassi, P.~Gambino and P.~P.~Giardino,
  arXiv:1411.7040 [hep-ph].

\bibitem{Hoang:2014oea}
  A.~H.~Hoang,
  arXiv:1412.3649 [hep-ph].

\bibitem{Abbiendi:2013hk}
  G.~Abbiendi {\it et al.}  [ALEPH and DELPHI and L3 and OPAL and LEP Collaborations],
  Eur.\ Phys.\ J.\ C {\bf 73} (2013) 2463
  [arXiv:1301.6065 [hep-ex]].

\bibitem{Aad:2013hla}
  G.~Aad {\it et al.}  [ATLAS Collaboration],
  Eur.\ Phys.\ J.\ C {\bf 73} (2013) 6,  2465
  [arXiv:1302.3694 [hep-ex]].

\bibitem{Aad:2014kga}
  G.~Aad {\it et al.}  [ATLAS Collaboration],
  JHEP {\bf 1503} (2015) 088
  [arXiv:1412.6663 [hep-ex]].

\bibitem{CMS:2014cdp}
  CMS Collaboration [CMS Collaboration],
  CMS-PAS-HIG-14-020.

\bibitem{CMS:2014kga}
  CMS Collaboration [CMS Collaboration],
  CMS-PAS-HIG-13-035.

\bibitem{Gutierrez:2010zz}
  P.~Gutierrez [CDF and D0 Collaborations],
  PoS CHARGED {\bf 2010} (2010) 004.

\bibitem{ATLASr}
  ATLAS Collaboration,
  ATLAS-CONF-2015-007.

\bibitem{Durieux:2014xla}
  G.~Durieux, F.~Maltoni and C.~Zhang,
  Phys.\ Rev.\ D {\bf 91} (2015) 7,  074017
  [arXiv:1412.7166 [hep-ph]].

\bibitem{Greljo:2014dka}
  A.~Greljo, J.~F.~Kamenik and J.~Kopp,
  JHEP {\bf 1407} (2014) 046
  [arXiv:1404.1278 [hep-ph]].

\bibitem{'tHooft:1978xw}
  G.~'t Hooft and M.~J.~G.~Veltman,
  Nucl.\ Phys.\ B {\bf 153} (1979) 365.

\bibitem{Passarino:1978jh}
  G.~Passarino and M.~J.~G.~Veltman,
  Nucl.\ Phys.\ B {\bf 160} (1979) 151.

\bibitem{Axelrod:1982yc}
  A.~Axelrod,
  Nucl.\ Phys.\ B {\bf 209} (1982) 349.

\end{thebibliography}
\end{document}